\documentclass[nolinenumbers]{aastex631}
\usepackage{dcolumn}
\usepackage{booktabs}
\usepackage{natbib}  
\usepackage{stix}
\usepackage{threeparttable}
\usepackage{multirow}
\usepackage{graphicx}
\usepackage{subfigure}
\newcolumntype{d}[1]{D{.}{.}{#1}}

\begin{document}

\title{Photometric Analysis of 30 Contact Binaries in M31}

\author[0009-0009-6364-0391]{Xiang Gao}
\author[0000-0003-3590-335X]{Kai Li}
\author[0009-0005-0485-418X]{Li-Heng Wang}
\affiliation{Shandong Key Laboratory of Optical Astronomy and Solar-Terrestrial Environment, School of Space Science and Technology, Institute of Space Sciences, 
Shandong University, Weihai, Shandong 264209, People's Republic of China; kaili@sdu.edu.cn}
\author[0000-0003-2471-2363]{Hai-bo Yuan}
\affiliation{Institute for Frontiers in Astronomy and Astrophysics,
Beijing Normal University, Beijing, 102206, China}
\affiliation{School of Physics and Astronomy, Beijing Normal University No.19, Xinjiekouwai St, Haidian District, Beijing, 100875, China}
\author[0009-0007-5610-6495]{Hong-rui Gu}
\affiliation{CAS Key Laboratory of Optical Astronomy, National Astronomical Observatories, Chinese Academy of Sciences, Beijing 100101, People's Republic of China}
\affiliation{School of Astronomy and Space Science, University of Chinese Academy of Sciences, Beijing 100049, People's Republic of China}
\author[0009-0008-3792-6444]{Ya-Ni Guo}
\affiliation{Shandong Key Laboratory of Optical Astronomy and Solar-Terrestrial Environment, School of Space Science and Technology, Institute of Space Sciences, 
Shandong University, Weihai, Shandong 264209, People's Republic of China; kaili@sdu.edu.cn}

\begin{abstract}
M31, as the largest galaxy in the Local Group, is of significant importance for the study of stellar formation and evolution. 
Based on the data of 5,859 targets observed in M31 by Gu et al (\citeyear{2024ApJS..273....9G}), we selected 30 contact binaries by visual inspection for further study. 
Using the PHOEBE software and employing Bayesian optimization and Markov Chain Monte Carlo sampling, we determined the physical parameters of these 30 systems. 
The results show that 10 systems exhibit the O'Connell effect, which is well explained by introducing a dark spot on the primary star. 
11 systems have mass ratios below 0.15, classifying them as extremely low mass ratio contact binaries, making them promising candidates for binary mergers. 
Six systems have primary star temperatures exceeding 10,000 K, classifying them as early-type contact binaries. 
The absolute physical parameters reveal that two contact binary systems contain massive stellar components, which may eventually evolve into compact binary star systems. 
To compare the effects of different galactic environments on the evolution of contact binaries, we constructed evolutionary diagrams for these 30 targets and for contact binaries in the Milky Way. 
The results show that, at the same mass, our targets have larger radii, higher luminosities, and lower orbital angular momenta than contact binaries in the Milky Way, indicating that they are at more advanced evolutionary stages. 
This may be attributed to the higher metallicity in M31 compared to the Milky Way.

\end{abstract}


\section{Introduction} \label{sec:intro}

Eclipsing binaries (EBs) provide a valuable means to determine fundamental stellar parameters, such as mass, radius and luminosity, through the analysis of their light curves and radial velocity curves (\citealp{2021MNRAS.501.2897G}; \citealp{2021NewA...8601565N}; \citealp{2006AJ....132..960H}). 
Among EBs, contact binaries are the most numerous. 
Based on the variable star catalog of the All-Sky Automated Survey for Supernovae (ASAS-SN; \citealp{2023MNRAS.519.5271C}), we found that the frequency of contact binaries accounts for approximately 59\% of EBs. 
Due to the presence of the common envelope, contact binaries show interaction between both primary and secondary components. This leads to many interesting physical phenomena, such as the fact that, although the two components of a contact binary often have significantly different masses, their effective temperatures are generally similar. 
The study of contact binaries will enhances our understandaning of the evolution of binary.

The Local Group is primarily composed of the Milky Way and the Andromeda galaxy (M31), which account for the majority of the mass in the Local Group \citep{2008MNRAS.384.1459L}. Therefore, as the nearest spiral galaxy to us, M31 is an ideal astrophysical laboratory, enabling us to explore various variable stars, including contact binary stars, at almost equal distances. So far, the majority of research on contact binaries has been focused within the Milky Way. 
This is because the study of contact binaries in M31 and other nearby galaxies is limited by photometric precision and the space resolution of dense stellar fields. Although projects such as the Local Group Galaxy Survey (LGGS; \citealp{2016AJ....152...62M}) and the Pan-STARRS1 Surveys (PS1; \citealp{2018AAS...23110201C}) have achieved breakthrough data in M31, the efficiency and accuracy of identifying contact binaries remain lower compared to that in the Milky Way. 
On the other hand, the identification proportion of contact binaries in M31 is lower than that in the Milky Way. 
For instance, Lee et al. (\citeyear{2014ApJ...797...22L}) identified 298 EBs using Pan-STARRS1 data, revealing that the number of contact binaries (22) was fewer than detached (120) and semi-detached (158) systems. 
This may be attributed to the combined effects of the intrinsic faintness of the contact binaries and the remoteness of M31. 
To date, only one contact binary system in M31, namely M31V J00450522+4138462, has been investigated through detailed light curve modeling (\citealp{2022ApJ...932...14L}). This system was found to be a deep-contact binary, with a mass ratio close to unity and a fill-out factor of 88\%, providing valuable insights into the formation and evolution of extragalactic contact binaries. In summary, studies of contact binaries in M31 are currently very limited.

Currently, as by-products of M31 microlensing searches, a large number of variable stars have been discovered, many of which have long periods, such as 100 days. For example, the POINT-AGAPE survey (\citealp{2004MNRAS.351.1071A}) and the WeCAPP project (\citealp{2006A&A...445..423F}) have both identified tens of thousands of variable stars. 
However, the search for short-period variables (period $\leq$ 1 day) is limited by non-uniform sampling and frequency aliasing effects. 
Contact binaries generally belong to short-period variable stars, so they are more difficult to be discovered in these projects.

Observations and studies of contact binaries in M31 will help verify the influence of different environmental factors on stellar evolution compared to the Milky Way. The metallicity gradient and star formation history of M31 differ from those of the Milky Way, and comparative studies can reveal how galactic environments regulate the evolutionary pathways of binary systems. 
Furthermore, the analyses of eclipsing binaries in M31 can be used to determine the galaxy’s distance, providing direct observational constraints for the Hubble constant (\citealp{2016ApJ...826...56R}). In the future, with the data releases from deep-space observations and wide-field survey facilities such as the James Webb Space Telescope (JWST; \citealp{2023PASP..135e8002M}) and the Legacy Survey of Space and Time (LSST; \citealp{2009arXiv0912.0201L}; \citealp{2019ApJ...873..111I}), studies of extragalactic contact binaries will become more extensive.

The structure of this paper is as follows: Section 2 provides an overview of our target selection process and the treatment of light curves. Section 3 describes the investigation of the light curves. Section 4 presents the results, calculating the absolute physical parameters and discussing the evolutionary states of these targets, along with comparisons to systems in the Milky Way. Finally, the research was summarized.

\section{Data Selection and Reducing} \label{sec:data}
Gu et al.(\citeyear{2024ApJS..273....9G}) utilized the Canada-France-Hawaii Telescope (CFHT) and the MegaCAM camera to conduct high-cadence observations of M31 over three consecutive nights. 
This program was originally designed to detect short-duration microlensing events caused by free-floating planets. 
The observations by MegaCAM consisted of consecutive 5-minute exposures in CFHT $g$ and $r$ bands, ultimately yielding 231 exposures. 
This dataset features both a high sampling frequency and a relatively long time span, helping to mitigate selection effects in identifying short-period variable stars. 
Ultimately, Gu et al. (\citeyear{2024ApJS..273....9G}) provided a catalog of 5,859 variable sources in M31, along with their photometric data.

Based on their published periods and photometric data, we obtained their light curves and visually inspected them to identify contact binaries.
Gu et al. (\citeyear{2024ApJS..273....9G}) provided periods for each target using the Phase Dispersion Minimization (PDM; \citealp{1978ApJ...224..953S}) algorithm. In addition, we recalculated the periods of all targets using the Generalized Lomb-Scargle Periodogram (GLS; \citealp{2018ascl.soft07019Z}) method. 
From their catalog, we identified contact binary candidates and selected 30 systems for further analysis.

We initially adopted the magnitude minimum as the zero point for phase-folding light curves. Using the periods calculated by us, we folded the light curves for all targets over both single and double periods. The double-period folding was specifically employed to prevent misclassification caused by the symmetric nature of contact binary light curves.
Through visual inspection, we initially identified 36 contact binaries. After excluding targets with less than 70\% phase coverage in their light curves and those missing eclipses, the final number of contact binaries was reduced to 30. 
Due to the sparse and dispersion of observations, we adopted the period shift method proposed by Li et al. (\citeyear{2020AJ....159..189L}) to determine the eclipsing times. 
Using the primary minimum  as zero point, we transformed the temporal data into phase data. 

Using the extinction coefficients in the g-band (0.205) and r-band (0.142) from  Schlafly \& Finkbeiner (\citeyear{2011ApJ...737..103S}), we calculated the g-r color index at the secondary minima for each target. By interpolating the color index with the Pecaut \& Mamajek's table (\citeyear{2013ApJS..208....9P}), we derived the effective temperatures of the primary components of all targets. 
The effective temperatures and other fundamental information of all 30 systems are tabulated in Table \ref{tab1: 30 targets information}. 
Subsequently, we converted the magnitude to flux and selected the median value of the flux around phase 0.25 as the normalization constant. The fluxes were then normalized using this constant to obtain the final light curves.

\begin{table}[htbp]
\caption{The information of 30 contact binaries in M31}
\label{tab1: 30 targets information}
\centering
\begin{threeparttable}
\begin{tabular}{lllcclcr}
\toprule
\text{Name} & \text{RA} & \text{DEC} & \text{g (mag)\textsuperscript{a}} & \text{r (mag)\textsuperscript{a}} & \text{Teff (K)} & \text{Period (day)\textsuperscript{b}} & \text{FitsID\textsuperscript{c}}\\
\midrule
M31\_CB\_1  & 10.36122  & 41.32866  & 22.39  & 21.88  & 5854  & 0.6030627  & 415  \\
M31\_CB\_2  & 10.30937  & 41.28408  & 22.14  & 22.06  & 7552  & 1.2515494  & 445  \\
M31\_CB\_3  & 10.19367  & 41.47958  & 19.77  & 19.00  & 5152  & 0.2430357  & 462  \\
M31\_CB\_4  & 10.25527  & 41.36872  & 22.45  & 22.14  & 6698  & 0.6384294  & 566  \\
M31\_CB\_5  & 11.07127  & 41.49139  & 21.11  & 21.18  & 8544  & 1.2708186  & 734  \\
M31\_CB\_6  & 11.27175  & 41.64620  & 20.73  & 20.66  & 7616  & 0.9177422  & 838  \\
M31\_CB\_7  & 11.24358  & 41.55811  & 21.75  & 21.37  & 6278  & 0.7090473  & 907  \\
M31\_CB\_8  & 10.93880  & 41.20141  & 22.66  & 21.84  & 5077  & 0.8886793  & 1226 \\
M31\_CB\_9	& 10.80530 	& 41.04921 	& 17.49  & 17.17  & 6642  & 0.2655759  & 1373 \\
M31\_CB\_10 & 10.78415  & 41.12773  & 21.85  & 22.40  & 33434 & 0.3520528  & 1442 \\
M31\_CB\_11	& 10.74698 	& 41.10799 	& 17.39  & 17.34  & 8070  &	0.2664163  & 1749 \\
M31\_CB\_12 & 10.58113  & 41.05935  & 22.08  & 21.51  & 5598  & 0.8924476  & 2000 \\
M31\_CB\_13 & 10.54441  & 41.26968  & 20.48  & 20.31  & 7212  & 0.3880803  & 2211 \\
M31\_CB\_14 & 10.27573  & 41.22203  & 21.10  & 21.30  & 10190 & 0.4459087  & 2627 \\
M31\_CB\_15 & 11.18023  & 41.72169  & 20.93  & 20.65  & 6783  & 0.3957422  & 2936 \\
M31\_CB\_16 & 11.08114  & 41.67401  & 20.25  & 19.80  & 5984  & 0.2698319  & 2955 \\
M31\_CB\_17 & 10.92796  & 40.82303  & 23.42  & 23.49  & 7609  & 0.5779466  & 3065 \\
M31\_CB\_18 & 10.57799  & 40.81091  & 21.31  & 21.49  & 9321  & 0.5688876  & 3478 \\
M31\_CB\_19 & 11.02212  & 41.28594  & 21.86  & 21.85  & 7850  & 0.3607007  & 3807 \\
M31\_CB\_20 & 10.32934  & 40.82152  & 21.58  & 21.52  & 7695  & 0.3608038  & 3822 \\
M31\_CB\_21 & 10.19456  & 40.97416  & 22.29  & 22.51  & 16033 & 0.9431018  & 3999 \\
M31\_CB\_22 & 11.03379  & 41.55100  & 21.81  & 21.66  & 7288  & 0.7937230  & 4311 \\
M31\_CB\_23 & 11.03586  & 41.55212  & 21.18  & 21.63  & 29102 & 0.3844181  & 4312 \\
M31\_CB\_24 & 10.99871  & 41.64080  & 21.11  & 21.21  & 8690  & 0.7160479  & 4400 \\
M31\_CB\_25 & 10.99755  & 41.61880  & 21.06  & 21.45  & 16377 & 0.4259067  & 4468 \\
M31\_CB\_26 & 10.69733  & 41.54861  & 21.58  & 21.81  & 10654 & 0.8030848  & 4673 \\
M31\_CB\_27 & 10.70587  & 41.69082  & 21.37  & 21.47  & 8740  & 0.8194765  & 4730 \\
M31\_CB\_28 & 10.71010  & 41.67691  & 22.92  & 22.77  & 7321  & 0.4985445  & 4766 \\
M31\_CB\_29 & 10.73234  & 41.67015  & 20.94  & 21.04  & 8740  & 0.8335401  & 4788 \\
M31\_CB\_30 & 10.93740  & 41.38453  & 21.61  & 21.70  & 7932  & 0.8071773  & 5446 \\
\bottomrule
\end{tabular}
\begin{tablenotes}
\raggedright
\item 
\textbf{Note}: 
\textsuperscript{a} Mean CFHT g and r magnitudes after $3\sigma$ clipping, from Gu et al. (\citeyear{2024ApJS..273....9G}); 
\textsuperscript{b} All the periods were calculated by us using the GLS method (\citealp{2018ascl.soft07019Z}); 
\textsuperscript{c} FitsID refers to the serial number of each FITS file in Gu et al. (\citeyear{2024ApJS..273....9G}).
\end{tablenotes}
\end{threeparttable}
\end{table}

\section{Light Curve Analysis} \label{sec:analysis}
PHysics Of Eclipsing BinariEs 2.4 (PHOEBE; A. Prša et al. \citeyear{2016ApJS..227...29P}; M. Horvat et al. \citeyear{2018ApJS..237...26H}; K. E. Conroy et al. \citeyear{2020ApJS..250...34C}; D. Jones et al. \citeyear{2020ApJS..247...63J}) was used to determine the physical parameters of 30 contact binaries. The contact configuration was adopted. The temperature of the primary components were adopted as the effective temperature derived from the color index in the previous section. 
Following the works of L. B. Lucy (\citeyear{1967ZA.....65...89L}), H. von Zeipel (\citeyear{1924MNRAS..84..665V}), and S. M. Ruciński (\citeyear{1973AcA....23...79R}), the gravity darkening and bolometric albedo coefficients were set to $g = 0.32$ and $A = 0.5$ respectively when the component's temperature is below 7200K. However, for the temperature exceeding 7200 K, we used $g = 1$ and $A = 1$.
The atmospheric model was chosen from F. Castelli \& R. L. Kurucz (\citeyear{2004A&A...419..725C}), and the limb-darkening coefficients were derived based on the logarithmic law.

\subsection{Initial Parameters Estimation of Contact Binaries} \label{subsec:phoebe}
For contact binary systems analyzed using PHOEBE, appropriate initial parameters are critical for subsequent Markov Chain Monte Carlo (MCMC; \citealp{gamerman2006markov}) calculation (\citealp{2024AJ....168..272P}; \citealp{2025MNRAS.537.3160P}). 
To efficiently determine reasonable initial parameters for modeling the light curves of our targets, we applied Bayesian optimization (\citealp{mockus1975bayesian}; \citealp{7352306}). 
Specifically, we defined the search spaces for key parameters such as mass ratio ($q$), secondary effective temperature ($T_2$), luminosity of primary component ($L_1$) and orbital inclination ($i$). 
Then, the loss function was defined as the sum of the negative log-likelihoods (NLL) of the residuals in both the g-band and r-band light curves. 
This approach allows for a robust evaluation of the model fit by quantifying the discrepancies between observed and synthetic light curves. 
Using the gp\_minimize algorithm from python package scikit-optimize\footnote{https://scikit-optimize.github.io/stable/}, which is based on Gaussian processes (\citealp{rasmussen2006gaussian}), the optimization iteratively predicts promising regions in parameter space and converges toward an initial solution that best reproduces the observed light curves.

To ensure multi-band consistency, the dataset-scaled mode (pblum\_mode = ‘‘dataset-scaled") is applied during light curve model initialization. 
The mass ratio $q$ is constrained within the interval [0, 1], the orbital inclination $i$ is restricted to the range $[30^\circ, 90^\circ]$, and the fill-out factor $f$ is limited to [0, 1], which are consistent with the common parameter ranges of contact binaries (\citealp{2025ApJS..277...51L}). 
The effective temperature of the secondary star $T_2$ is set to $[ T_1 - 1500 K, T_1 + 1500 K ]$ due to the approximately same depths of the two eclipsing minima. 
For targets exhibiting the O’Connell effect (unequal heights in two maxima of light curves; \citealp{1951PRCO....2...85O}; \citealp{2009SASS...28..107W}), a spot is added to the primary component, with the spot latitude fixed at $90^\circ$ to expedite spot parameter estimation. 
The spot longitude $\lambda$ is set within $[0^\circ, 360^\circ]$ to cover the complete surface. 
The spot radius $r_s$ is set to $[4^\circ, 60^\circ]$, and the spot temperature factor $T_s$ is set to [0.6, 1.4], which conform to the general settings for spots (\citealp{2025ApJS..277...51L}). 
Such settings balance physical constraints and the freedom of parameter exploration, facilitating the effective convergence of the Bayesian optimization algorithm with limited computational resources. 
Once the values of $q$, $i$, $T_2$, $f$, $\lambda$, $r_s$ and $T_s$ are obtained, the luminosity of the primary component $L_1$ can be automatically determined using PHOEBE.

This approach that uses Bayesian optimization avoids the inefficiency and instability of manual parameter guessing. 
Moreover, compared to the traditional one-dimensional q-search method, the multi-parameter search yields more reliable results. 
Following the framework outlined by Conroy et al. (\citeyear{2020ApJS..250...34C}), the parameter space in such systems is highly complex and non-orthogonal, with strong correlations between key parameters (e.g., mass ratio, inclination, fill-out factor, temperature). As indicated in their analysis, reliable solutions require exploring a large number of forward models across multiple parameters simultaneously. Therefore, we perform Bayesian optimization or sampling in multiple parameters rather than restricting the search to a single parameter such as the mass ratio $q$. 
After iterative optimization, the optimal parameters are obtained. 
The optimal values of these parameters will then be used as inputs for the subsequent MCMC calculations. 

After the modeling of PHOEBE, we found that the fitting results of some targets were not satisfactory, with significant differences between the theoretical curves and the observed curves of the two bands. Therefore, we introduced the third light for all targets to conduct refitting. By comparing and adopting the better goodness-of-fit, we used the third-light models results for 14 targets. 

\subsection{MCMC Calculation} \label{subsec:MCMC}
\begin{figure}[htbp]
    \centering 
    \includegraphics[width=0.618\textwidth]{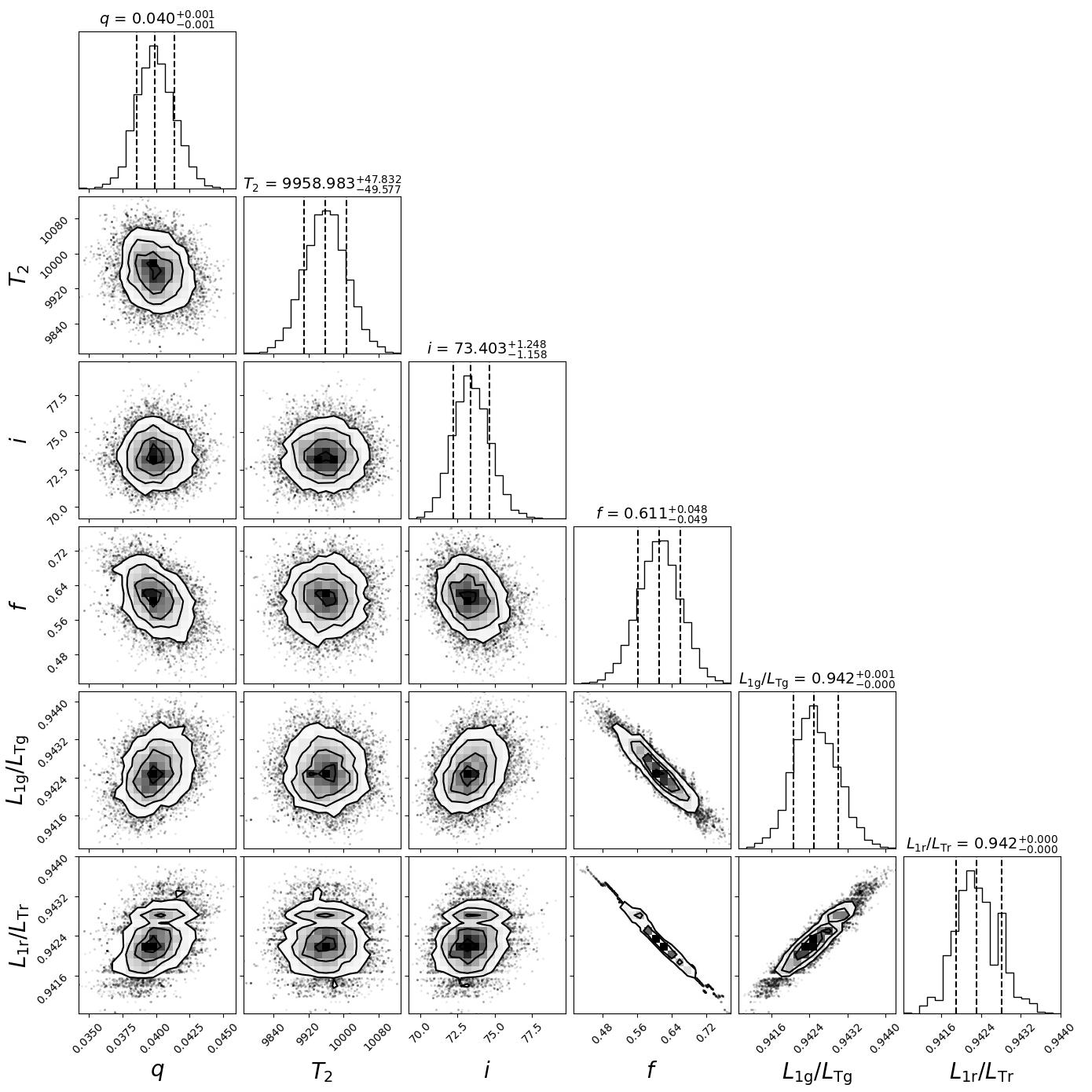}
    \caption{Probability distributions of $q$, $T_2$, $i$, $f$, $L_{1g}/L_{Tg}$ and $L_{1r}/L_{Tr}$ determined by the MCMC modeling of M31\_CB\_14.}
    \label{fig1}
\end{figure}

\begin{figure}[htbp]
    \centering 
    \includegraphics[width=0.95\textwidth]{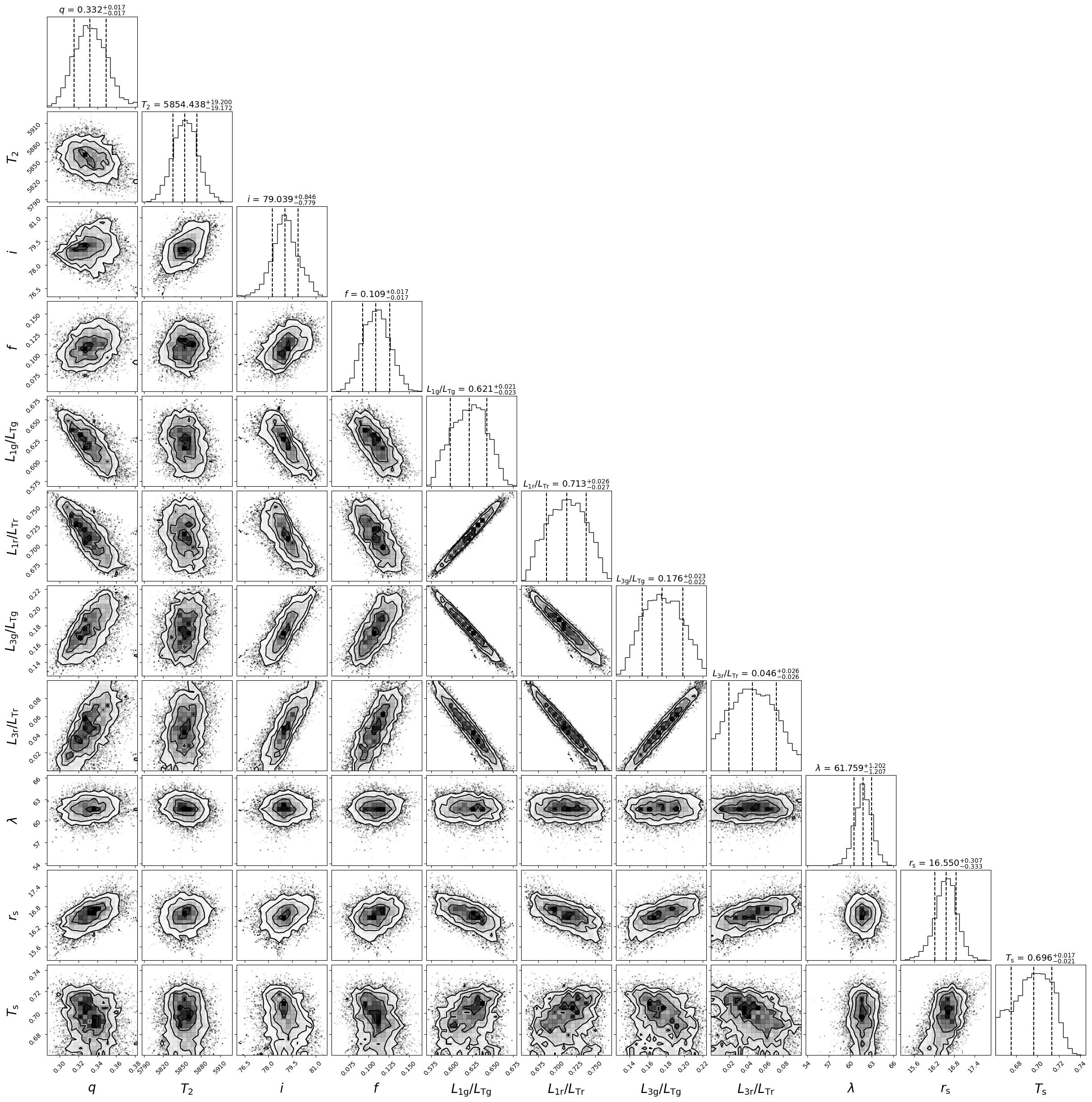}
    \caption{Probability distributions of $q$, $T_2$, $i$, $f$, $L_{1g}/L_{Tg}$, $L_{1r}/L_{Tr}$, $L_{3g}/L_{Tg}$, $L_{3r}/L_{Tr}$, $\lambda$, $r_s$ and $T_s$ determined by the MCMC modeling of M31\_CB\_16.}
    \label{fig2}
\end{figure}

\begin{figure}[htbp]
    \centering 
    \includegraphics[width=0.95\textwidth]{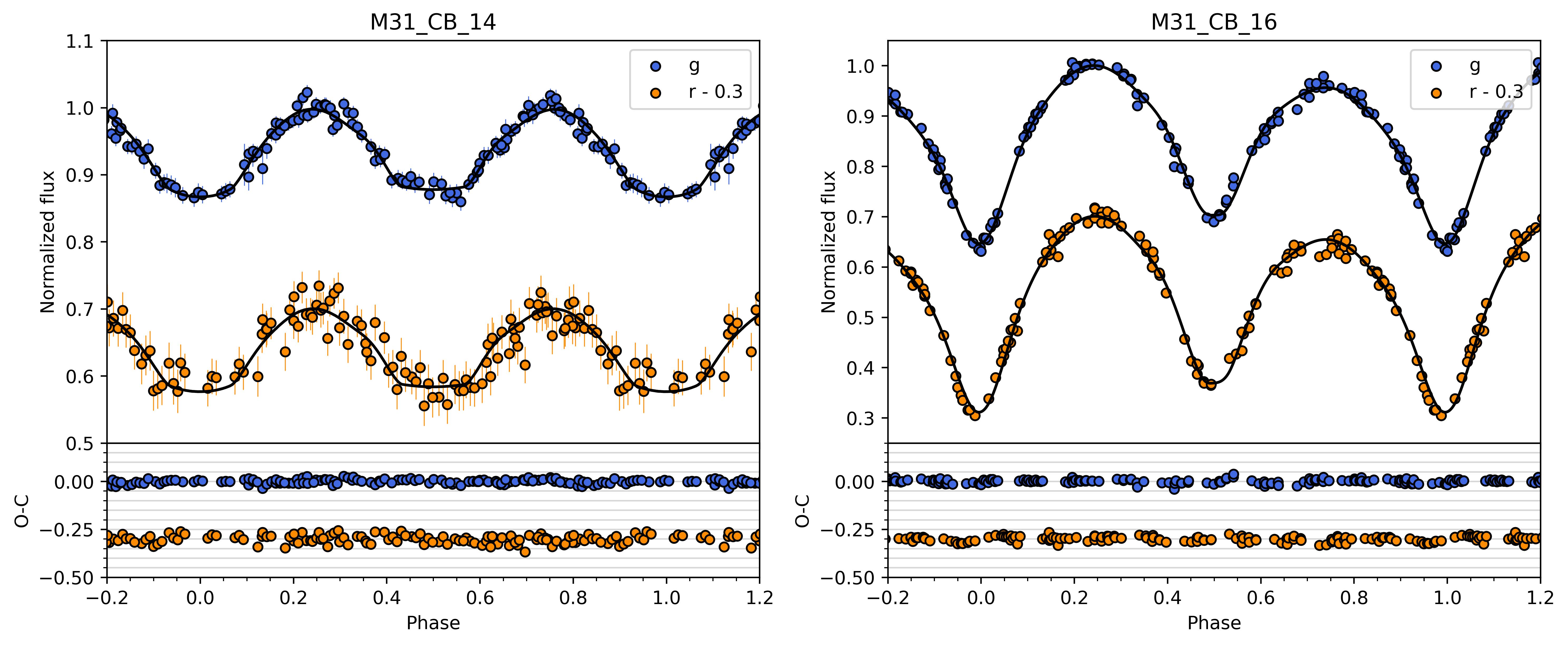}
    \caption{The fitting results and residuals of M31\_CB\_14 and M31\_CB\_16. In each panel, the upper shows the fitting results, with blue dots for g-band data and orange dots for r-band data, and the black curve indicates the fitting curve. The bottom of the panel shows the O-C residuals of the two bands respectively. The observational errors of M31\_CB\_16 are smaller than the size of the symbols.}
    \label{fig3}
\end{figure}

For MCMC sampling, Gaussian distributions are assigned to all parameters, with means set to the initial values. We configure $n_{walkers}=20$ and set the initial number of iterations as 2000. 
Following K. E. Conroy et al. 
(\citeyear{2020ApJS..250...34C}), convergence is ensured by requiring the iteration number for each parameter to exceed 10 times its autocorrelation time. 
For targets that do not converge within the initial 2000 iterations, the number of iterations is increased until convergence is achieved. 
Finally, we obtained the MCMC sampling results for 30 targets and generated their corresponding corner plots.
As an example, the posterior distributions of the physical parameters for two targets (one without spot and one with a spot and third light) are shown in Figures \ref{fig1} and \ref{fig2}, respectively. 
Figure \ref{fig3} shows the observed and fitted curves, and the O–C (observation minus calculation) residuals of the two targets (Other targets' figures similar to Figure \ref{fig3} are provided in the Figure \ref{figA1}). 
Based on the results of MCMC, the physical parameters of all targets are listed in Table \ref{tab2: 30 targets's parameters}. The primary star is defined as the more massive one regardless of the eclipse depth, this is the reason that all mass ratios in Table \ref{tab2: 30 targets's parameters} are less than 1. It should be mentioned that the errors associated with these physical parameters are all underestimated (\citealp{2005ApJ...628..426P}; \citealp{2016ApJS..227...29P}). 
Regarding the 14 targets with third light, we list their third-light ratios in Table \ref{tab2: 30 targets's parameters}, and provide a comparison of these 14 targets without and with the third light in Figure \ref{figA2}.

\begin{table}[htbp]
\caption{Physical Parameters of the 30 Targets}
\label{tab2: 30 targets's parameters}
\centering
\resizebox{\textwidth}{!}{
\begin{threeparttable}
\begin{tabular}{lcccccccccccc}
\toprule
\multirow{2}{*}{Parameter} & \multirow{2}{*}{$q$} & \multirow{2}{*}{$i$} & \multirow{2}{*}{$T_2$} & \multirow{2}{*}{$\Omega_1=\Omega_2$} & \multirow{2}{*}{$f$} & \multirow{2}{*}{$\lambda$} & \multirow{2}{*}{$r_s$} & \multirow{2}{*}{$T_s^{\text{a}}$} & \multirow{2}{*}{$r_1^{\text{b}}$} & \multirow{2}{*}{$r_2^{\text{b}}$} & \multirow{2}{*}{$L_{3g}/L_{Tg}$} & \multirow{2}{*}{$L_{3r}/L_{Tr}$} \\
 &  &  &  &  &  &  &  &  &  &  &  & \\
 &  & (deg) & (K) &  & (\%) & (deg) & (deg) &  &  &  &  & \\
\midrule
M31\_CB\_1  & $0.114^{+0.011}_{-0.010}$ & $70.0^{+0.7}_{-0.6}$ & $5681 ^{+149}_{-145}$ & $1.996^{+0.031}_{-0.031}$ & $7.4 ^{+1.9}_{-1.3}$ & $...^{   }_{   }$ & $...^{  }_{  }$ & $... ^{     }_{     }$ & $0.573^{+0.008}_{-0.007}$ & $0.216^{+0.006}_{-0.007}$ & $0.188^{+0.029}_{-0.025}$ & $0.532^{+0.020}_{-0.029}$ \\
M31\_CB\_2  & $0.826^{+0.050}_{-0.063}$ & $86.9^{+0.7}_{-0.9}$ & $6127 ^{+63 }_{-80 }$ & $3.433^{+0.082}_{-0.106}$ & $6.0 ^{+0.8}_{-0.5}$ & $269^{+1 }_{-1 }$ & $29 ^{+1}_{-1}$ & $0.65^{+0.02}_{-0.02}$ & $0.402^{+0.007}_{-0.005}$ & $0.369^{+0.005}_{-0.007}$ & $0.350^{+0.021}_{-0.018}$ & $0.482^{+0.018}_{-0.022}$ \\
M31\_CB\_3  & $0.772^{+0.009}_{-0.006}$ & $35.4^{+0.4}_{-0.4}$ & $4741 ^{+11 }_{-10 }$ & $3.155^{+0.003}_{-0.001}$ & $48.9^{+3.2}_{-3.0}$ & $272^{+1 }_{-1 }$ & $20 ^{+1}_{-1}$ & $0.91^{+0.01}_{-0.01}$ & $0.448^{+0.002}_{-0.002}$ & $0.403^{+0.005}_{-0.004}$ & $...  ^{      }_{      }$ & $...  ^{      }_{      }$ \\
M31\_CB\_4  & $0.094^{+0.005}_{-0.005}$ & $67.8^{+1.2}_{-1.1}$ & $7004 ^{+49 }_{-48 }$ & $1.916^{+0.012}_{-0.011}$ & $39.4^{+4.6}_{-4.9}$ & $...^{   }_{   }$ & $...^{  }_{  }$ & $... ^{     }_{     }$ & $0.596^{+0.002}_{-0.002}$ & $0.215^{+0.005}_{-0.005}$ & $...  ^{      }_{      }$ & $...  ^{      }_{      }$ \\
M31\_CB\_5  & $0.525^{+0.001}_{-0.001}$ & $83.4^{+0.1}_{-0.1}$ & $7982 ^{+3  }_{-4  }$ & $2.908^{+0.001}_{-0.001}$ & $5.0 ^{+0.1}_{-0.1}$ & $...^{   }_{   }$ & $...^{  }_{  }$ & $... ^{     }_{     }$ & $0.441^{+0.001}_{-0.001}$ & $0.329^{+0.001}_{-0.001}$ & $0.720^{+0.005}_{-0.008}$ & $0.800^{+0.009}_{-0.006}$ \\
M31\_CB\_6  & $0.785^{+0.017}_{-0.014}$ & $85.3^{+0.3}_{-0.3}$ & $7104 ^{+33 }_{-27 }$ & $3.135^{+0.025}_{-0.020}$ & $57.8^{+1.4}_{-1.2}$ & $311^{+5 }_{-4 }$ & $17 ^{+0}_{-0}$ & $0.63^{+0.03}_{-0.02}$ & $0.456^{+0.002}_{-0.002}$ & $0.414^{+0.002}_{-0.002}$ & $0.450^{+0.004}_{-0.004}$ & $0.522^{+0.005}_{-0.004}$ \\
M31\_CB\_7  & $0.145^{+0.003}_{-0.005}$ & $70.6^{+0.5}_{-0.6}$ & $6462 ^{+15 }_{-17 }$ & $2.084^{+0.009}_{-0.013}$ & $5.4 ^{+0.2}_{-0.3}$ & $...^{   }_{   }$ & $...^{  }_{  }$ & $... ^{     }_{     }$ & $0.553^{+0.003}_{-0.002}$ & $0.231^{+0.001}_{-0.002}$ & $0.211^{+0.017}_{-0.016}$ & $0.580^{+0.007}_{-0.010}$ \\
M31\_CB\_8  & $0.385^{+0.063}_{-0.052}$ & $75.3^{+1.8}_{-2.0}$ & $4799 ^{+46 }_{-46 }$ & $2.628^{+0.123}_{-0.107}$ & $7.8 ^{+1.5}_{-1.7}$ & $...^{   }_{   }$ & $...^{  }_{  }$ & $... ^{     }_{     }$ & $0.470^{+0.013}_{-0.013}$ & $0.305^{+0.013}_{-0.012}$ & $0.296^{+0.033}_{-0.026}$ & $0.583^{+0.010}_{-0.012}$ \\
M31\_CB\_9  & $0.070^{+0.000}_{-0.000}$ & $68.7^{+0.0}_{-0.1}$ & $6699 ^{+6  }_{-7  }$ & $1.835^{+0.000}_{-0.000}$ & $57.4^{+0.7}_{-0.6}$ & $230^{+0 }_{-0 }$ & $17 ^{+0}_{-0}$ & $0.99^{+0.00}_{-0.00}$ & $0.620^{+0.000}_{-0.000}$ & $0.202^{+0.000}_{-0.000}$ & $...  ^{      }_{      }$ & $...  ^{      }_{      }$ \\
M31\_CB\_10 & $0.785^{+0.010}_{-0.010}$ & $39.5^{+1.2}_{-1.2}$ & $33073^{+48 }_{-48 }$ & $3.236^{+0.005}_{-0.006}$ & $34.9^{+4.8}_{-4.6}$ & $...^{   }_{   }$ & $...^{  }_{  }$ & $... ^{     }_{     }$ & $0.433^{+0.004}_{-0.003}$ & $0.391^{+0.006}_{-0.006}$ & $...  ^{      }_{      }$ & $...  ^{      }_{      }$ \\
M31\_CB\_11 & $0.583^{+0.000}_{-0.001}$ & $73.9^{+0.1}_{-0.0}$ & $7428 ^{+18 }_{-10 }$ & $2.991^{+0.000}_{-0.000}$ & $12.3^{+0.1}_{-0.6}$ & $...^{   }_{   }$ & $...^{  }_{  }$ & $... ^{     }_{     }$ & $0.437^{+0.000}_{-0.000}$ & $0.343^{+0.000}_{-0.001}$ & $...  ^{      }_{      }$ & $...  ^{      }_{      }$ \\
M31\_CB\_12 & $0.144^{+0.020}_{-0.017}$ & $69.3^{+1.1}_{-0.9}$ & $5743 ^{+71 }_{-71 }$ & $2.081^{+0.052}_{-0.046}$ & $6.5 ^{+0.9}_{-0.8}$ & $...^{   }_{   }$ & $...^{  }_{  }$ & $... ^{     }_{     }$ & $0.553^{+0.010}_{-0.010}$ & $0.231^{+0.008}_{-0.008}$ & $0.105^{+0.030}_{-0.042}$ & $0.569^{+0.016}_{-0.020}$ \\
M31\_CB\_13 & $0.028^{+0.001}_{-0.000}$ & $57.5^{+0.5}_{-0.5}$ & $6196 ^{+46 }_{-50 }$ & $1.690^{+0.003}_{-0.001}$ & $24.4^{+4.9}_{-3.4}$ & $...^{   }_{   }$ & $...^{  }_{  }$ & $... ^{     }_{     }$ & $0.671^{+0.000}_{-0.001}$ & $0.143^{+0.002}_{-0.001}$ & $...  ^{      }_{      }$ & $...  ^{      }_{      }$ \\
M31\_CB\_14 & $0.040^{+0.001}_{-0.001}$ & $73.4^{+1.3}_{-1.2}$ & $9959 ^{+48 }_{-50 }$ & $1.732^{+0.004}_{-0.004}$ & $61.1^{+4.9}_{-5.0}$ & $...^{   }_{   }$ & $...^{  }_{  }$ & $... ^{     }_{     }$ & $0.655^{+0.001}_{-0.001}$ & $0.169^{+0.003}_{-0.003}$ & $...  ^{      }_{      }$ & $...  ^{      }_{      }$ \\
M31\_CB\_15 & $0.077^{+0.005}_{-0.004}$ & $60.1^{+0.6}_{-0.6}$ & $6409 ^{+41 }_{-44 }$ & $1.871^{+0.013}_{-0.012}$ & $28.1^{+4.3}_{-4.7}$ & $85 ^{+1 }_{-1 }$ & $22 ^{+1}_{-1}$ & $0.88^{+0.01}_{-0.02}$ & $0.607^{+0.003}_{-0.003}$ & $0.199^{+0.005}_{-0.005}$ & $...  ^{      }_{      }$ & $...  ^{      }_{      }$ \\
M31\_CB\_16 & $0.332^{+0.017}_{-0.017}$ & $79.0^{+0.8}_{-0.8}$ & $5854 ^{+19 }_{-19 }$ & $2.514^{+0.020}_{-0.018}$ & $10.9^{+1.7}_{-1.7}$ & $62 ^{+1 }_{-1 }$ & $17 ^{+0}_{-0}$ & $0.70^{+0.02}_{-0.02}$ & $0.485^{+0.002}_{-0.002}$ & $0.295^{+0.002}_{-0.002}$ & $0.176^{+0.023}_{-0.022}$ & $0.047^{+0.026}_{-0.026}$ \\
M31\_CB\_17 & $0.339^{+0.001}_{-0.001}$ & $89.9^{+0.1}_{-0.1}$ & $8796 ^{+4  }_{-4  }$ & $2.371^{+0.001}_{-0.001}$ & $85.8^{+0.1}_{-0.1}$ & $...^{   }_{   }$ & $...^{  }_{  }$ & $... ^{     }_{     }$ & $0.531^{+0.001}_{-0.001}$ & $0.350^{+0.001}_{-0.001}$ & $0.000^{+0.000}_{-0.000}$ & $0.246^{+0.001}_{-0.001}$ \\
M31\_CB\_18 & $0.229^{+0.007}_{-0.007}$ & $59.3^{+0.7}_{-0.6}$ & $8213 ^{+46 }_{-46 }$ & $2.194^{+0.007}_{-0.007}$ & $76.1^{+4.7}_{-4.5}$ & $...^{   }_{   }$ & $...^{  }_{  }$ & $... ^{     }_{     }$ & $0.549^{+0.001}_{-0.000}$ & $0.304^{+0.006}_{-0.006}$ & $...  ^{      }_{      }$ & $...  ^{      }_{      }$ \\
M31\_CB\_19 & $0.067^{+0.003}_{-0.003}$ & $61.5^{+0.7}_{-0.8}$ & $8343 ^{+47 }_{-47 }$ & $1.824^{+0.008}_{-0.009}$ & $60.9^{+4.7}_{-5.0}$ & $...^{   }_{   }$ & $...^{  }_{  }$ & $... ^{     }_{     }$ & $0.624^{+0.002}_{-0.002}$ & $0.200^{+0.004}_{-0.005}$ & $...  ^{      }_{      }$ & $...  ^{      }_{      }$ \\
M31\_CB\_20 & $0.041^{+0.004}_{-0.002}$ & $60.3^{+0.6}_{-0.7}$ & $6683 ^{+48 }_{-47 }$ & $1.747^{+0.014}_{-0.008}$ & $23.8^{+4.7}_{-3.6}$ & $...^{   }_{   }$ & $...^{  }_{  }$ & $... ^{     }_{     }$ & $0.647^{+0.003}_{-0.005}$ & $0.162^{+0.006}_{-0.004}$ & $...  ^{      }_{      }$ & $...  ^{      }_{      }$ \\
M31\_CB\_21 & $0.744^{+0.126}_{-0.110}$ & $83.0^{+1.3}_{-1.2}$ & $16846^{+225}_{-236}$ & $3.121^{+0.212}_{-0.182}$ & $45.9^{+9.3}_{-9.0}$ & $...^{   }_{   }$ & $...^{  }_{  }$ & $... ^{     }_{     }$ & $0.448^{+0.016}_{-0.017}$ & $0.397^{+0.012}_{-0.014}$ & $0.359^{+0.022}_{-0.020}$ & $0.069^{+0.032}_{-0.032}$ \\
M31\_CB\_22 & $0.563^{+0.040}_{-0.042}$ & $79.1^{+1.8}_{-1.4}$ & $6736 ^{+67 }_{-68 }$ & $2.962^{+0.071}_{-0.079}$ & $10.0^{+3.3}_{-3.2}$ & $119^{+12}_{-10}$ & $16 ^{+3}_{-2}$ & $0.83^{+0.07}_{-0.11}$ & $0.439^{+0.007}_{-0.006}$ & $0.338^{+0.007}_{-0.006}$ & $0.208^{+0.027}_{-0.026}$ & $0.417^{+0.026}_{-0.024}$ \\
M31\_CB\_23 & $0.530^{+0.009}_{-0.009}$ & $49.7^{+0.7}_{-0.6}$ & $28206^{+48 }_{-51 }$ & $2.750^{+0.000}_{-0.000}$ & $58.4^{+4.3}_{-4.7}$ & $...^{   }_{   }$ & $...^{  }_{  }$ & $... ^{     }_{     }$ & $0.482^{+0.003}_{-0.003}$ & $0.374^{+0.006}_{-0.006}$ & $...  ^{      }_{      }$ & $...  ^{      }_{      }$ \\
M31\_CB\_24 & $0.222^{+0.006}_{-0.007}$ & $32.8^{+0.7}_{-0.6}$ & $10029^{+49 }_{-51 }$ & $2.192^{+0.006}_{-0.007}$ & $66.6^{+4.5}_{-4.9}$ & $270^{+1 }_{-1 }$ & $19 ^{+1}_{-1}$ & $0.71^{+0.05}_{-0.05}$ & $0.546^{+0.000}_{-0.000}$ & $0.295^{+0.005}_{-0.006}$ & $...  ^{      }_{      }$ & $...  ^{      }_{      }$ \\
M31\_CB\_25 & $0.053^{+0.003}_{-0.003}$ & $59.3^{+0.7}_{-0.7}$ & $15379^{+49 }_{-49 }$ & $1.778^{+0.008}_{-0.010}$ & $60.4^{+4.7}_{-4.6}$ & $...^{   }_{   }$ & $...^{  }_{  }$ & $... ^{     }_{     }$ & $0.639^{+0.003}_{-0.002}$ & $0.185^{+0.005}_{-0.005}$ & $...  ^{      }_{      }$ & $...  ^{      }_{      }$ \\
M31\_CB\_26 & $0.504^{+0.001}_{-0.001}$ & $87.5^{+0.1}_{-0.1}$ & $10768^{+7  }_{-7  }$ & $2.869^{+0.003}_{-0.002}$ & $5.0 ^{+0.1}_{-0.1}$ & $...^{   }_{   }$ & $...^{  }_{  }$ & $... ^{     }_{     }$ & $0.445^{+0.001}_{-0.001}$ & $0.325^{+0.001}_{-0.001}$ & $0.453^{+0.001}_{-0.001}$ & $0.530^{+0.001}_{-0.001}$ \\
M31\_CB\_27 & $0.830^{+0.109}_{-0.238}$ & $74.4^{+1.7}_{-1.3}$ & $7260 ^{+4  }_{-4  }$ & $3.432^{+0.179}_{-0.420}$ & $7.9 ^{+3.3}_{-1.8}$ & $...^{   }_{   }$ & $...^{  }_{  }$ & $... ^{     }_{     }$ & $0.403^{+0.031}_{-0.011}$ & $0.371^{+0.011}_{-0.028}$ & $0.508^{+0.017}_{-0.015}$ & $0.567^{+0.013}_{-0.018}$ \\
M31\_CB\_28 & $0.707^{+0.010}_{-0.009}$ & $66.5^{+1.5}_{-1.6}$ & $6329 ^{+40 }_{-40 }$ & $3.182^{+0.002}_{-0.003}$ & $18.1^{+4.7}_{-4.4}$ & $...^{   }_{   }$ & $...^{  }_{  }$ & $... ^{     }_{     }$ & $0.426^{+0.003}_{-0.003}$ & $0.365^{+0.005}_{-0.005}$ & $...  ^{      }_{      }$ & $...  ^{      }_{      }$ \\
M31\_CB\_29 & $0.198^{+0.008}_{-0.005}$ & $65.5^{+0.3}_{-0.4}$ & $8066 ^{+38 }_{-40 }$ & $2.219^{+0.018}_{-0.010}$ & $6.4 ^{+1.9}_{-1.4}$ & $240^{+1 }_{-1 }$ & $24 ^{+1}_{-1}$ & $0.98^{+0.01}_{-0.01}$ & $0.527^{+0.001}_{-0.003}$ & $0.253^{+0.004}_{-0.002}$ & $...  ^{      }_{      }$ & $...  ^{      }_{      }$ \\
M31\_CB\_30 & $0.864^{+0.065}_{-0.060}$ & $89.2^{+0.4}_{-0.4}$ & $7744 ^{+61 }_{-57 }$ & $3.497^{+0.107}_{-0.101}$ & $5.9 ^{+0.2}_{-0.3}$ & $340^{+2 }_{-3 }$ & $56 ^{+2}_{-4}$ & $0.95^{+0.01}_{-0.01}$ & $0.398^{+0.006}_{-0.006}$ & $0.372^{+0.006}_{-0.007}$ & $0.477^{+0.004}_{-0.004}$ & $0.579^{+0.008}_{-0.011}$ \\
\bottomrule
\end{tabular}
\begin{tablenotes}
\raggedright
\item 
\textbf{Note}: \textsuperscript{a} $T_s$ refers to the ratio of the spot temperature to the surface temperature of the primary component; 
\textsuperscript{b} $r_1$ and $r_2$ represent the volume-equivalent radii of the two components, respectively.
\end{tablenotes}
\end{threeparttable}
}
\end{table}


\section{Discussions and Conclusions} \label{sec:discussions&conclusions}
Based on our analysis, the results reveal that 10 systems exhibit a significant O'Connell effect, all of which can be well explained by introducing a dark spot on the primary component. 
11 systems have mass ratios less than 0.15, classifying them as extremely low mass ratio contact binaries (ELMRCBs; \citealp{2022AJ....164..202L}). 
Among these, some systems display total eclipse features. According to the results of Terrell et al. (\citeyear{2005Ap&SS.296..221T}) and Latković et al. (\citeyear{2021ApJS..254...10L}), the photometric mass ratios of totally eclipsing contact binaries agree well with spectroscopic mass ratios, confirming the reliability of their photometric mass ratios. 
Additionally, six systems have primary star temperatures exceeding 10,000 K, suggesting that their primary components are early-type stars.

\subsection{Absolute Physical Parameters} \label{subsec:Absolute}
The absolute physical parameters of contact binaries are crucial for understanding their formation, structure, and evolutionary states (A. Shokry et al. \citeyear{2018NewA...59....8S}; K. Li et al. \citeyear{2021AJ....162...13L}). 
According to Gu et al.(\citeyear{2024ApJS..273....9G}), all targets are members of M31. Therefore, we adopted the distance of M31  and the brightest apparent magnitude to calculate the luminosities of these targets. The absolute magnitude $M$ of each system was calculated by $ M = m - glog_{10}D + 5 - 0.205$, where $D = 761 kpc$ is the distance of M31 (\citealp{2021ApJ...920...84L}), and 0.205 is the extinction coefficient in the SDSS g-band (\citealp{2011ApJ...737..103S}). 
Subsequently, the total luminosity $L_T$ was derived by $\frac{L_T}{L_\odot}=10^{(M_\odot-M)/2.5}$. Finally, the luminosities of the two components $L_1$ and $L_2$, were obtained using the luminosity ratios listed in Table \ref{tab2: 30 targets's parameters}.
Then, we estimated the primary star masses ($M_1$) by applying the temperature-mass relation from the updated online tables of Pecaut \& Mamajek (\citeyear{2013ApJS..208....9P}). The secondary star masses ($M_2$) were then calculated using the mass ratio ($q$) obtained from PHOEBE. Subsequently, we determined the semi-major axis ($a$) by Kepler’s third law ($M_1 + M_2 = \frac{0.0134a^3}{P^2}$). The radius $R_{1,2}$ was given by $R_{1,2} = a \times r_{1,2}$. The errors of the above parameters were calculated with the error transfer formula. The absolute parameters of all targets are shown in Table \ref{tab3: Ab parameters of 30}.

\begin{table}[htbp]
\caption{Absolute Parameters of the 30 Targets}
\label{tab3: Ab parameters of 30}
\centering
\resizebox{\textwidth}{!}{
\begin{threeparttable}
\begin{tabular}{lcccccccc}
\toprule
\multirow{2}{*}{Parameter} & \multirow{2}{*}{$M_1$} & \multirow{2}{*}{$M_2$} & \multirow{2}{*}{$a$} & \multirow{2}{*}{$R_1$} & \multirow{2}{*}{$R_2$} & \multirow{2}{*}{$L_1$} & \multirow{2}{*}{$L_2$} & \multirow{2}{*}{$J_{\text{orb}}$} \\
& & & & & & & & \\
 & (M$_\odot$) & (M$_\odot$) & (R$_\odot$) & (R$_\odot$) & (R$_\odot$) & (L$_\odot$) & (L$_\odot$) & (log) \\
\midrule
M31\_CB\_1  &	$1.03 \pm 0.09$ & $0.12 \pm 0.02$ & $3.14 \pm 0.09$ & $1.80 \pm 0.05$ & $0.68 \pm 0.03$ & $686.57   \pm 41.37   $ & $82.24    \pm 36.11  $ & $51.08$ \\
M31\_CB\_2  & $1.80 \pm 0.01$ & $1.48 \pm 0.10$ & $7.27 \pm 0.08$ & $2.92 \pm 0.05$ & $2.68 \pm 0.05$ & $633.24   \pm 36.41   $ & $178.16   \pm 35.21  $ & $52.38$ \\
M31\_CB\_3  & $0.85 \pm 0.00$ & $0.66 \pm 0.01$ & $1.88 \pm 0.00$ & $0.84 \pm 0.00$ & $0.76 \pm 0.01$ & $7135.21  \pm 220.55  $ & $3172.44  \pm 118.55 $ & $51.58$ \\
M31\_CB\_4  & $1.40 \pm 0.08$ & $0.13 \pm 0.01$ & $3.60 \pm 0.06$ & $2.14 \pm 0.04$ & $0.77 \pm 0.02$ & $774.78   \pm 37.56   $ & $125.61   \pm 11.96  $ & $51.23$ \\
M31\_CB\_5  & $1.87 \pm 0.01$ & $0.98 \pm 0.00$ & $7.01 \pm 0.01$ & $3.09 \pm 0.01$ & $2.30 \pm 0.01$ & $597.84   \pm 18.98   $ & $255.78   \pm 21.42  $ & $52.24$ \\
M31\_CB\_6  & $1.80 \pm 0.01$ & $1.42 \pm 0.03$ & $5.87 \pm 0.02$ & $2.67 \pm 0.01$ & $2.43 \pm 0.01$ & $1609.29  \pm 52.28   $ & $941.16   \pm 39.20  $ & $52.32$ \\
M31\_CB\_7  & $1.21 \pm 0.04$ & $0.17 \pm 0.01$ & $3.73 \pm 0.04$ & $2.06 \pm 0.02$ & $0.86 \pm 0.01$ & $1117.65  \pm 48.30   $ & $228.04   \pm 39.26  $ & $51.32$ \\
M31\_CB\_8  & $0.82 \pm 0.04$ & $0.31 \pm 0.05$ & $4.05 \pm 0.08$ & $1.91 \pm 0.06$ & $1.24 \pm 0.05$ & $409.98   \pm 33.63   $ & $115.80   \pm 32.03  $ & $51.47$ \\
M31\_CB\_9	& $1.37 \pm 0.00$ & $0.10 \pm 0.00$ & $1.98 \pm 0.00$ & $1.22 \pm 0.00$ & $0.40 \pm 0.00$ & $78787.01 \pm 2278.91 $ & $8505.25  \pm 253.74 $ & $50.96$ \\
M31\_CB\_10 & $20.43\pm 0.28$ & $16.03\pm 0.29$ & $6.96 \pm 0.03$ & $3.01 \pm 0.03$ & $2.72 \pm 0.04$ & $835.29   \pm 34.30   $ & $665.61   \pm 27.84  $ & $53.94$ \\
M31\_CB\_11	& $1.87 \pm 0.00$ & $1.09 \pm 0.00$ & $2.50 \pm 0.00$ & $1.10 \pm 0.00$ & $0.86 \pm 0.00$ & $74919.69 \pm 2190.17 $ & $31532.22 \pm 967.15 $ & $52.06$ \\
M31\_CB\_12 & $0.97 \pm 0.03$ & $0.14 \pm 0.02$ & $4.04 \pm 0.04$ & $2.24 \pm 0.05$ & $0.93 \pm 0.03$ & $939.83   \pm 59.85   $ & $191.65   \pm 62.69  $ & $51.19$ \\
M31\_CB\_13 & $1.61 \pm 0.01$ & $0.04 \pm 0.00$ & $2.65 \pm 0.01$ & $1.78 \pm 0.00$ & $0.38 \pm 0.00$ & $5120.14  \pm 155.68  $ & $107.55   \pm 8.56   $ & $50.74$ \\
M31\_CB\_14 & $2.53 \pm 0.02$ & $0.10 \pm 0.00$ & $3.39 \pm 0.01$ & $2.22 \pm 0.01$ & $0.57 \pm 0.01$ & $2837.44  \pm 88.60   $ & $172.16   \pm 10.19  $ & $51.24$ \\
M31\_CB\_15 & $1.45 \pm 0.01$ & $0.11 \pm 0.01$ & $2.63 \pm 0.00$ & $1.60 \pm 0.01$ & $0.52 \pm 0.01$ & $3327.40  \pm 105.00  $ & $264.18   \pm 25.85  $ & $51.10$ \\
M31\_CB\_16 & $1.08 \pm 0.01$ & $0.36 \pm 0.02$ & $1.98 \pm 0.01$ & $0.96 \pm 0.01$ & $0.59 \pm 0.01$ & $4575.74  \pm 210.88  $ & $1495.07  \pm 383.22 $ & $51.44$ \\
M31\_CB\_17 & $1.81 \pm 0.06$ & $0.61 \pm 0.02$ & $3.92 \pm 0.03$ & $2.08 \pm 0.02$ & $1.37 \pm 0.01$ & $271.77   \pm 26.93   $ & $200.03   \pm 19.82  $ & $51.93$ \\
M31\_CB\_18 & $2.06 \pm 0.02$ & $0.47 \pm 0.01$ & $3.94 \pm 0.01$ & $2.16 \pm 0.01$ & $1.20 \pm 0.02$ & $2156.21  \pm 72.54   $ & $415.11   \pm 25.48  $ & $51.86$ \\
M31\_CB\_19 & $1.81 \pm 0.03$ & $0.12 \pm 0.01$ & $2.65 \pm 0.01$ & $1.66 \pm 0.01$ & $0.53 \pm 0.01$ & $1346.71  \pm 53.56   $ & $172.58   \pm 13.41  $ & $51.19$ \\
M31\_CB\_20 & $1.79 \pm 0.01$ & $0.07 \pm 0.01$ & $2.62 \pm 0.01$ & $1.70 \pm 0.01$ & $0.43 \pm 0.01$ & $1851.72  \pm 64.37   $ & $58.83    \pm 6.60   $ & $50.97$ \\
M31\_CB\_21 & $4.89 \pm 0.09$ & $3.64 \pm 0.58$ & $8.27 \pm 0.19$ & $3.70 \pm 0.16$ & $3.29 \pm 0.13$ & $378.81   \pm 31.68   $ & $300.90   \pm 37.56  $ & $53.03$ \\
M31\_CB\_22 & $1.66 \pm 0.07$ & $0.94 \pm 0.08$ & $4.96 \pm 0.07$ & $2.18 \pm 0.04$ & $1.68 \pm 0.04$ & $995.68   \pm 56.41   $ & $396.57   \pm 64.01  $ & $52.11$ \\
M31\_CB\_23 & $14.92\pm 0.12$ & $7.92 \pm 0.15$ & $6.32 \pm 0.02$ & $3.04 \pm 0.02$ & $2.36 \pm 0.04$ & $1837.20  \pm 62.40   $ & $1026.48  \pm 38.26  $ & $53.57$ \\
M31\_CB\_24 & $1.91 \pm 0.02$ & $0.42 \pm 0.01$ & $4.47 \pm 0.02$ & $2.44 \pm 0.01$ & $1.32 \pm 0.03$ & $2135.14  \pm 74.85   $ & $886.39   \pm 43.37  $ & $51.83$ \\
M31\_CB\_25 & $5.09 \pm 0.02$ & $0.27 \pm 0.02$ & $4.17 \pm 0.01$ & $2.66 \pm 0.01$ & $0.77 \pm 0.02$ & $2920.83  \pm 92.81   $ & $211.32   \pm 14.78  $ & $51.86$ \\
M31\_CB\_26 & $2.74 \pm 0.02$ & $1.38 \pm 0.01$ & $5.83 \pm 0.01$ & $2.59 \pm 0.01$ & $1.90 \pm 0.01$ & $724.49   \pm 24.65   $ & $389.32   \pm 13.39  $ & $52.43$ \\
M31\_CB\_27 & $1.94 \pm 0.02$ & $1.61 \pm 0.34$ & $5.63 \pm 0.18$ & $2.27 \pm 0.14$ & $2.09 \pm 0.13$ & $875.34   \pm 67.48   $ & $332.03   \pm 73.65  $ & $52.38$ \\
M31\_CB\_28 & $1.69 \pm 0.13$ & $1.19 \pm 0.10$ & $3.77 \pm 0.07$ & $1.60 \pm 0.03$ & $1.38 \pm 0.03$ & $469.62   \pm 31.51   $ & $164.13   \pm 12.77  $ & $52.14$ \\
M31\_CB\_29 & $1.94 \pm 0.03$ & $0.38 \pm 0.01$ & $4.94 \pm 0.02$ & $2.60 \pm 0.02$ & $1.25 \pm 0.02$ & $3141.93  \pm 99.97   $ & $540.73   \pm 31.09  $ & $51.82$ \\
M31\_CB\_30 & $1.84 \pm 0.04$ & $1.59 \pm 0.12$ & $5.50 \pm 0.07$ & $2.19 \pm 0.04$ & $2.05 \pm 0.04$ & $604.29   \pm 27.55   $ & $475.25   \pm 25.16  $ & $52.35$ \\
\bottomrule
\end{tabular}
\end{threeparttable}
}
\end{table}

\subsection{Evolutionary State} \label{subsec:Evolutionary}
\begin{figure}[htbp]
    \centering 
    \begin{subfigure}
        \centering
        \includegraphics[width=0.46\textwidth]{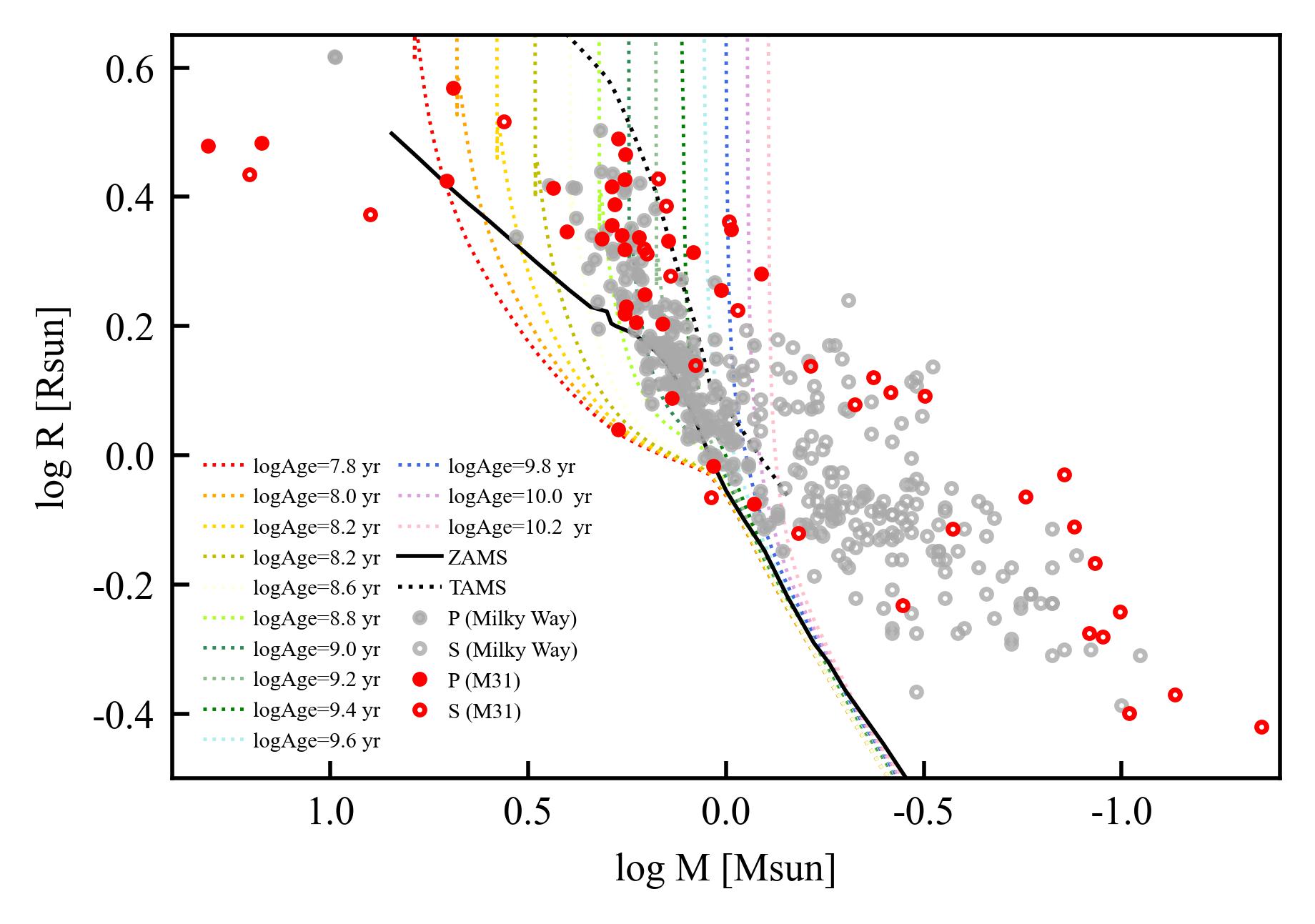}
    \end{subfigure}
    \begin{subfigure}
        \centering
        \includegraphics[width=0.45\textwidth]{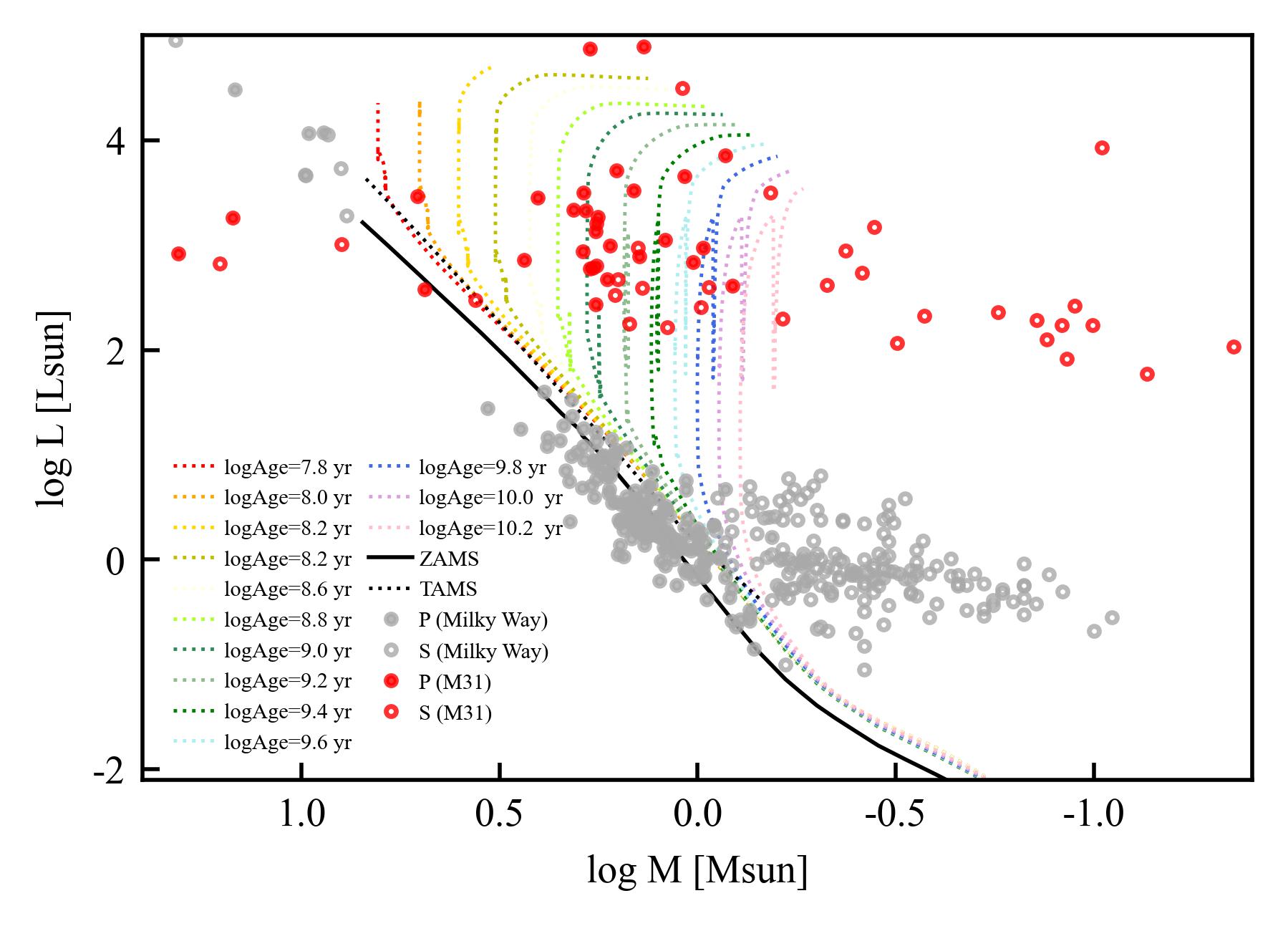}
    \end{subfigure}
    \begin{subfigure}
        \centering
        \includegraphics[width=0.45\textwidth]{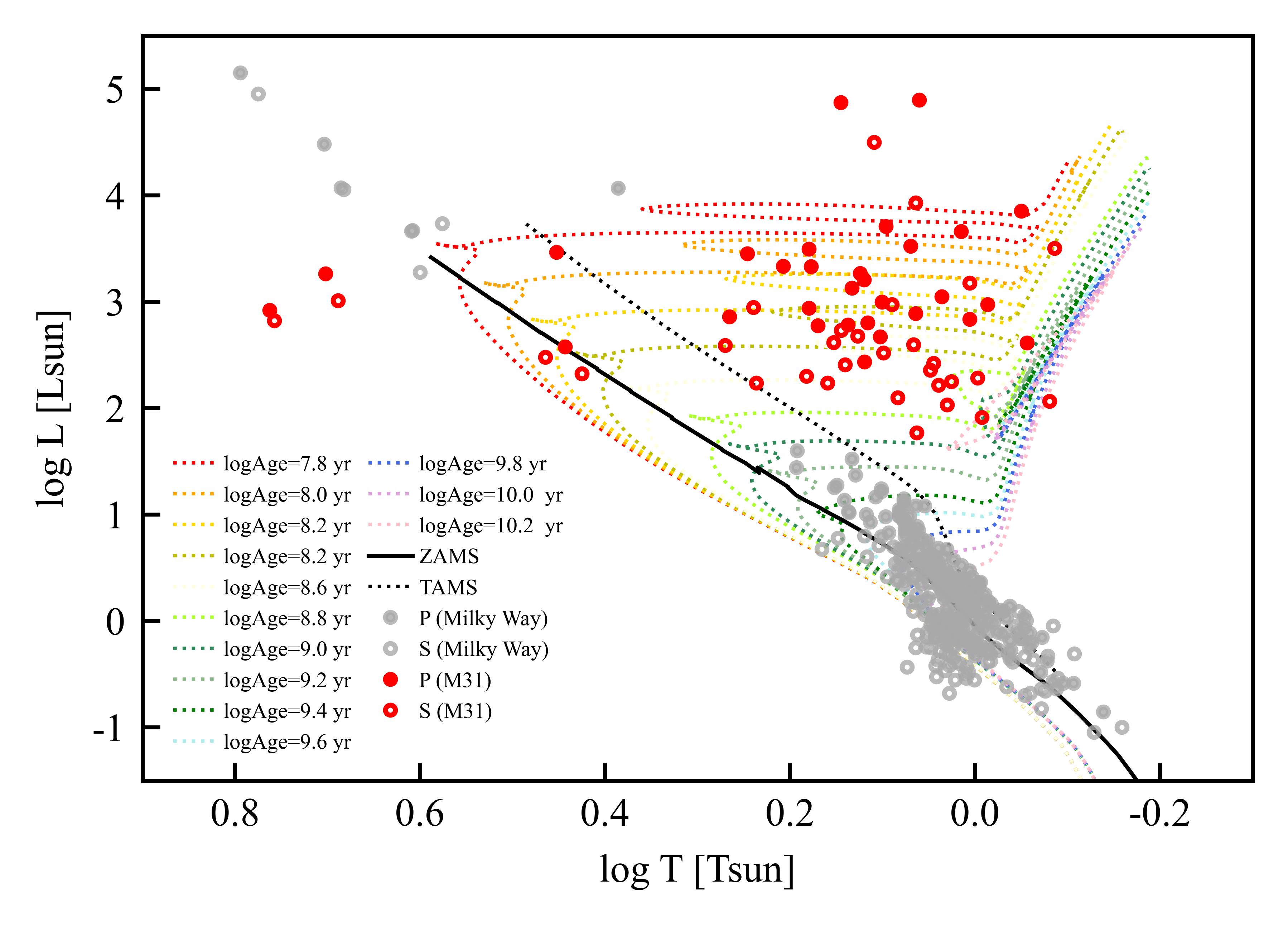}
    \end{subfigure}  
    \caption{The M–R, M–L and T-L distributions for contact binaries. The colored dotted lines indicate the evolutionary tracks for solar chemical compositions. The solid and dashed lines indicate the ZAMS and the TAMS, respectively. The gray filled and open circles represent the primary and secondary components of the contact binaries in the Milky Way, while the red filled and open circles represent those of the 30 contact binaries in M31.}
    \label{fig4}
\end{figure}

We constructed the mass-radius (M-R), the mass-luminosity (M-L) and the temperature-luminosity (T-L) distributions in Figure \ref{fig4} to investigate evolutionary states and compare them with the contact binaries in the Milky Way. 
The evolutionary tracks for solar chemical compositions (colored dashed lines) were adopted from L. Girardi et al. (\citeyear{2000A&AS..141..371G}). 
The zero-age main sequence (ZAMS) and the terminal-age main sequence (TAMS) are plotted as solid and dashed lines, respectively. 
The data for 173 Galactic contact binaries in Figure \ref{fig4} are from Li et al. (\citeyear{2021AJ....162...13L}), which provide reliable physical parameters based on both light curves and radial velocity measurements.

As depicted in M-R distribution of Figure \ref{fig4}, the majority of our target systems exhibit that primary components remain on the main sequence, while the secondary components have evolved off the main sequence. 
This configuration is analogous to contact binary systems in the Milky Way. 
However, as shown in Figures \ref{fig4}, our targets exhibit larger radii and higher luminosities compared to Galactic ones of similar masses or temperature. 
This indicates that our targets are more evolved compared to contact binaries in the Milky Way, which may result from differences in metallicity between the two galaxies as well as their distinct evolutionary histories. 
According to Williams et al. (\citeyear{2017ApJ...846..145W}), M31 demonstrates an overall higher metallicity than the Milky Way, particularly in its central regions. 
In the disk regions, the average metallicity ([Fe/H]) of M31 is approximately 0.1–0.3 dex higher than that of the Milky Way at comparable galactocentric radii. 
As noted by Ekström et al. (\citeyear{2012A&A...537A.146E}), stars with higher metallicity experience shortened main-sequence lifetimes due to enhanced core burning efficiency, which accelerates fuel consumption. Consequently, contact binaries in M31 may evolve faster than those of the same mass in the Milky Way, potentially accounting for the observed larger radii and higher luminosities. 

\begin{figure}[htbp]
    \centering 
    \includegraphics[width=0.6\textwidth]{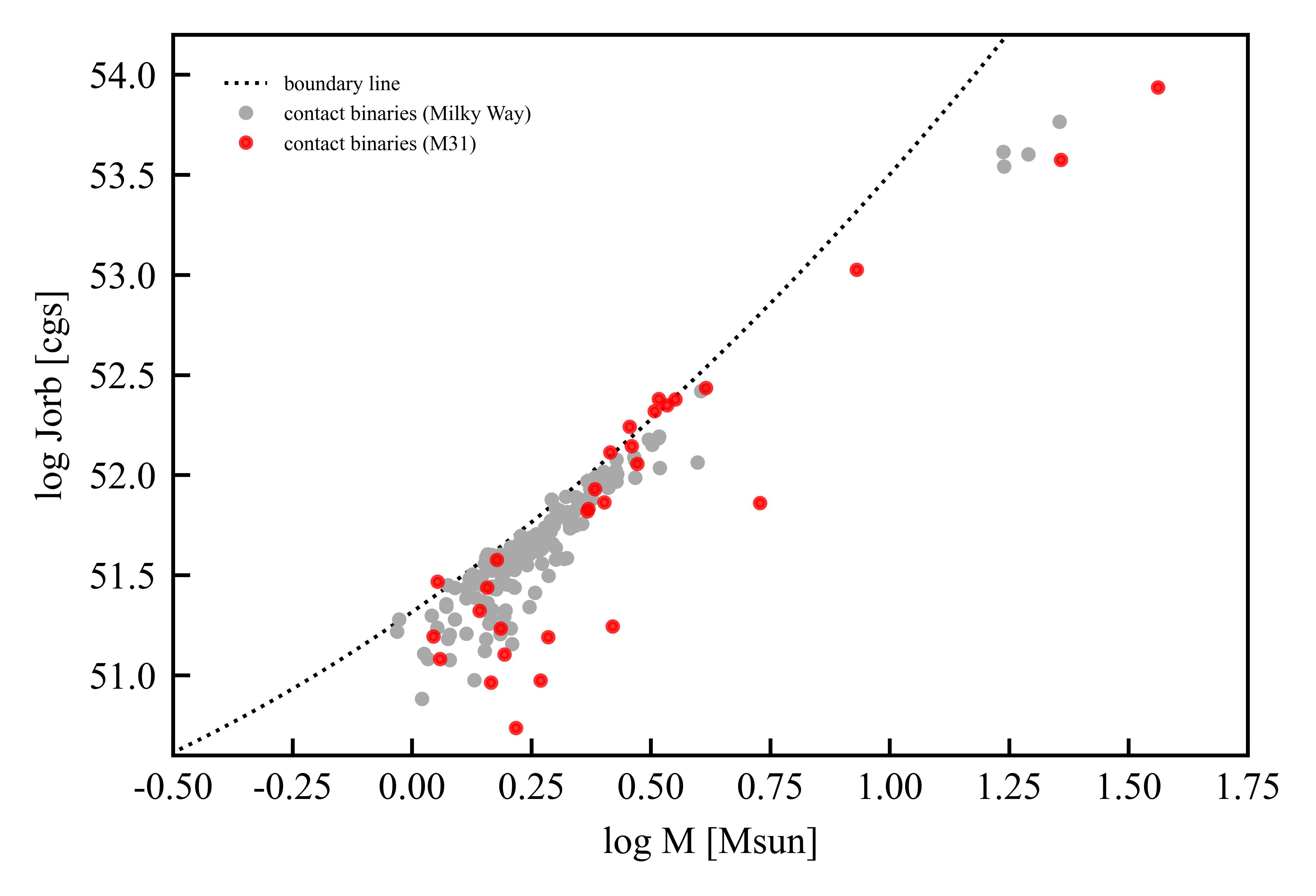}
    \caption{The relation between orbital angular momentum and total mass for contact binaries. The dashed line indicates the boundary line between the detached and contact binaries. The gray filled and open circles represent the primary and secondary components of the contact binaries in the Milky Way, while the red filled and open circles represent those of the 30 contact binaries in M31.}
    \label{fig5}
\end{figure}

To further investigate the evolutionary states of these targets, we calculated the orbital angular momenta of all systems and constructed a diagram of orbital angular momentum versus total mass of the binary systems. 
The orbital angular momenta of these targets were calculated using the following equation (\citealp{2006MNRAS.373.1483E}):
\begin{equation}
  \label{eq:Jorb}  
J_{orb} = \frac{q}{(1+q)^2}\sqrt[3]{\frac{G^2}{2\pi}M_T^5P}
\end{equation}
where $M_T$ is the total mass of the binary system, $P$ is the orbital period, and $q$ is the mass ratio. 
The relationship between orbital angular momentum and total mass of contact binaries for our targets are shown in Figure \ref{fig5}. 
The dashed line in the figure represents the boundary between detached and contact binaries (Z. Eker et al. \citeyear{2006MNRAS.373.1483E}), and the Milky Way contact binary targets are also from Li et al. (\citeyear{2021AJ....162...13L}).
Contact binaries are generally considered to be formed from short period detached binary systems through angular momentum loss. 
As angular momentum continues to be lost, the binary orbit shrinks, and contact binaries may eventually merge (\citealp{2014MNRAS.438..859J}; \citealp{1995MNRAS.274.1019S}; \citealp{2008MNRAS.390.1577G}). 
Therefore, contact binary systems with low orbital angular momentum evolved faster. 
From Figure \ref{fig5}, several targets such as M31\_CB\_2, M31\_CB\_5, M31\_CB\_8, M31\_CB\_22, M31\_CB\_27, and M31\_CB\_30, lie above or on the boundary line between detached and contact binaries. Meanwhile, Table \ref{tab2: 30 targets's parameters} shows that these targets have relatively low fill-out factors. Consequently, they are likely newly formed contact binaries. 
Apart from these targets, compared with contact binaries of similar mass in the Milky Way, most of our targets exhibit lower orbital angular momentum. This further indicates that our targets have evolved faster than those in the Milky Way.

\subsection{Several interesting targets} \label{subsec:interesting}
According to Li et al. (\citeyear{2022AJ....164..202L}), contact binaries with a mass ratio below 0.15 are classified as ELMRCBs, which are considered good candidates for binary mergers. 
11 of our targets have a mass ratio below 0.15 and are thus classified as ELMRCBs. These 11 targets are valuable candidates for studying binary merger events in M31. 
The mass ratios of M31\_CB\_13, M31\_CB\_14, and M31\_CB\_20 are 0.028, 0.040, and 0.041, respectively, all of which are lower than the derived theoretically minimum mass ratio of contact binaries ($0.042 < q_{min} < 0.044$; \citealp{2024SerAJ.208....1A}). 
In particular, the mass ratio of M31\_CB\_13 is lower than that of the currently known lowest mass ratio contact binary, TYC-3801-1529-1 ($q = 0.0356$; Li et al., \citeyear{2024A&A...692L...4L}). 
However, due to the limitations of the data, high-precision photometric and spectroscopic data may be required in the future for further research on these targets.

Massive stars are generally defined as those with masses exceeding approximately 8 $M_\odot$, which is the minimum threshold for a star to undergo core-collapse at the end of its life \citep{Kippenhahn2012, Langer2012, Maeder2009}. 
Such core-collapse events typically result in supernova explosions (Type II, Ib, or Ic) and leave compact remnants in the form of neutron stars or black holes. 
Specifically, progenitor stars with initial masses between 8 and 25 $M_\odot$ are expected to form neutron stars. However, under certain conditions—such as inefficient explosion mechanisms or fallback accretion—some stars within this mass range may directly collapse into black holes \citep{2025A&A...693A.283P}. 
For massive contact binary stars, their evolution is significantly different from that of single stars (\citealp{2024ARA&A..62...21M}), and their final fates are determined not only by their initial masses but also by factors such as mass ratio, surface temperature, luminosity, and evolution. 
Based on the absolute physical parameters listed in Table \ref{tab3: Ab parameters of 30}, two contact binary systems in our sample, M31\_CB\_10 and M31\_CB\_23, contain massive stellar components.

For M31\_CB\_10, the primary component has a mass of 20.43 $M_\odot$ and the secondary component has a mass of 16.03  $M_\odot$, with the mass ratio of 0.785 and the orbital period of 0.352 day. 
Both the primary and secondary components have a temperature of approximately 30,000 K, and their luminosities satisfy $log L/L_\odot\geq5$. 
According to Marchant \& Bodensteiner et al.(\citeyear{2024ARA&A..62...21M}), M31\_CB\_10 meets the conditions for chemically homogeneous evolution (CHE). 
During its evolution, there is no significant mass loss, and it will eventually form a double black hole system.

For M31\_CB\_23, the primary component has a mass of 14.92 $M_\odot$ and the secondary component has a mass of 7.92 $M_\odot$, with the mass ratio of 0.53 and the orbital period of 0.384 day. 
Both the primary and secondary components have a temperature of approximately 29,000 K, and their luminosities also satisfy $log L/L_\odot\geq5$. 
According to Marchant \& Bodensteiner et al.(\citeyear{2024ARA&A..62...21M}), M31\_CB\_23 meets the conditions for stable mass transfer. 
The primary component will first undergo rapid Case A Roche lobe overflow (RLOF), stripping down to a helium star (subdwarf O-type B star or Wolf–Rayet star) with a mass of approximately 4–5 $M_\odot$, and the secondary component will evolve into an O-type Be star after accreting material. 
Eventually, the primary component may collapse into a black hole, while the secondary component may evolve into a neutron star, forming a black hole-neutron star binary system. 

Systems like M31\_CB\_10 and M31\_CB\_23 are considered potential progenitors of gravitational wave sources detectable by facilities such as Laser Interferometer Gravitational-Wave Observatory (LIGO; \citealp{1992Sci...256..325A}) and Virgo Interferometer (Virgo; \citealp{1996dmcq.conf..347H}) \citep{Abbott2016, Abbott2020}.

\subsection{Conclusions} \label{subsec:conclusions}
In conclusion, 30 contact binaries in M31 were analyzed. After analyzing these targets with PHOEBE, we derived their physical parameters and calculated their absolute physical parameters. 
The results show that 10 targets exhibit a significant O’Connell effect, which we successfully explained by introducing a spot on the primary component. 
11 of the targets are ELMRCBs and are good candidates for future studies of binary mergers. 
Six targets have primary star temperatures above 10,000 K, suggesting that their primary components are early-type stars. 
Analysis of the components masses indicates that the massive contact binary system M31\_CB\_10, may eventually undergo a core-collapse and evolve into a double black hole binaries. 
While the the massive contact binary system M31\_CB\_23 may eventually evolve into a black hole-neutron star binary. These two systems are potential candidates for future gravitational wave sources.

To investigate the evolutionary states of the 30 targets and compare them with those in the Milky Way, we constructed the M-R and M-L diagrams. The results show that our targets have larger radii and higher luminosities than contact binaries in the Milky Way of similar mass. This more-evolved state may be due to differences in metallicity between M31 and the Milky Way. We also calculated the orbital angular momentum for all targets and plotted its relationship with total mass. The results show that our targets have lower orbital angular momentum than contact binaries in the Milky Way of similar mass, again indicating a more-evolved state. Further high-precision photometric and spectroscopic observations are required in the future to confirm these results.

\begin{acknowledgments}
This work is supported by the National Natural Science Foundation of China (NSFC; No. 12273018; No. 12222301), the Joint Research Fund in Astronomy (No. U1931103) under cooperative agreement between NSFC and Chinese Academy of Sciences (CAS), the Taishan Scholars Young Expert Program of Shandong Province, the Qilu Young Researcher Project of Shandong University, the Young Data Scientist Project of the National Astronomical Data Center, and by the Cultivation Project for LAMOST Scientific Payoff and Research Achievement of CAMS-CAS. The calculations in this work were carried out at the Supercomputing Center of Shandong University, Weihai. This work was completed at the Shandong Key Laboratory of Space Environment and Exploration Technology, Institute of Space Sciences, School of Space Science and Technology, Shandong University, Shandong, China.
\end{acknowledgments}

\appendix
\counterwithin{figure}{section}
\renewcommand{\thefigure}{\Alph{section}\arabic{figure}}
\section{Fitted curves of other targets}\label{appendix.A}
Figure \ref{figA1} presents the final fitting and residuals plots for all targets except M31\_CB\_14 and M31\_CB\_16. 
Figure \ref{figA2} shows the comparison results of 14 targets without and with third light. For each target, the left shows the fitting result without the third light, and the right shows the fitting result with the third light.

\begin{figure}[htbp]
    \centering 
    \begin{subfigure}
        \centering
        \includegraphics[width=0.29\textwidth]{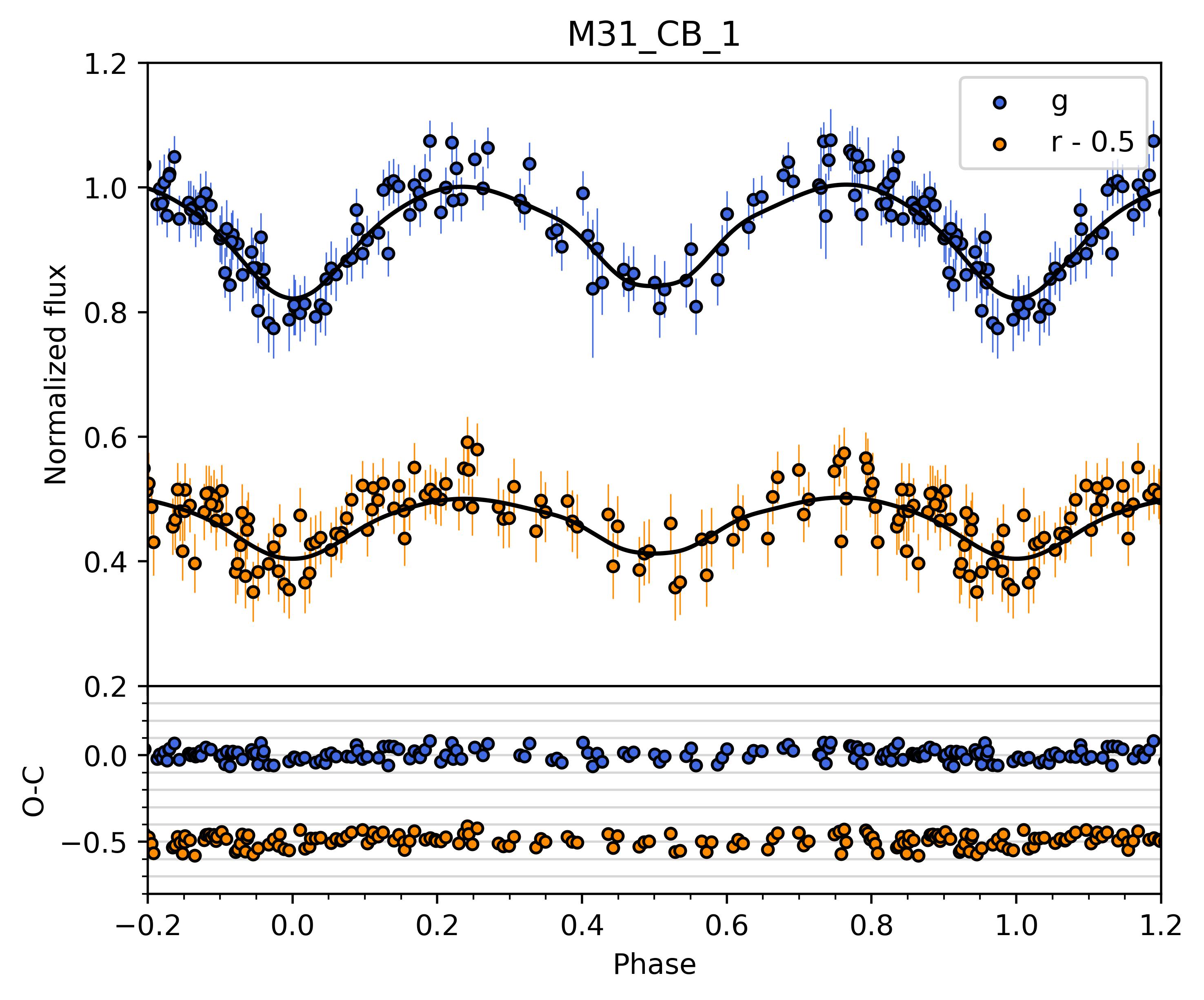}
    \end{subfigure}
    \begin{subfigure}
        \centering
        \includegraphics[width=0.29\textwidth]{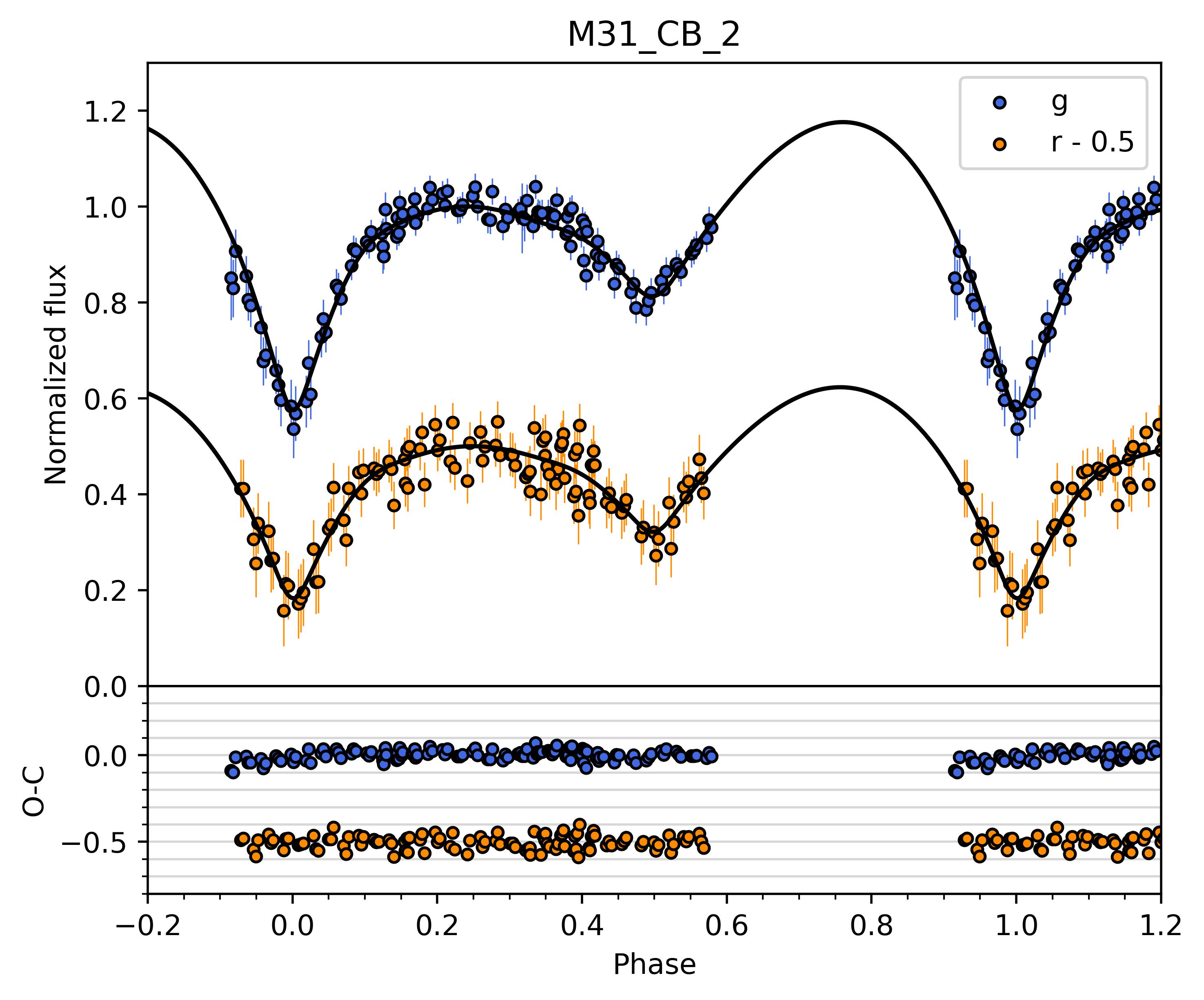}
    \end{subfigure}
    \begin{subfigure}
        \centering
        \includegraphics[width=0.29\textwidth]{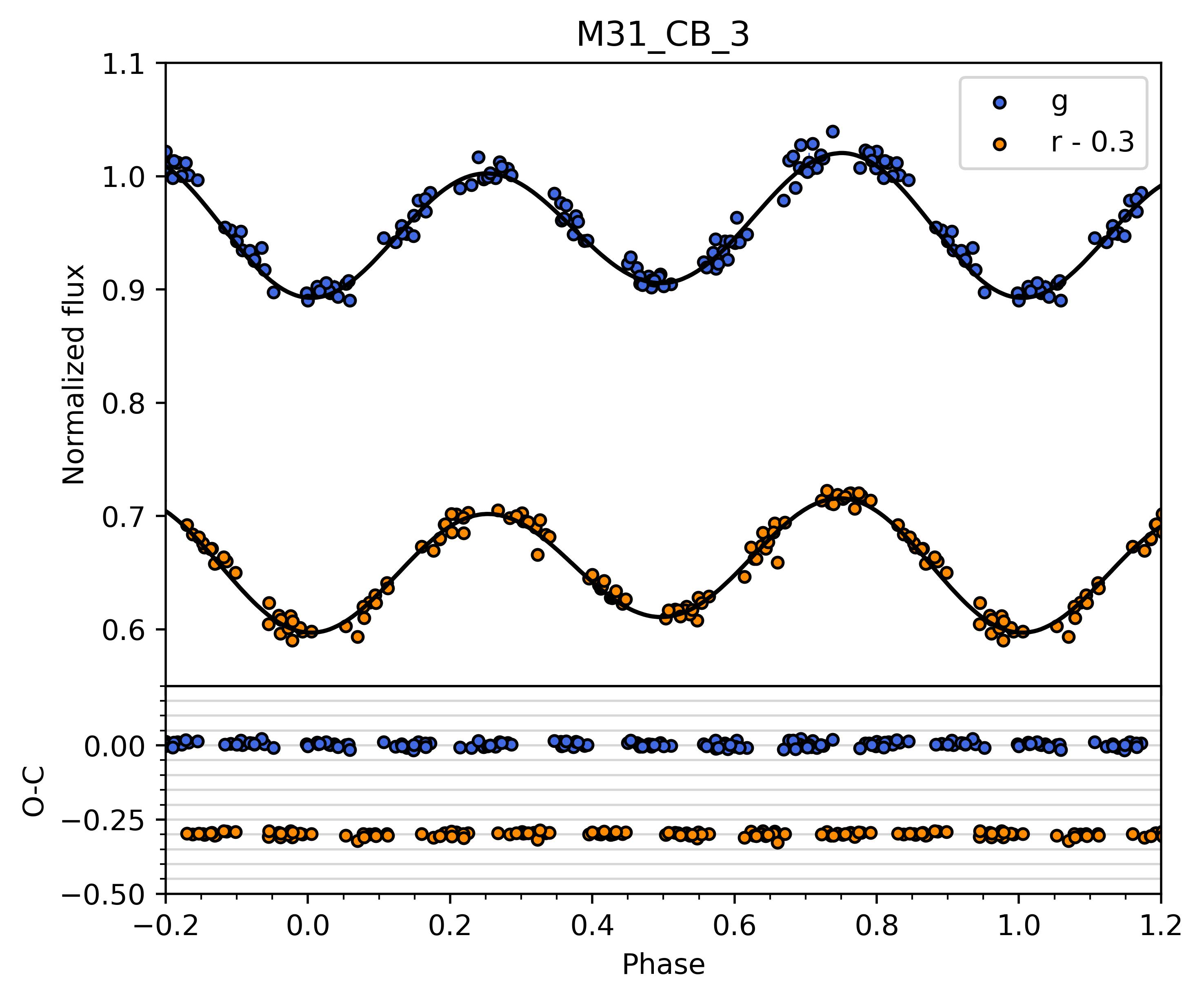}
    \end{subfigure}  

    \begin{subfigure}
        \centering
        \includegraphics[width=0.29\textwidth]{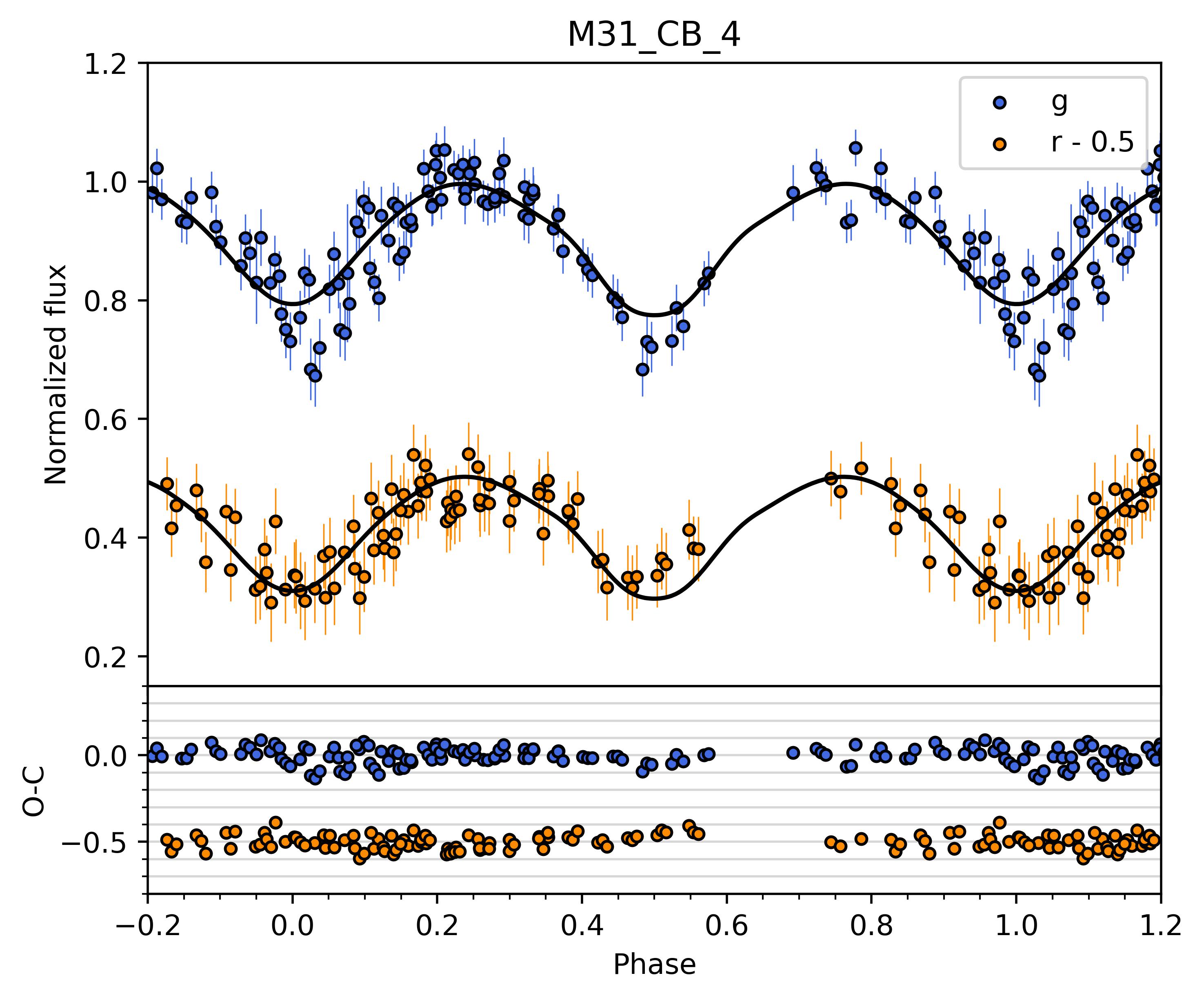}
    \end{subfigure}
    \begin{subfigure}
        \centering
        \includegraphics[width=0.29\textwidth]{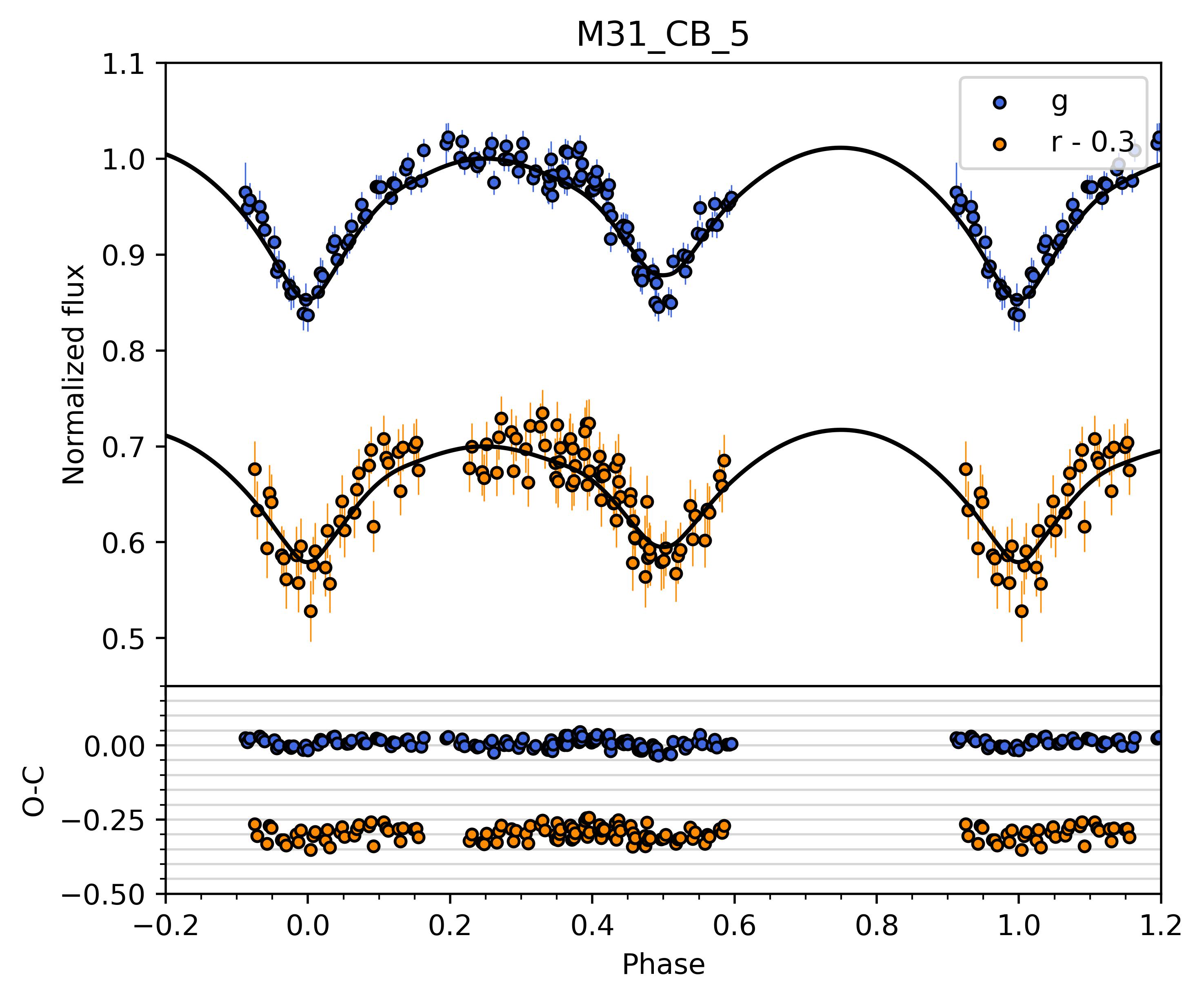}
    \end{subfigure}
    \begin{subfigure}
        \centering
        \includegraphics[width=0.29\textwidth]{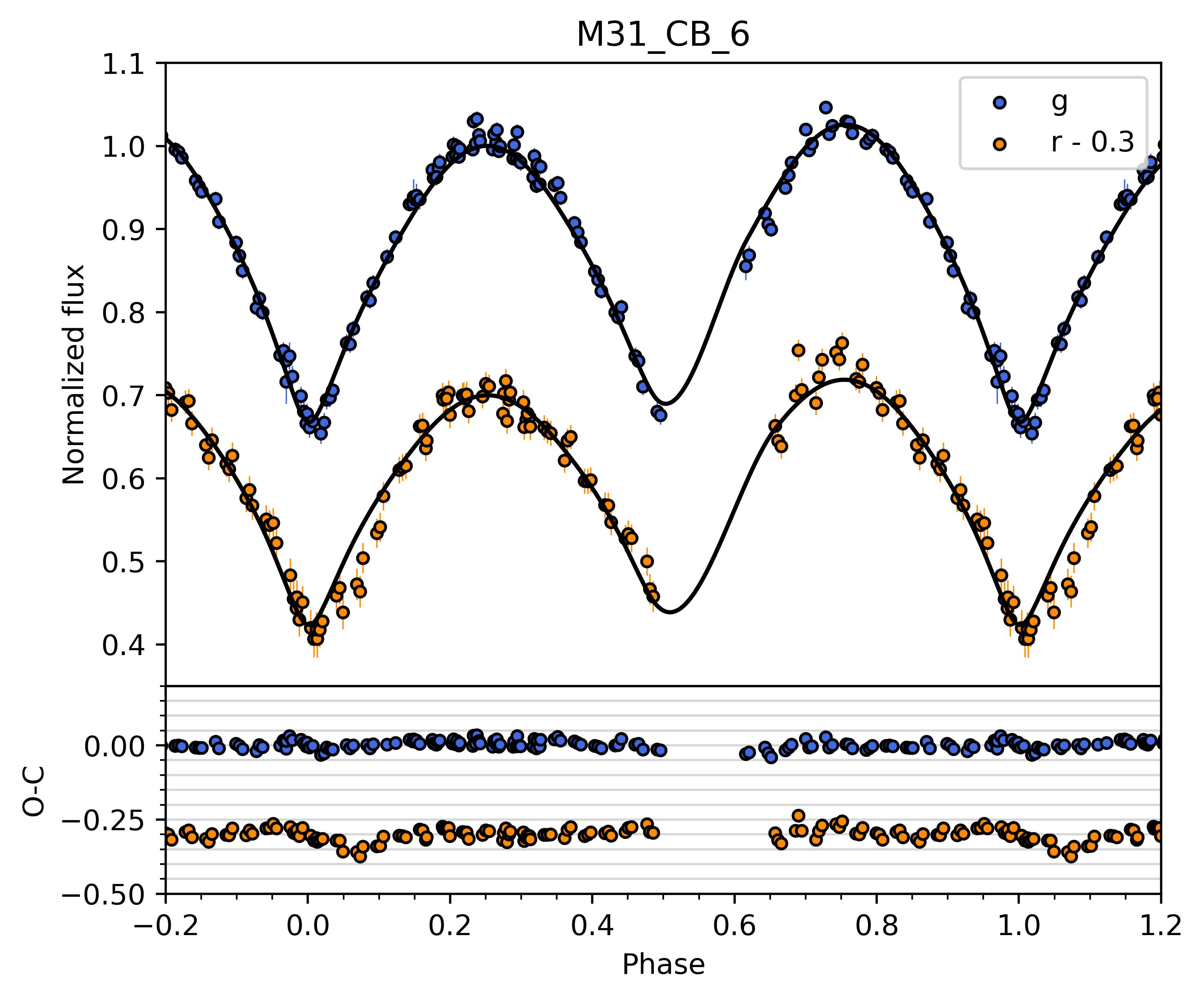}
    \end{subfigure}  

    \begin{subfigure}
        \centering
        \includegraphics[width=0.29\textwidth]{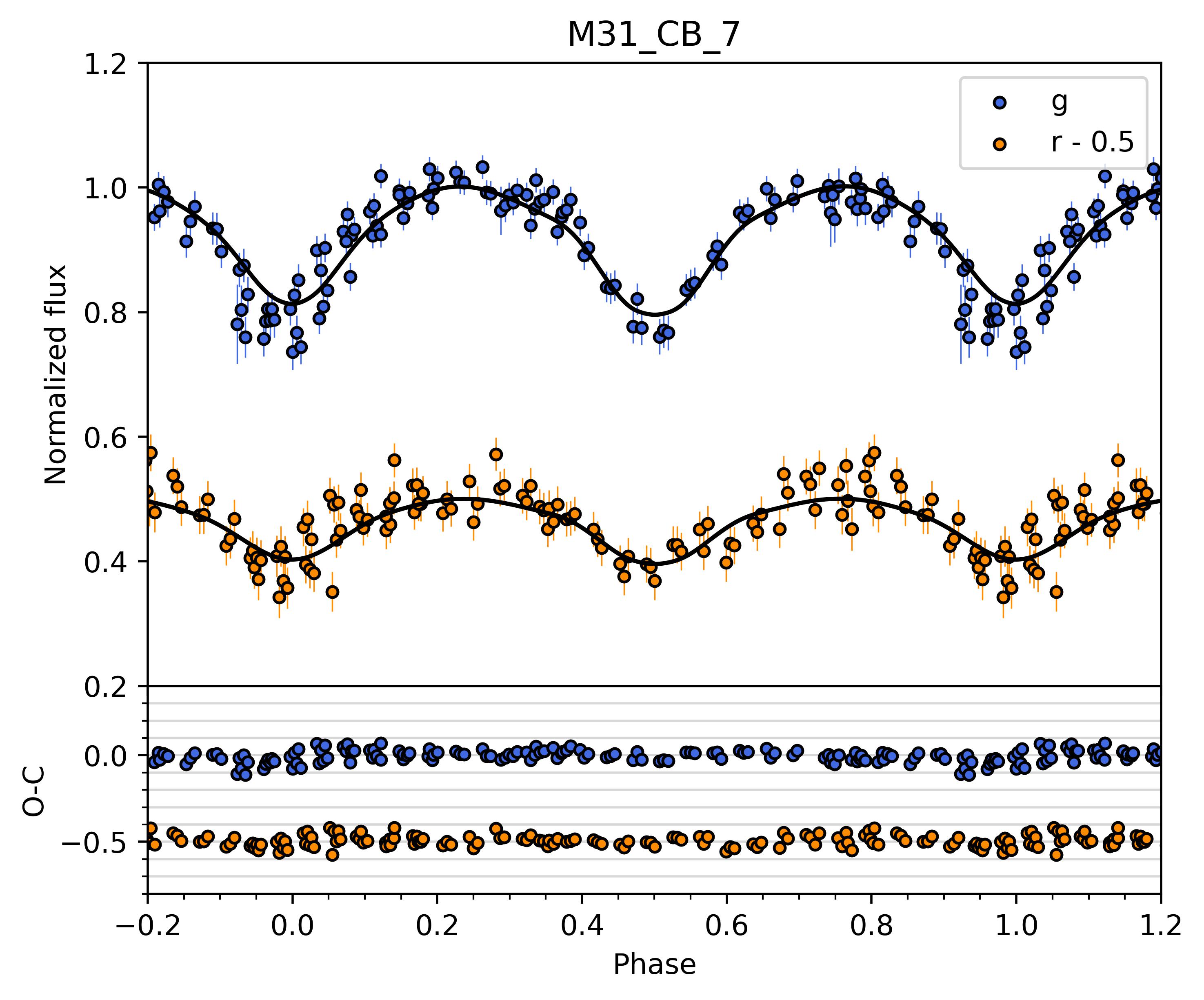}
    \end{subfigure}
    \begin{subfigure}
        \centering
        \includegraphics[width=0.29\textwidth]{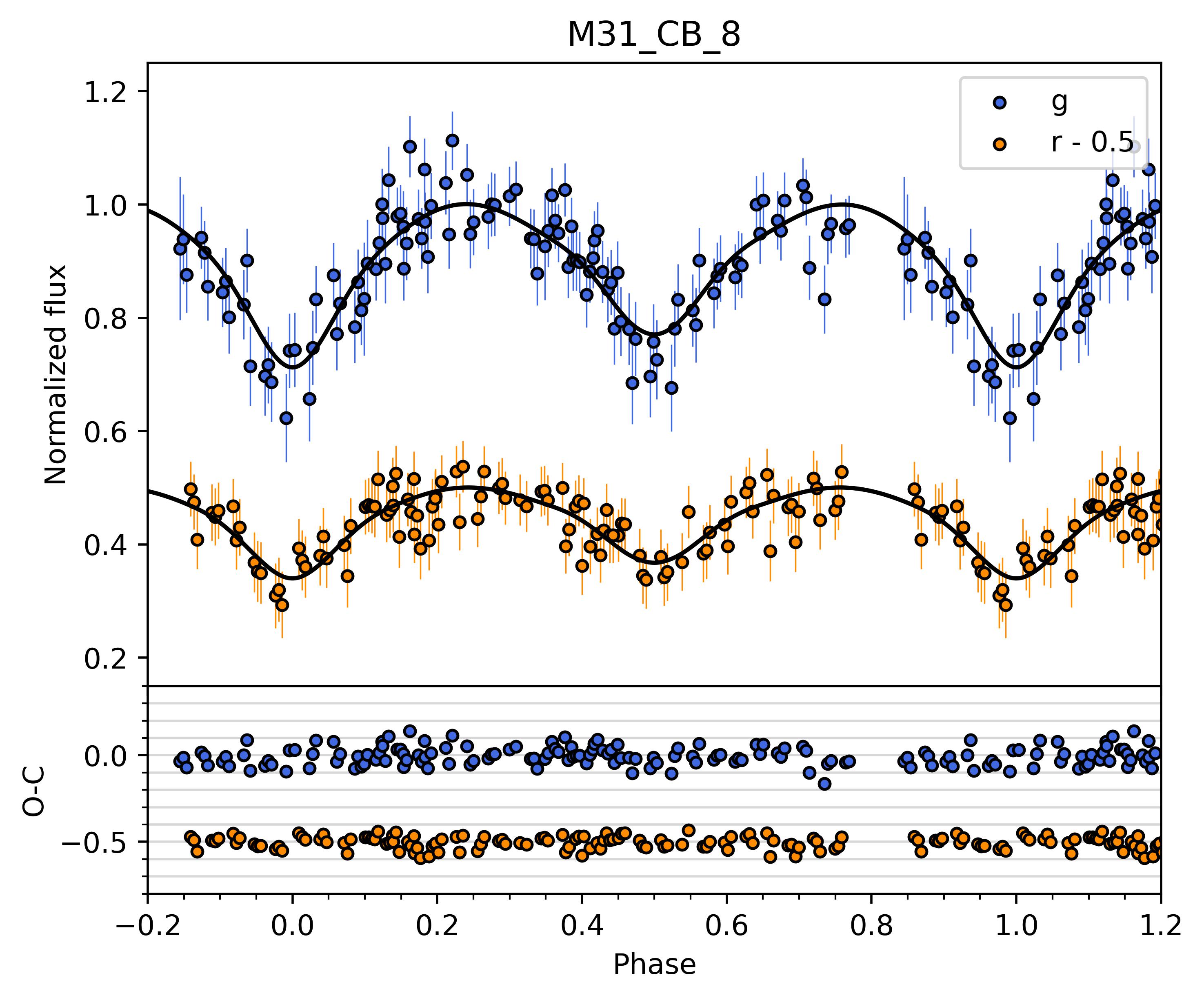}
    \end{subfigure}
    \begin{subfigure}
        \centering
        \includegraphics[width=0.29\textwidth]{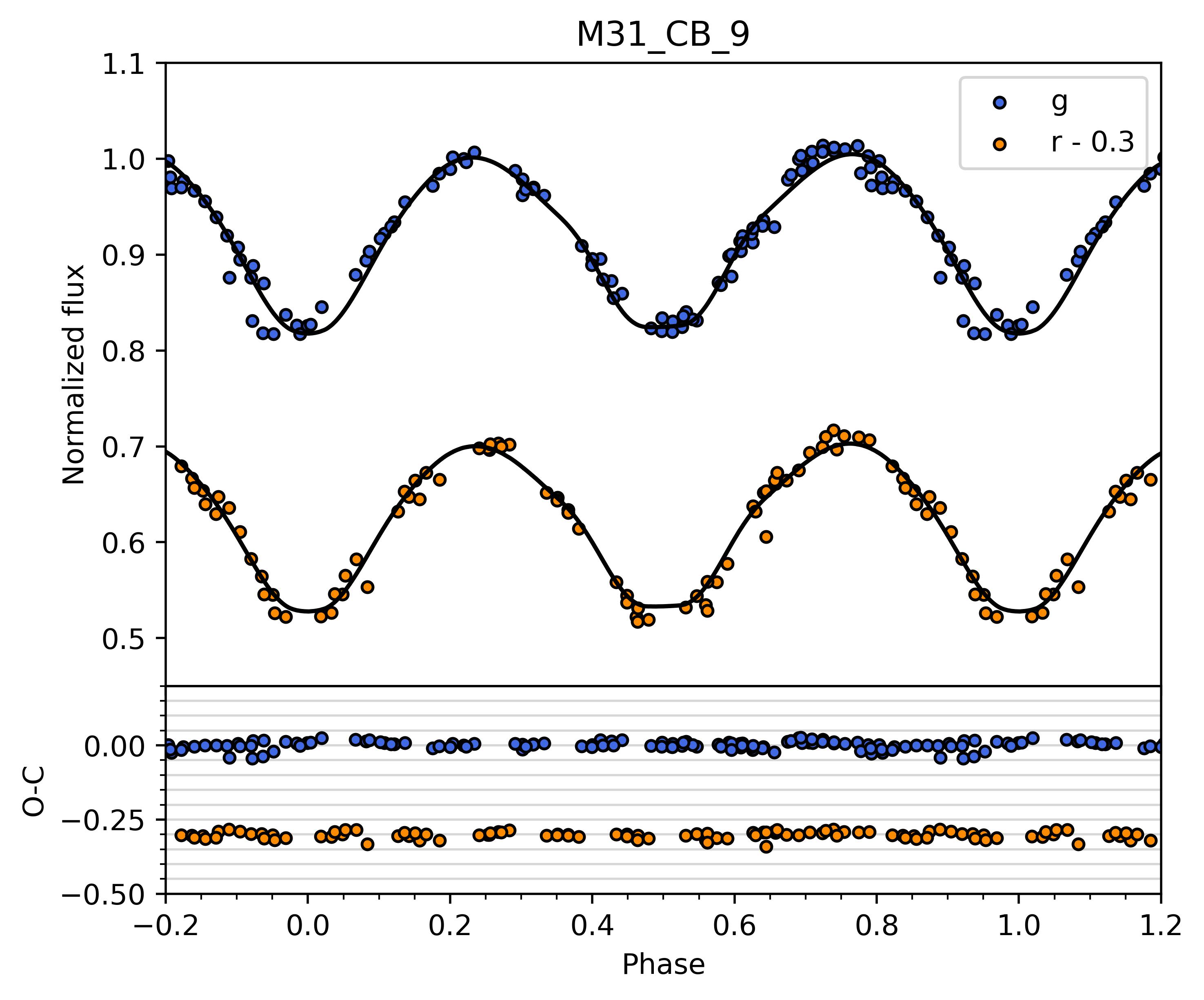}
    \end{subfigure}  

    \caption{Fitted light curves and residuals plots of other targets. In each panel, the upper shows the fitting results, with blue dots for g-band data and orange dots for r-band data, and the black curve indicates the fitting curve. The bottom of the panel shows the O-C residuals.}
    \label{figA1}
\end{figure}

\begin{figure}[htbp]
    \centering 
    \addtocounter{figure}{-1}  
    \begin{subfigure}
        \centering
        \includegraphics[width=0.29\textwidth]{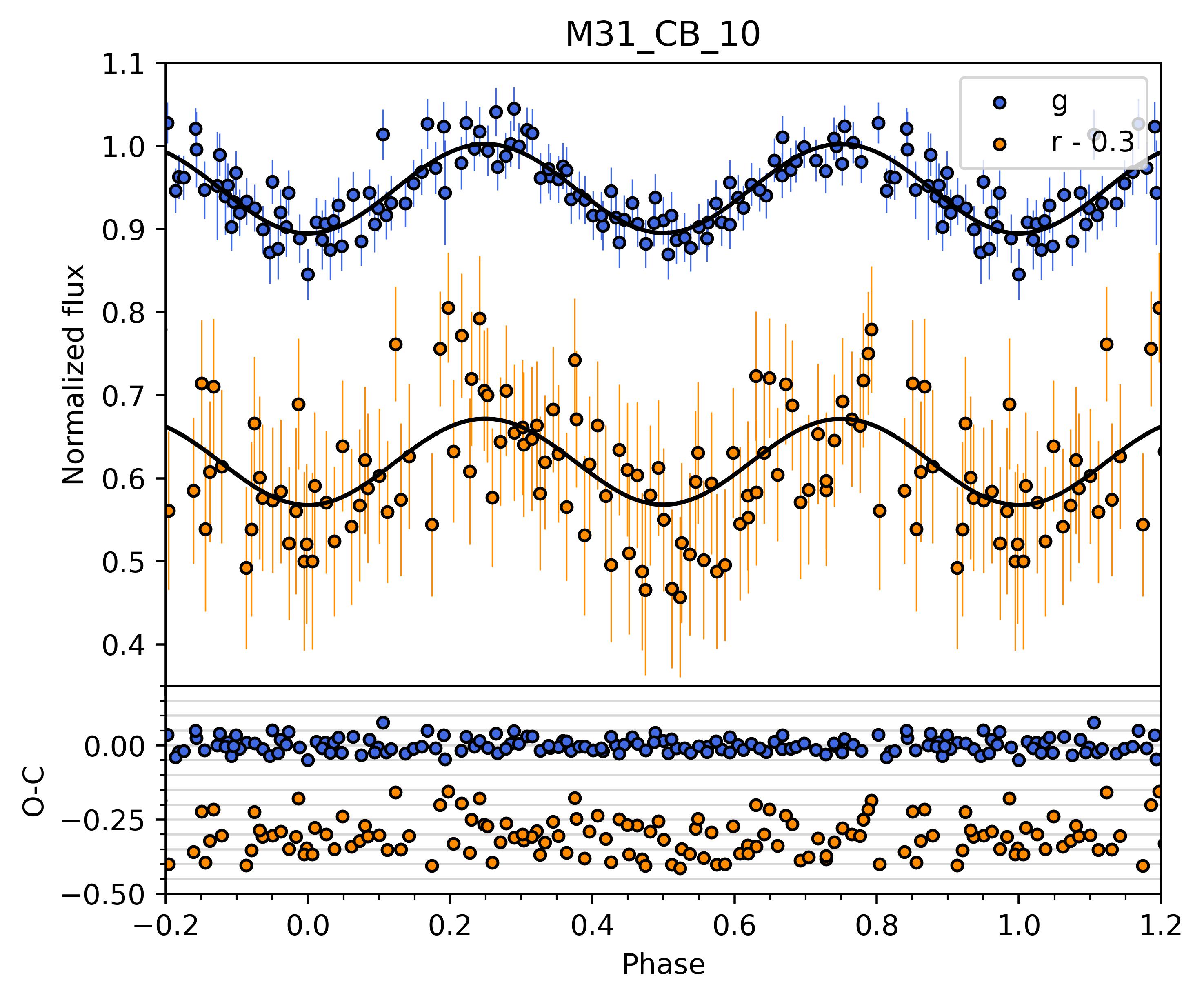}
    \end{subfigure}
    \begin{subfigure}
        \centering
        \includegraphics[width=0.29\textwidth]{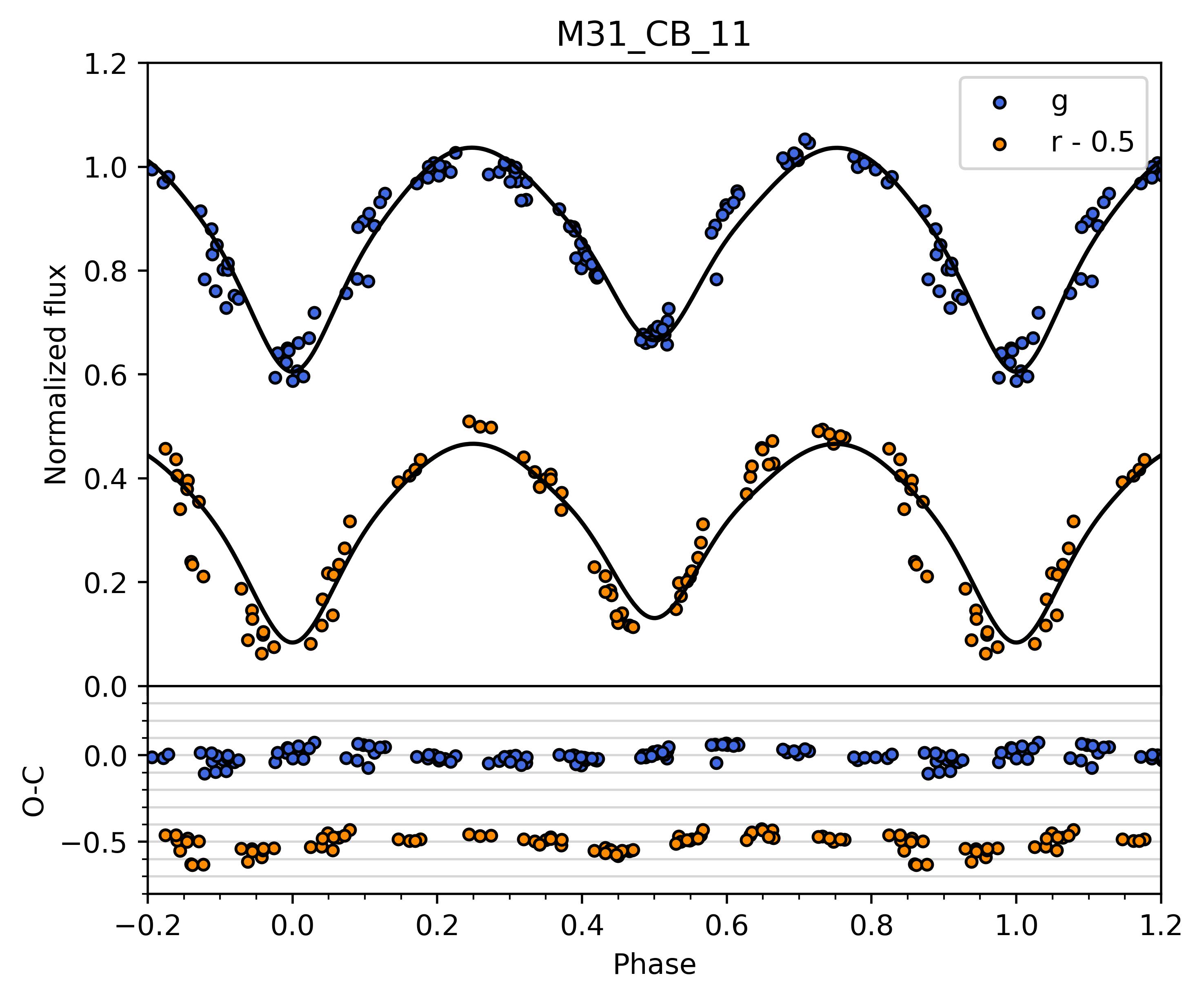}
    \end{subfigure}
    \begin{subfigure}
        \centering
        \includegraphics[width=0.29\textwidth]{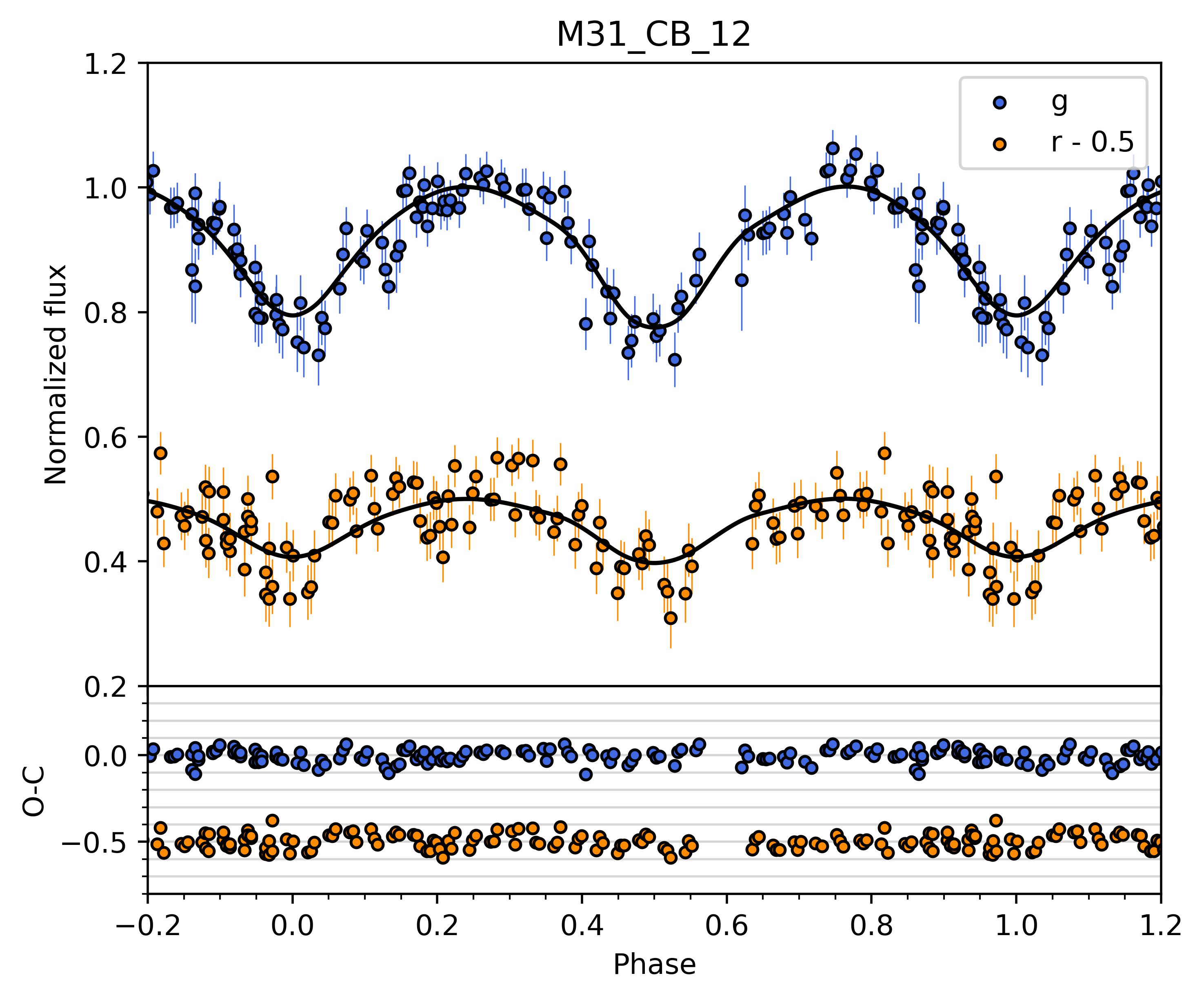}
    \end{subfigure}  
    
    \begin{subfigure}
        \centering
        \includegraphics[width=0.29\textwidth]{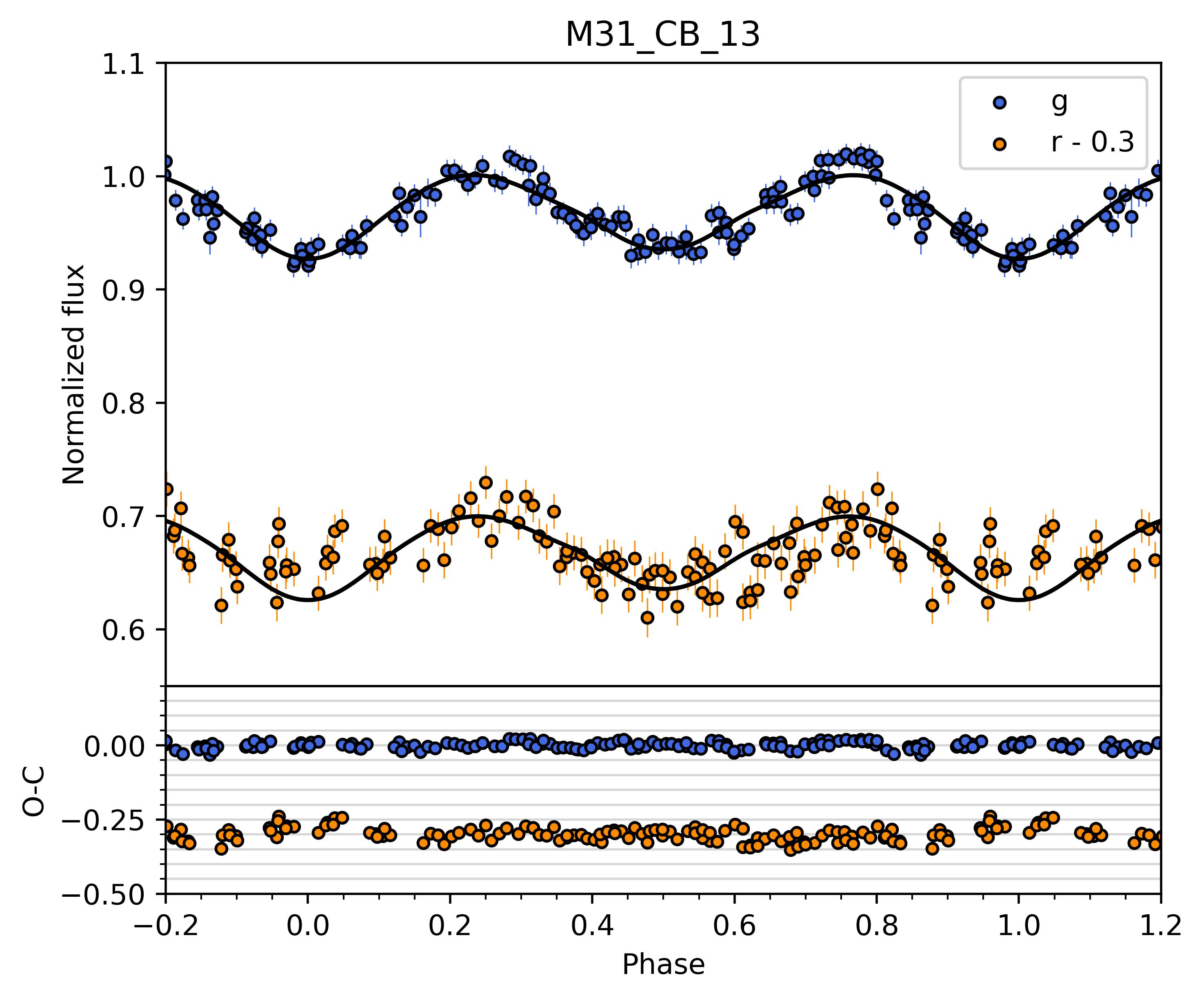}
    \end{subfigure}
    \begin{subfigure}
        \centering
        \includegraphics[width=0.29\textwidth]{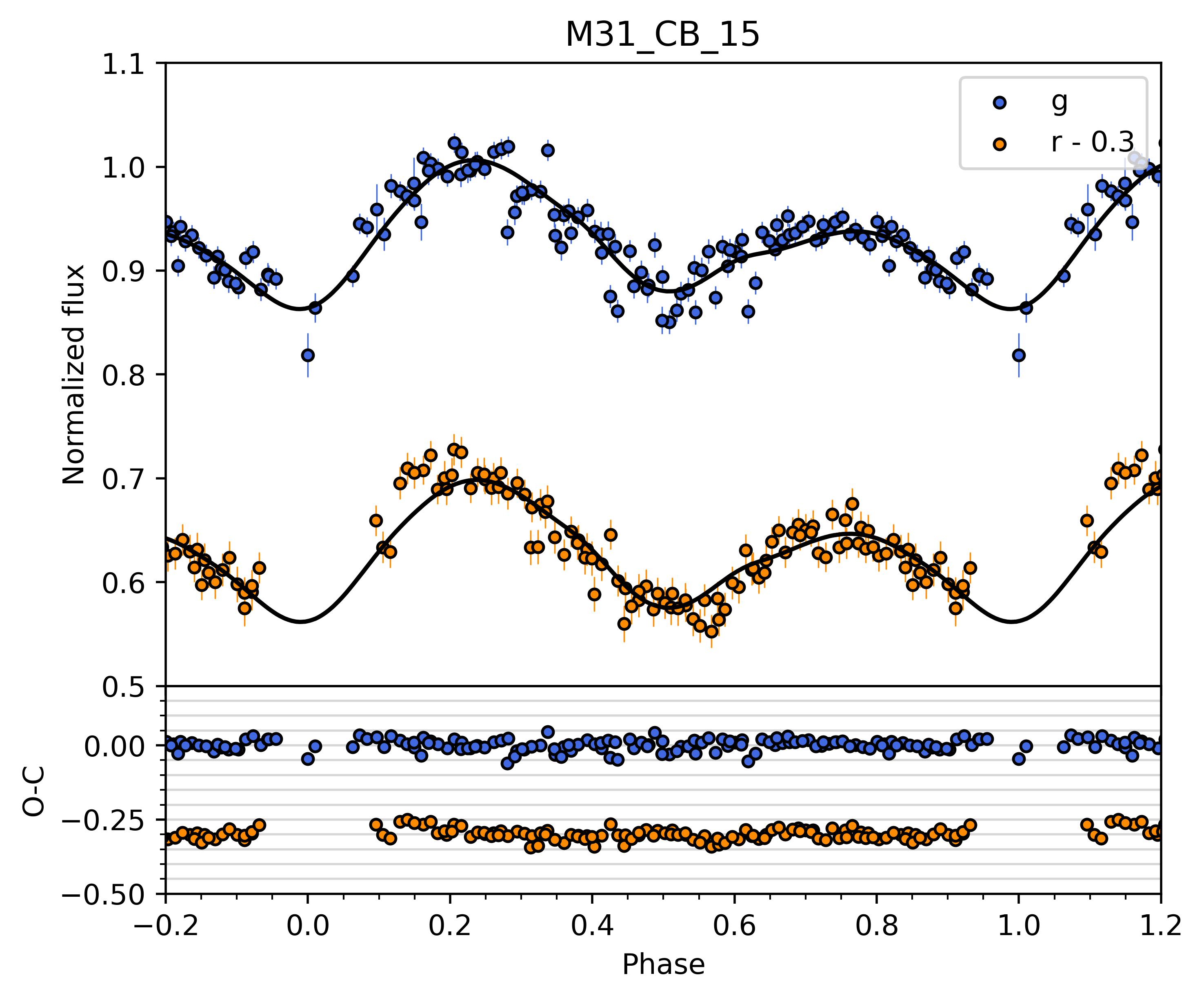}
    \end{subfigure}
    \begin{subfigure}
        \centering
        \includegraphics[width=0.29\textwidth]{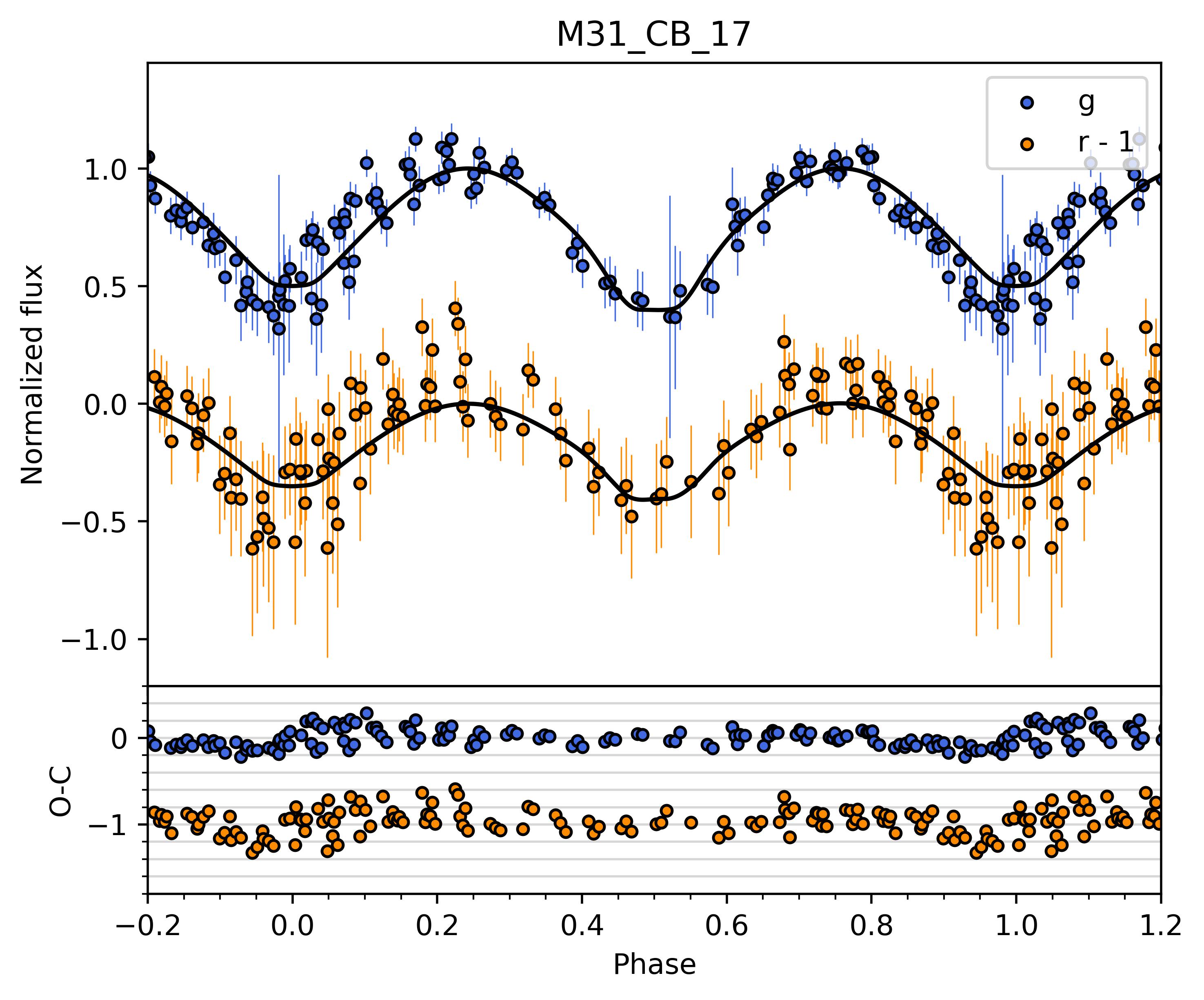}
    \end{subfigure} 
    
    \begin{subfigure}
        \centering
        \includegraphics[width=0.29\textwidth]{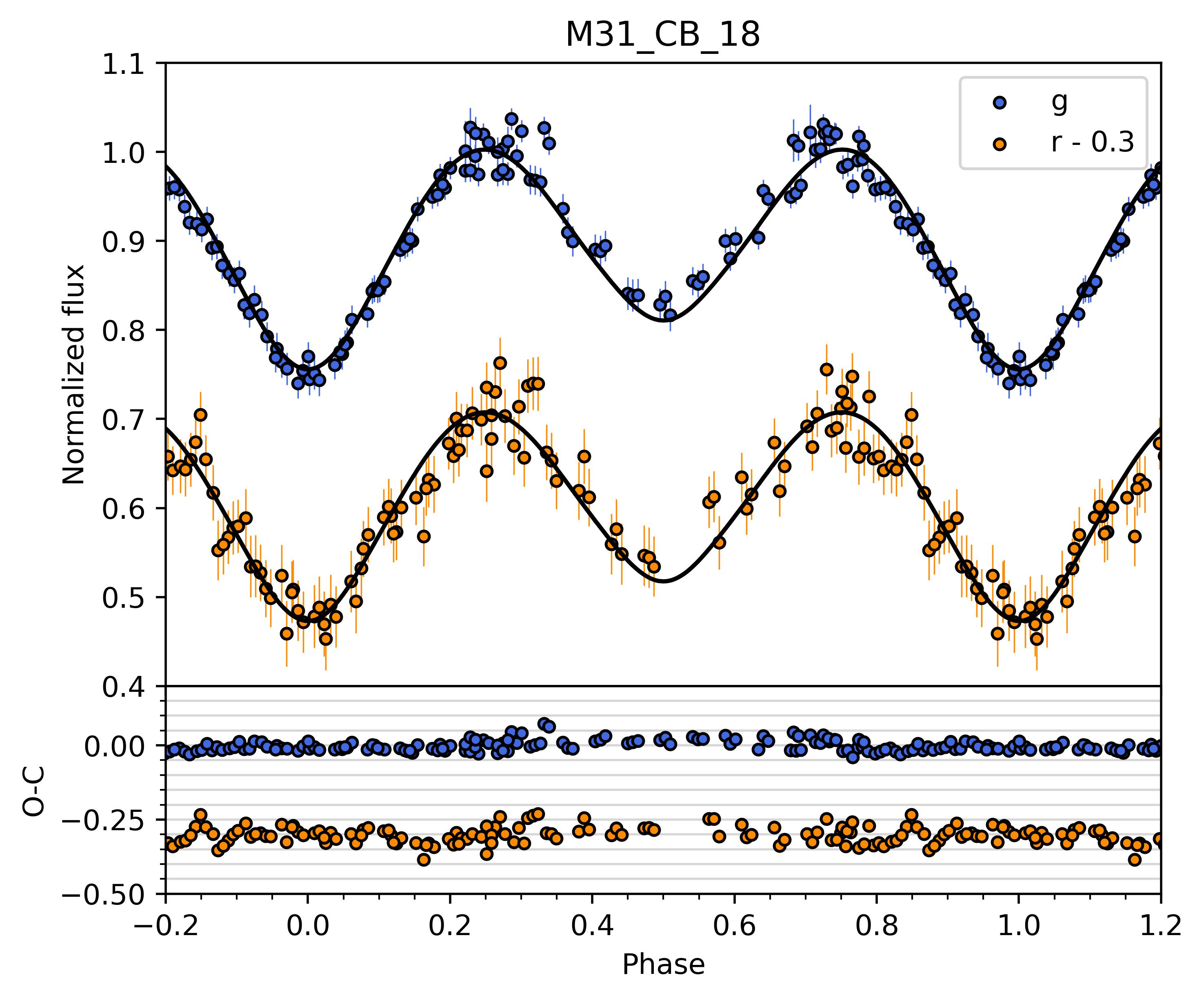}
    \end{subfigure}
    \begin{subfigure}
        \centering
        \includegraphics[width=0.29\textwidth]{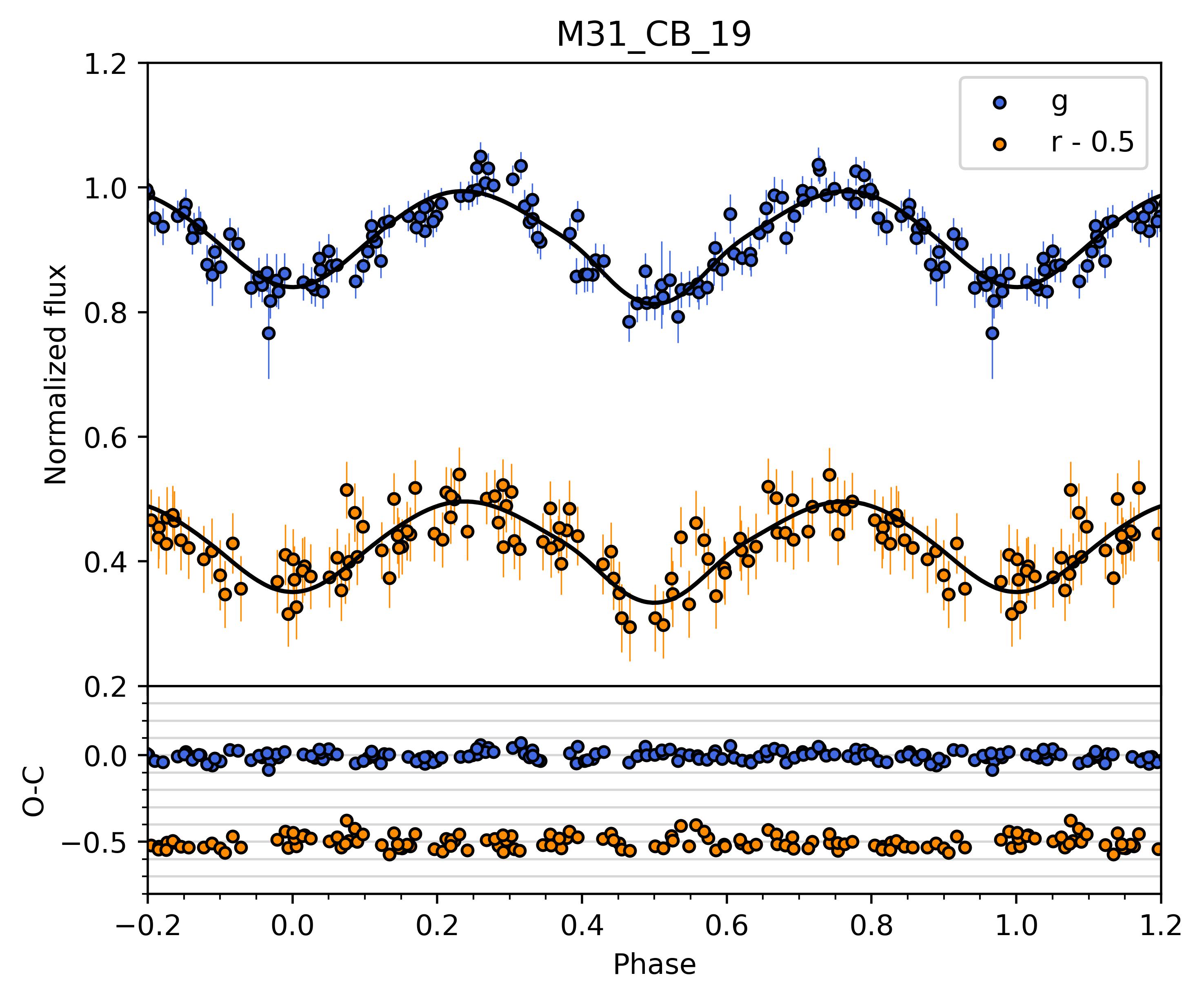}
    \end{subfigure}
    \begin{subfigure}
        \centering
        \includegraphics[width=0.29\textwidth]{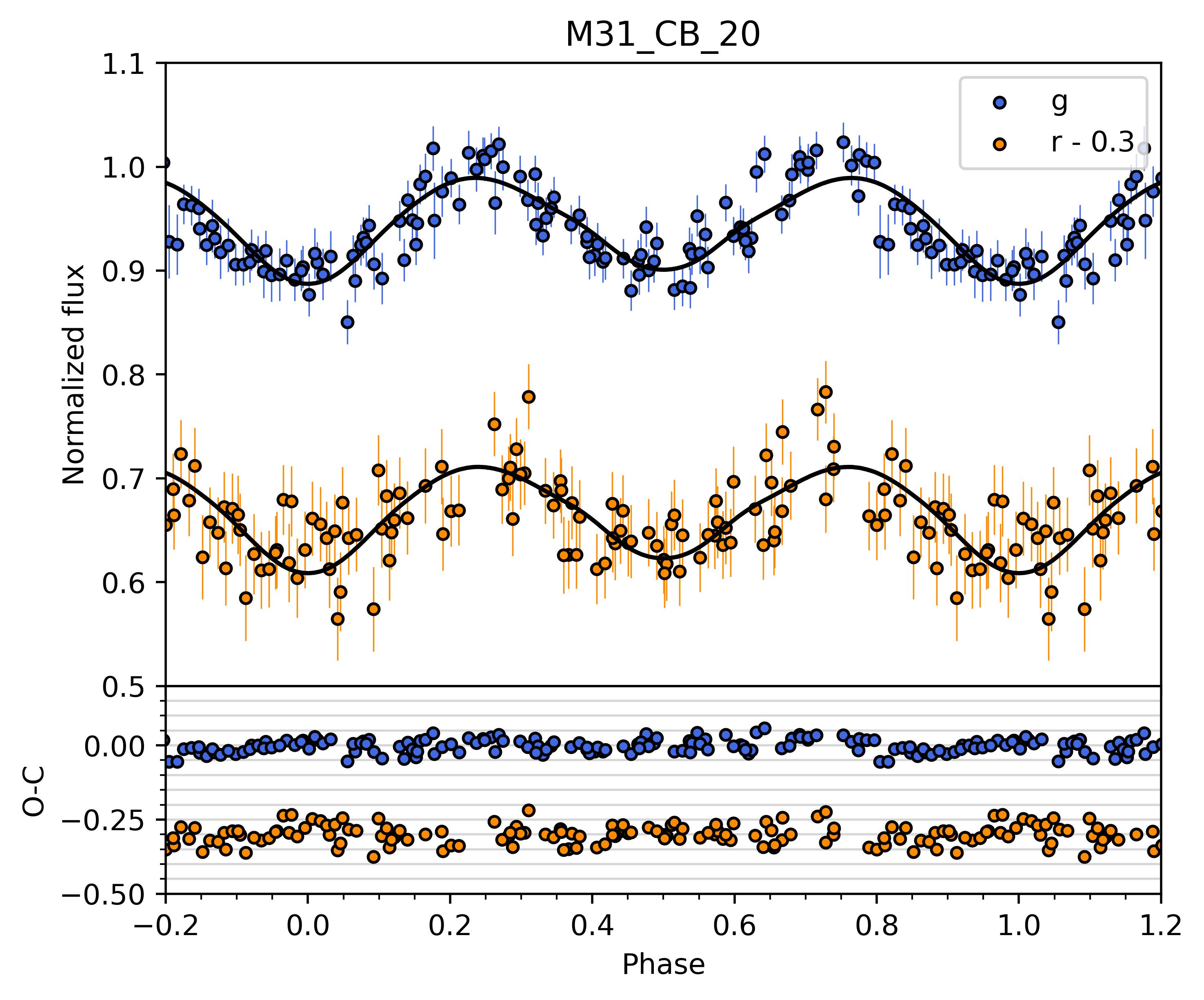}
    \end{subfigure}  

    \begin{subfigure}
        \centering
        \includegraphics[width=0.29\textwidth]{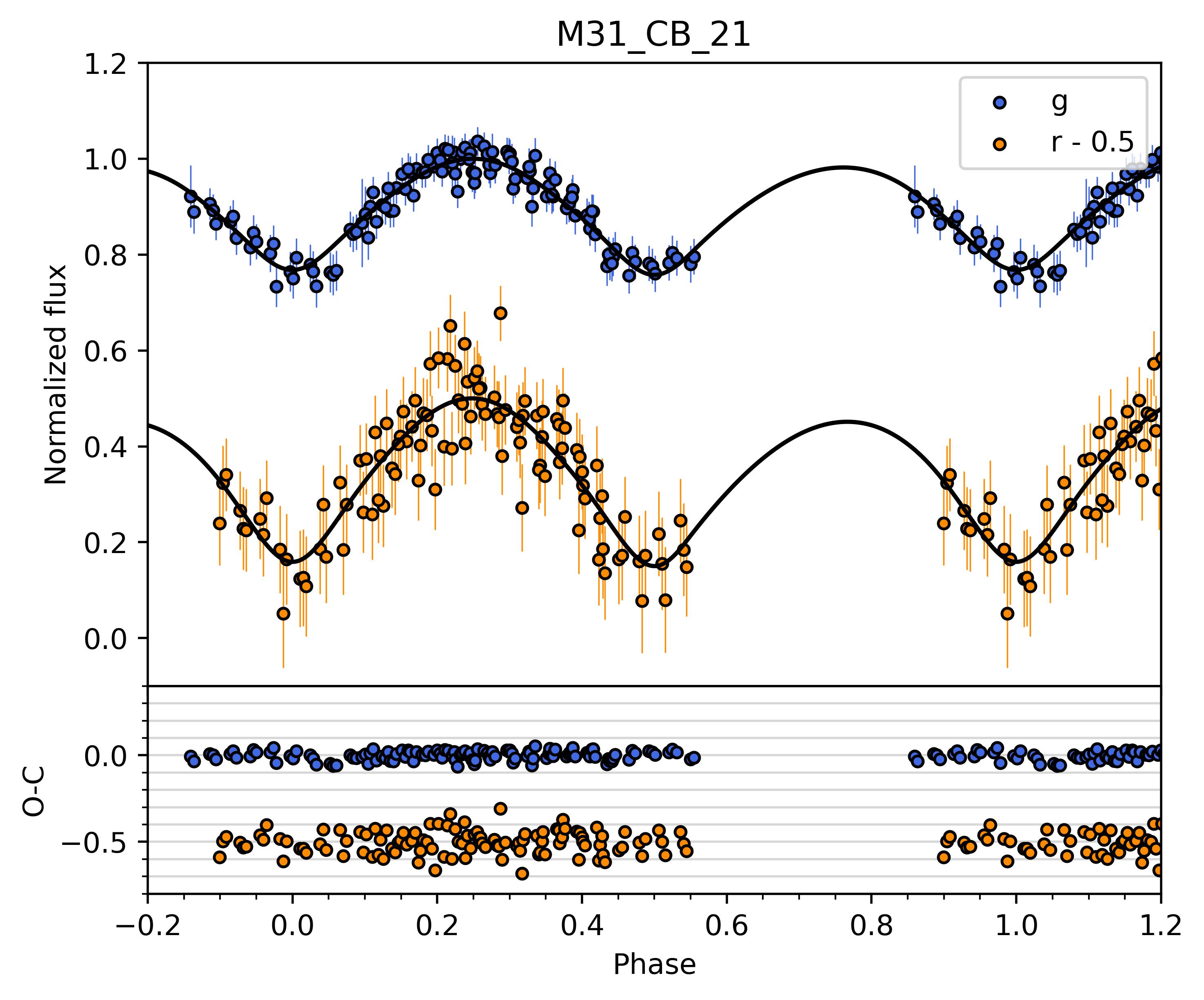}
    \end{subfigure}
    \begin{subfigure}
        \centering
        \includegraphics[width=0.29\textwidth]{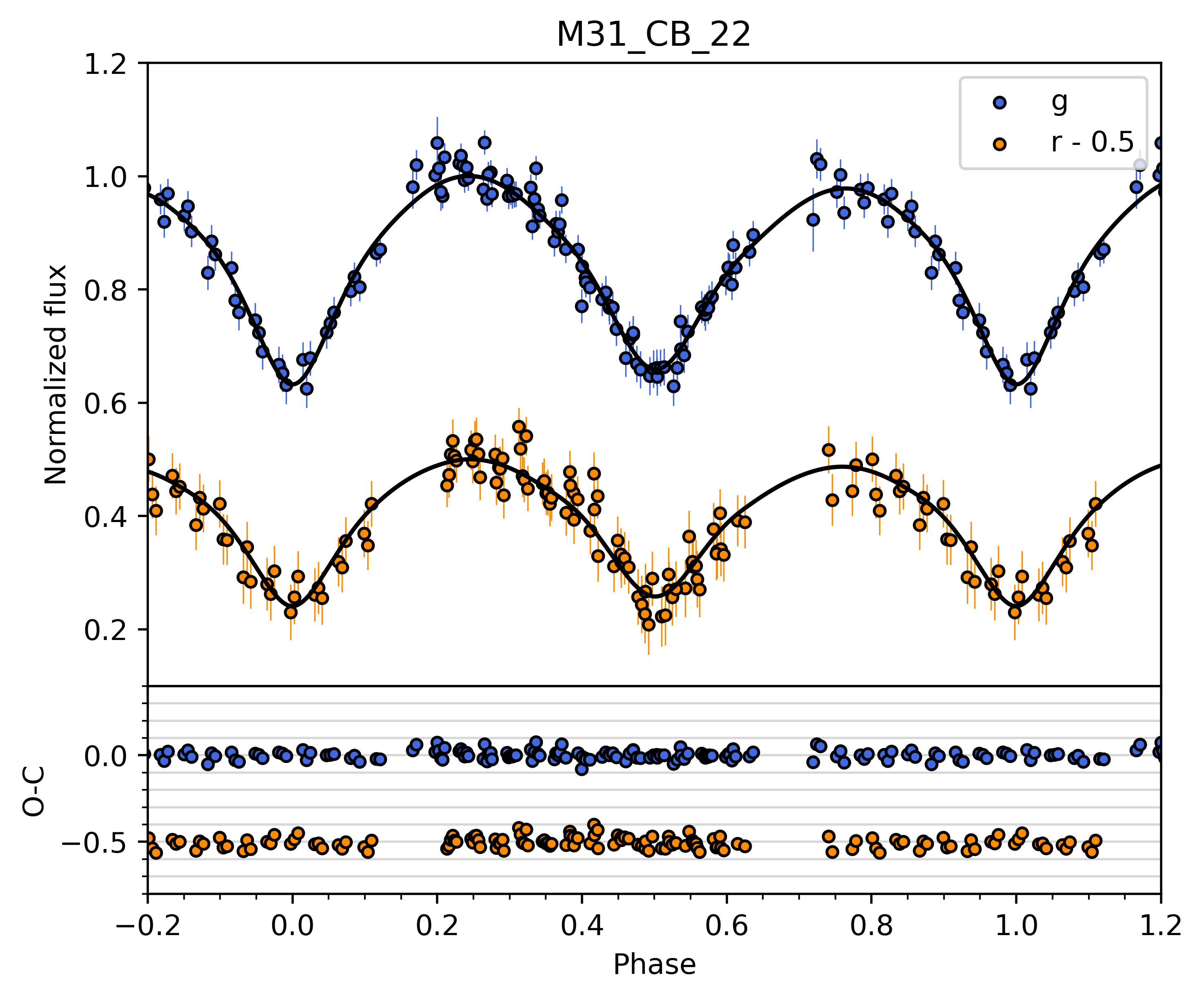}
    \end{subfigure}
    \begin{subfigure}
        \centering
        \includegraphics[width=0.29\textwidth]{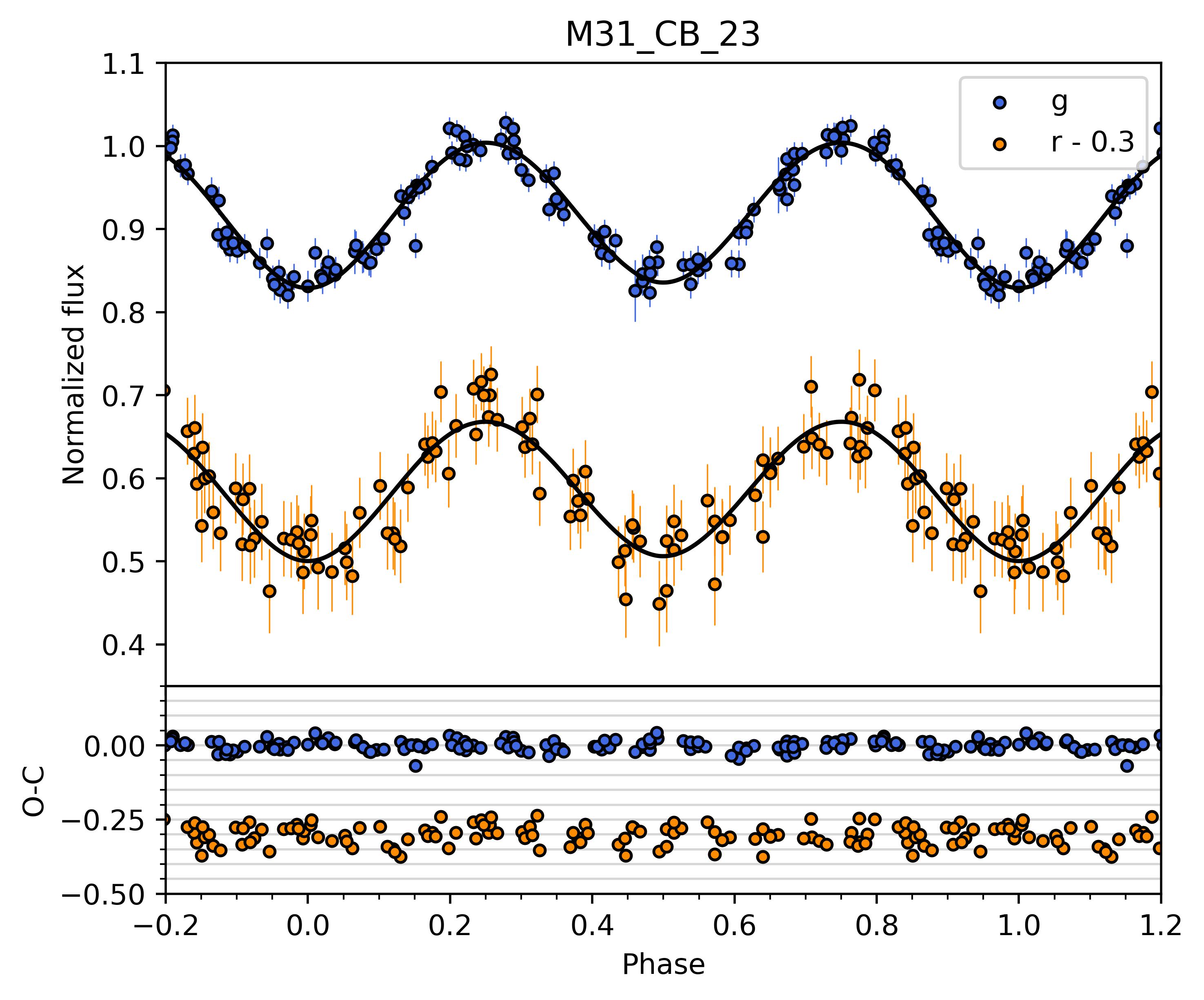}
    \end{subfigure}  

    \begin{subfigure}
        \centering
        \includegraphics[width=0.29\textwidth]{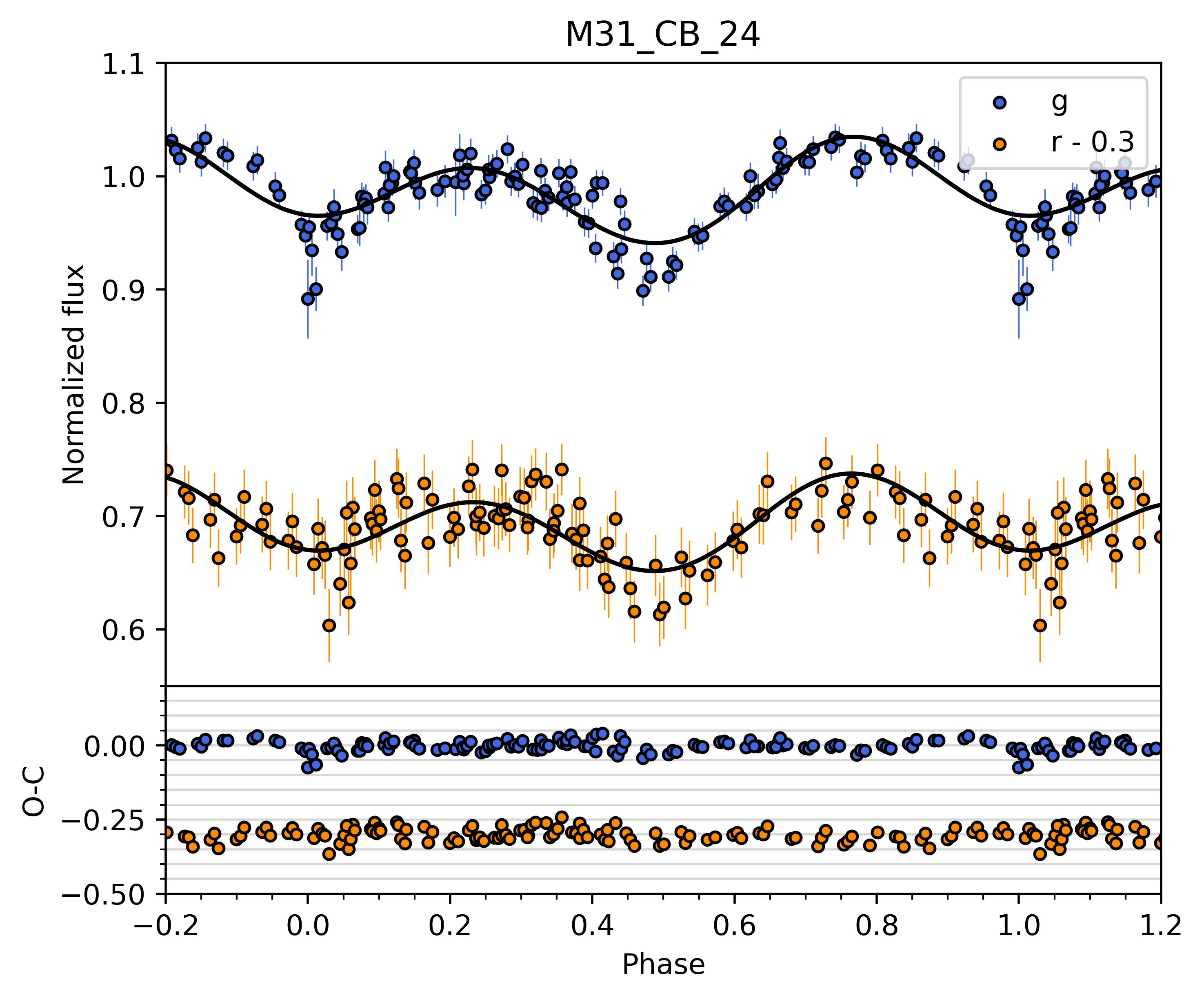}
    \end{subfigure}
    \begin{subfigure}
        \centering
        \includegraphics[width=0.29\textwidth]{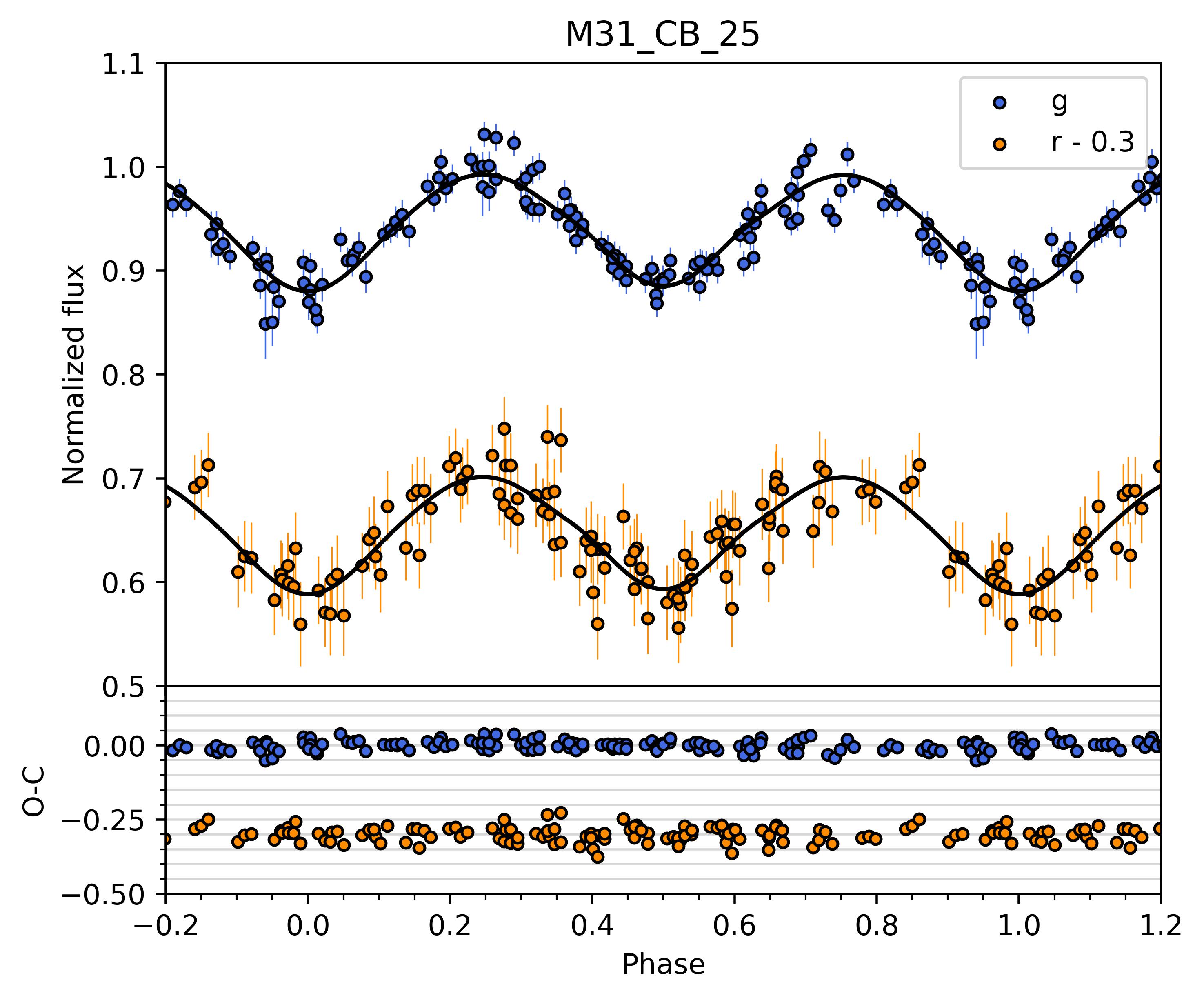}
    \end{subfigure}
    \begin{subfigure}
        \centering
        \includegraphics[width=0.29\textwidth]{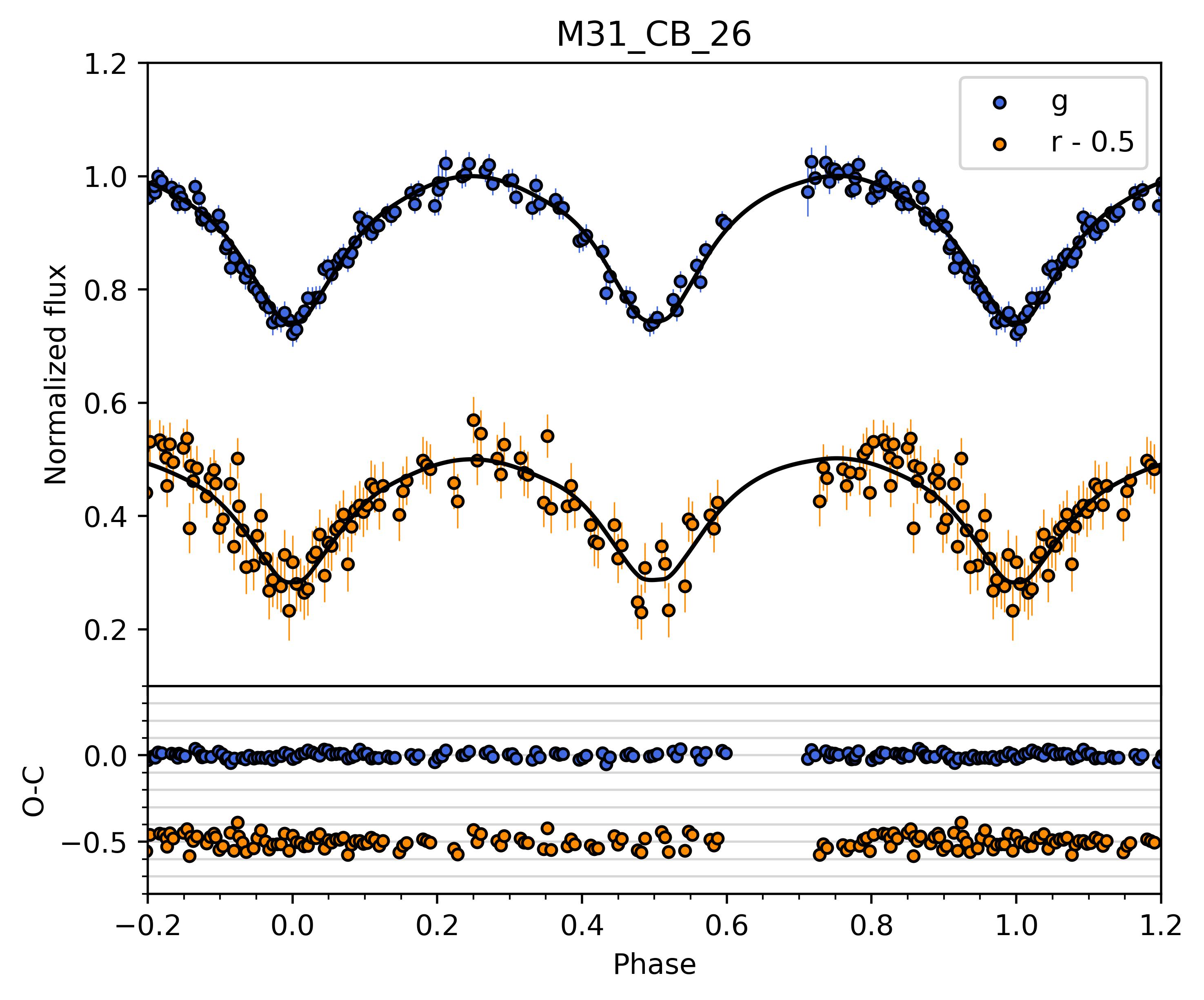}
    \end{subfigure}  

    \caption{(continued)}
    \label{figA1-1}
\end{figure}

\begin{figure}[htbp]
    \centering 
    \addtocounter{figure}{-1}  
    
    \begin{subfigure}
        \centering
        \includegraphics[width=0.29\textwidth]{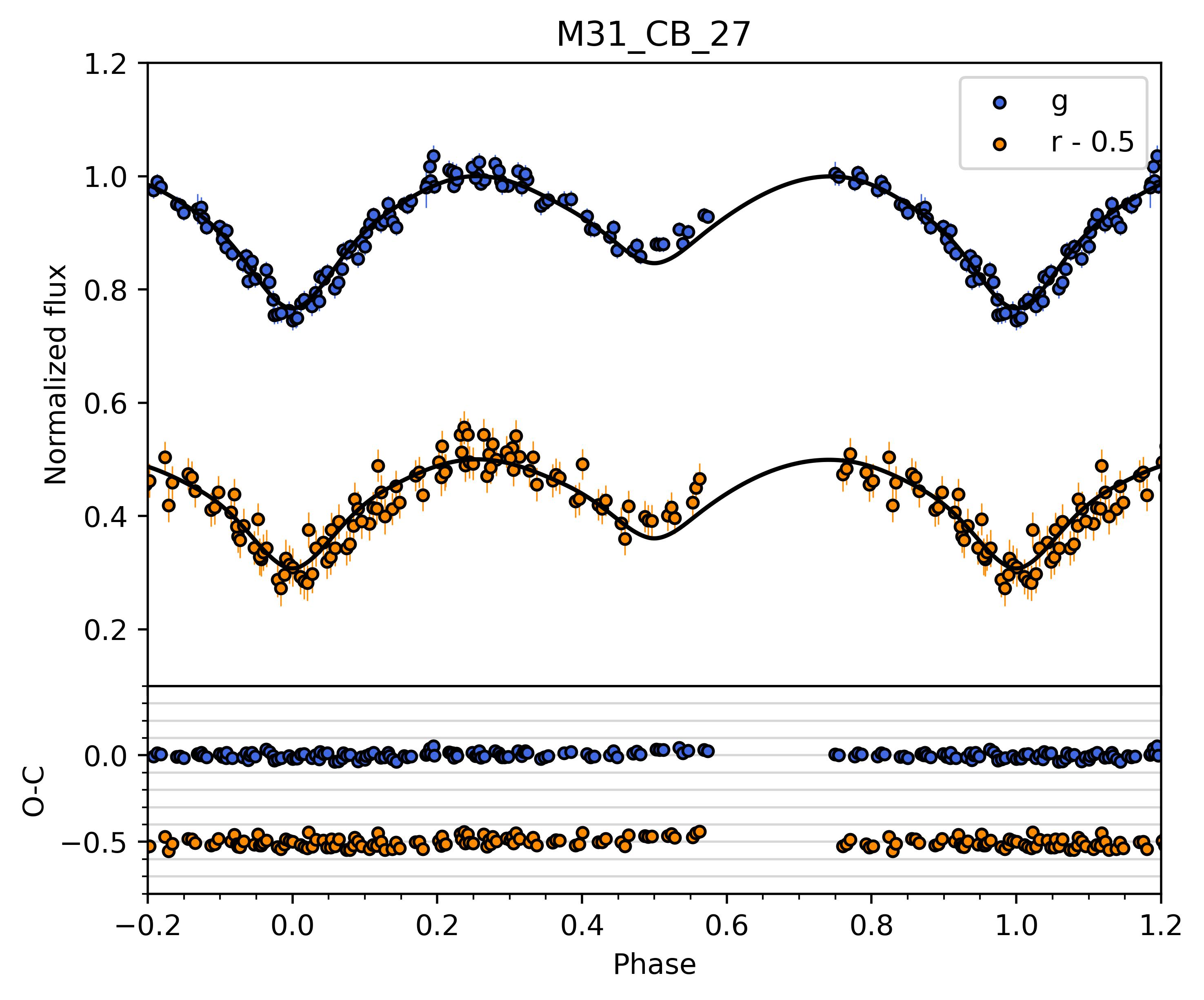}
    \end{subfigure}
    \begin{subfigure}
        \centering
        \includegraphics[width=0.29\textwidth]{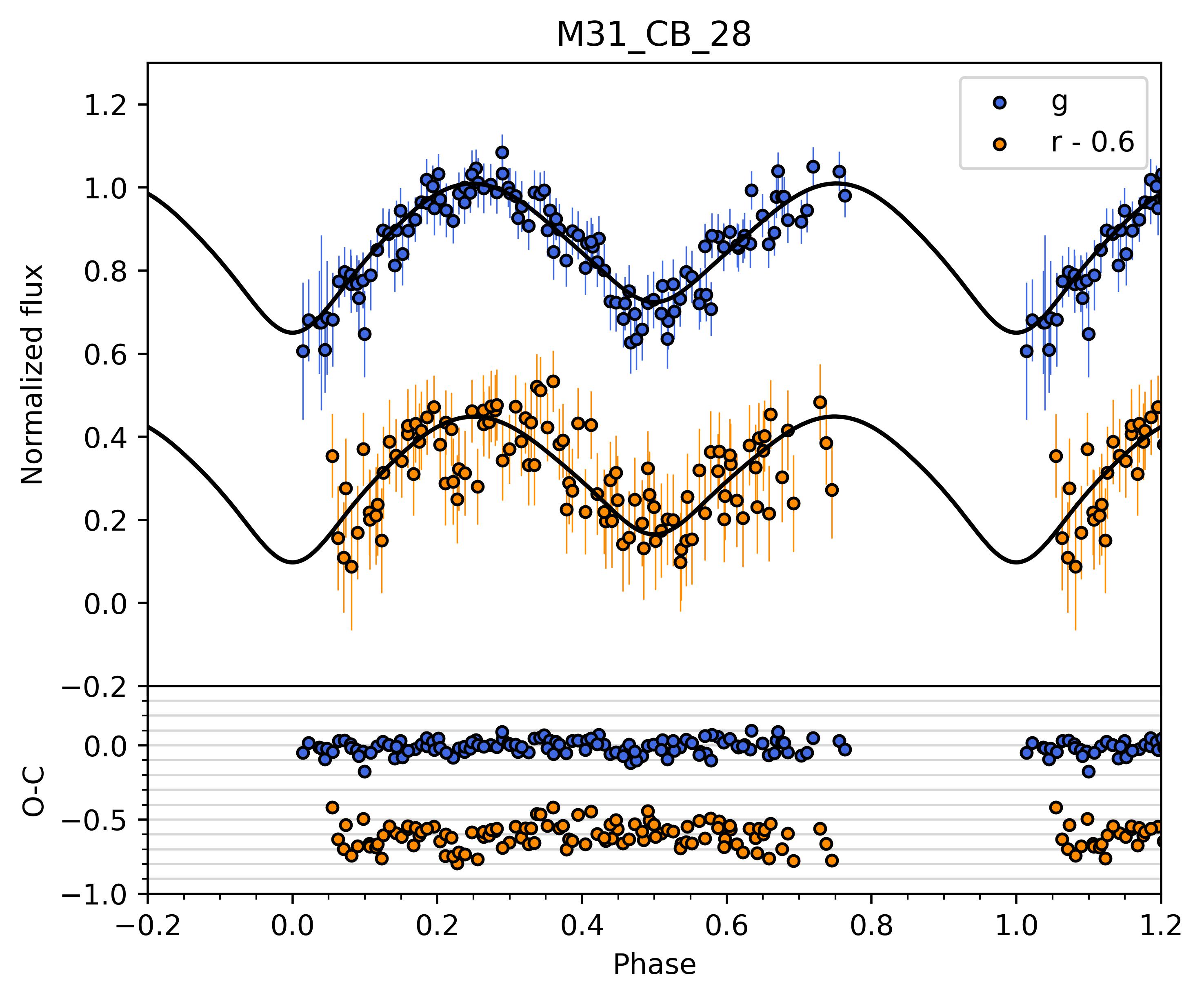}
    \end{subfigure}
    \begin{subfigure}
        \centering
        \includegraphics[width=0.29\textwidth]{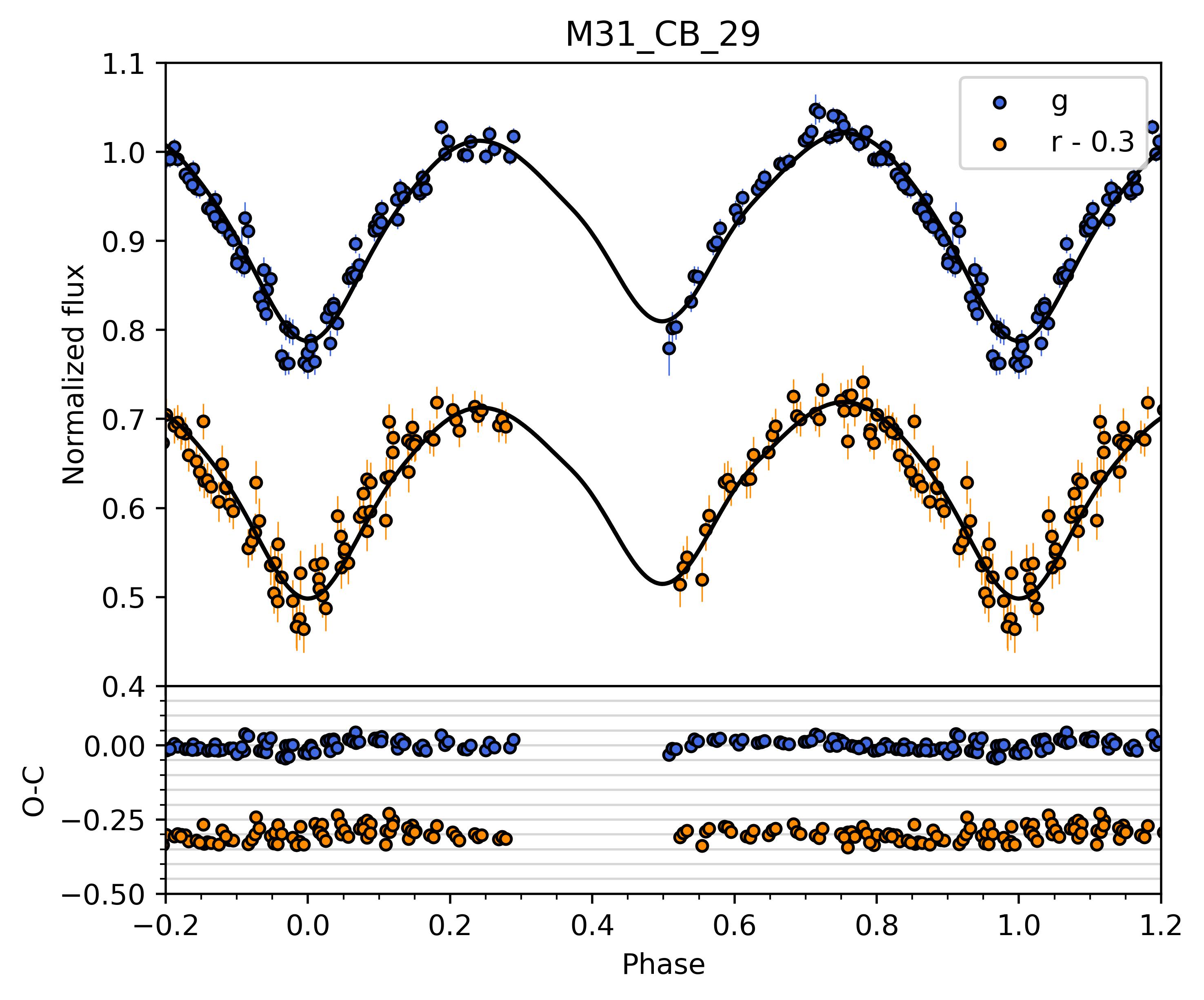}
    \end{subfigure}  
    
    \begin{subfigure}
        \centering
        \includegraphics[width=0.29\textwidth]{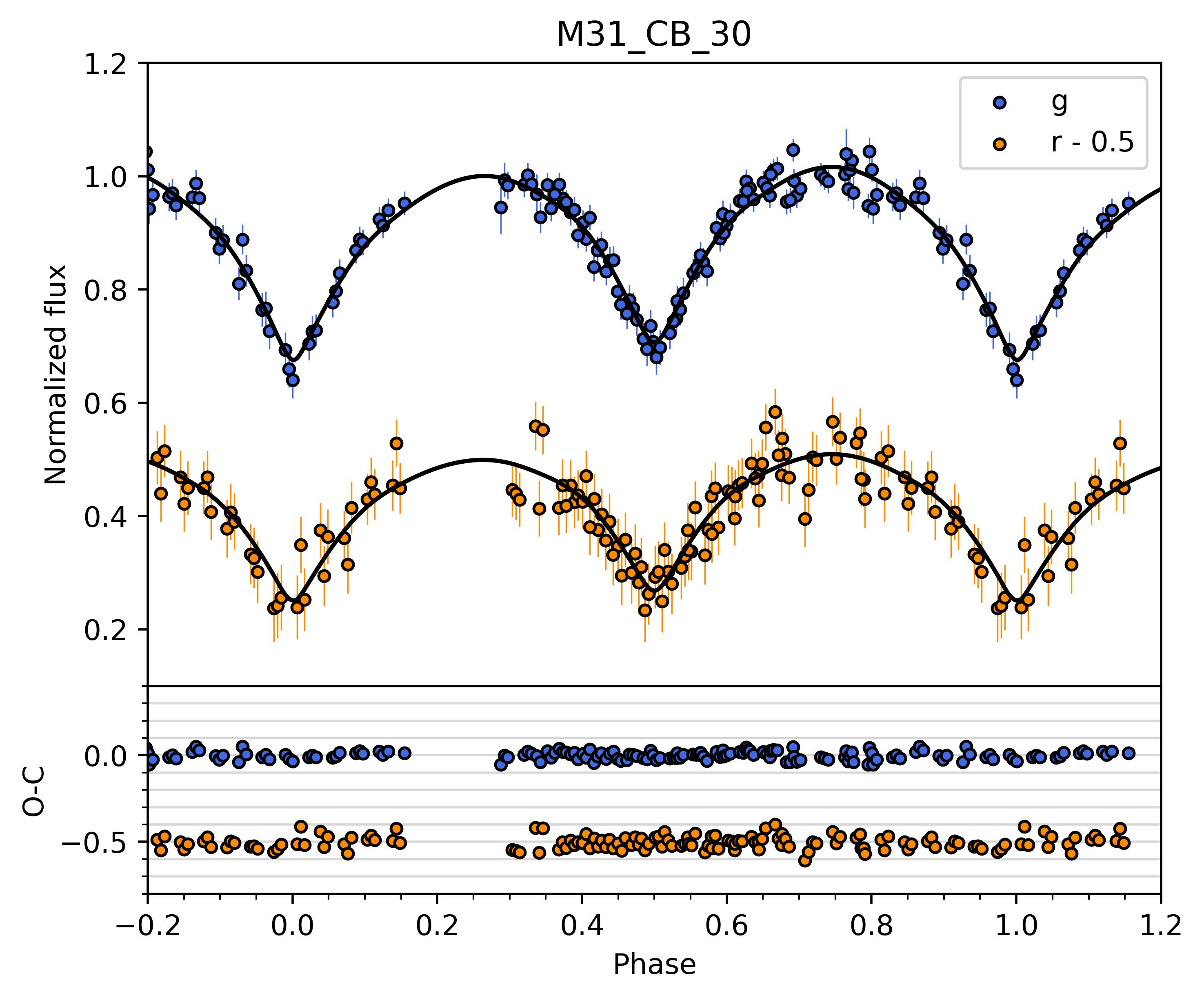}
    \end{subfigure}
    \caption{(continued)}
    \label{figA1-2}
\end{figure}

\begin{figure}[htbp]
    \centering 
    
    \begin{subfigure}
        \centering
        \includegraphics[width=0.24\textwidth]{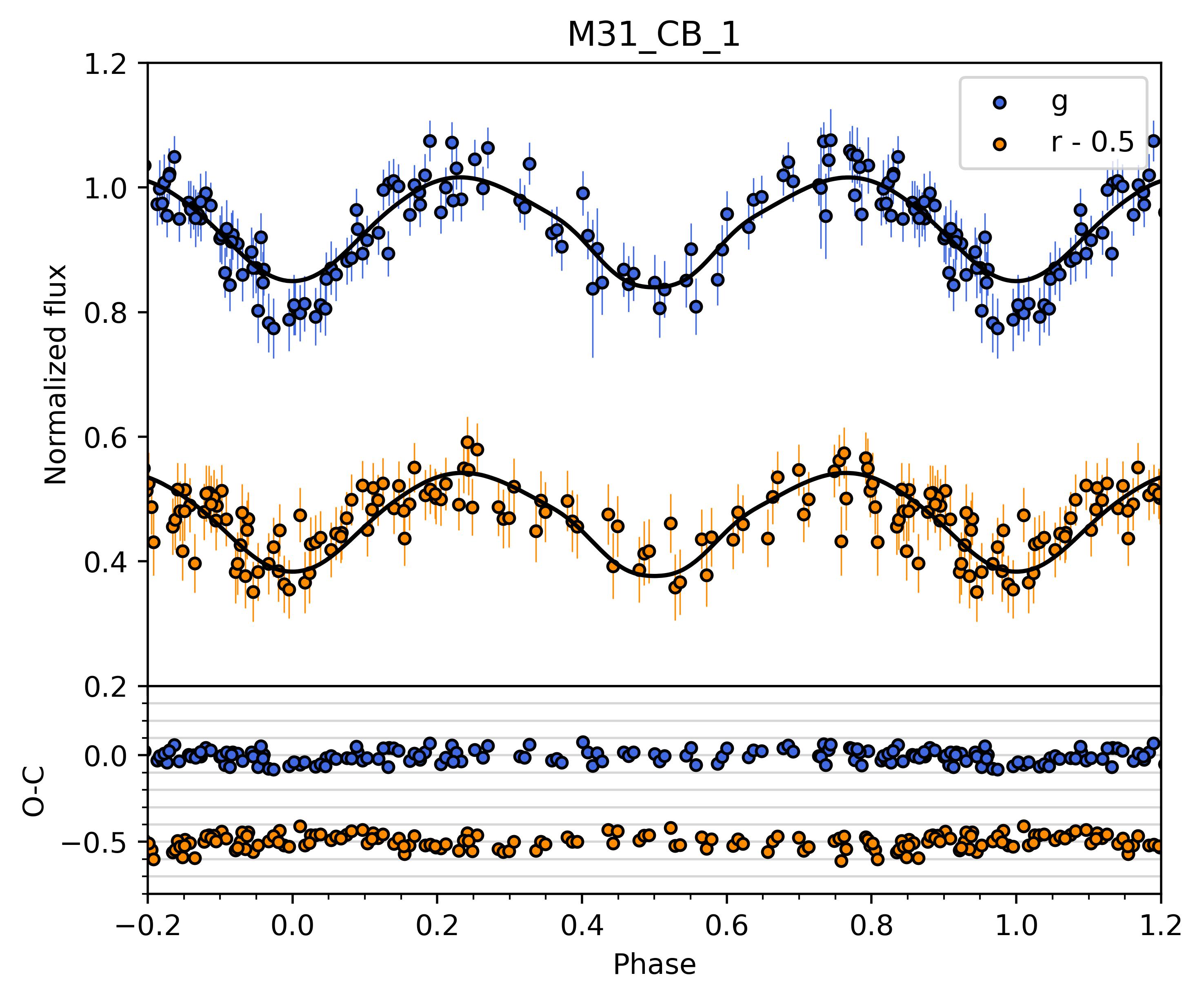}
    \end{subfigure}
    \begin{subfigure}
        \centering
        \includegraphics[width=0.24\textwidth]{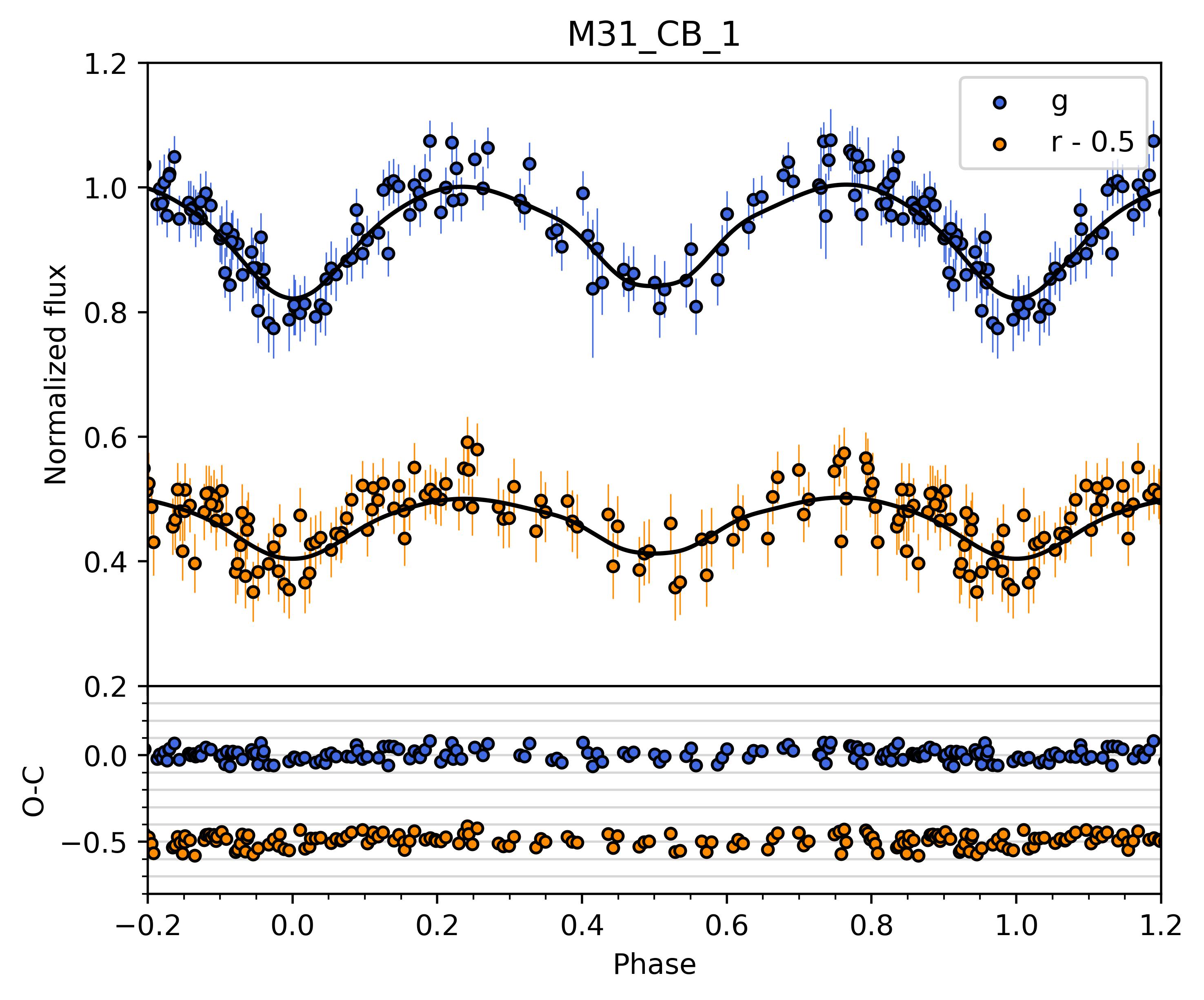}
    \end{subfigure}
    \begin{subfigure}
        \centering
        \includegraphics[width=0.24\textwidth]{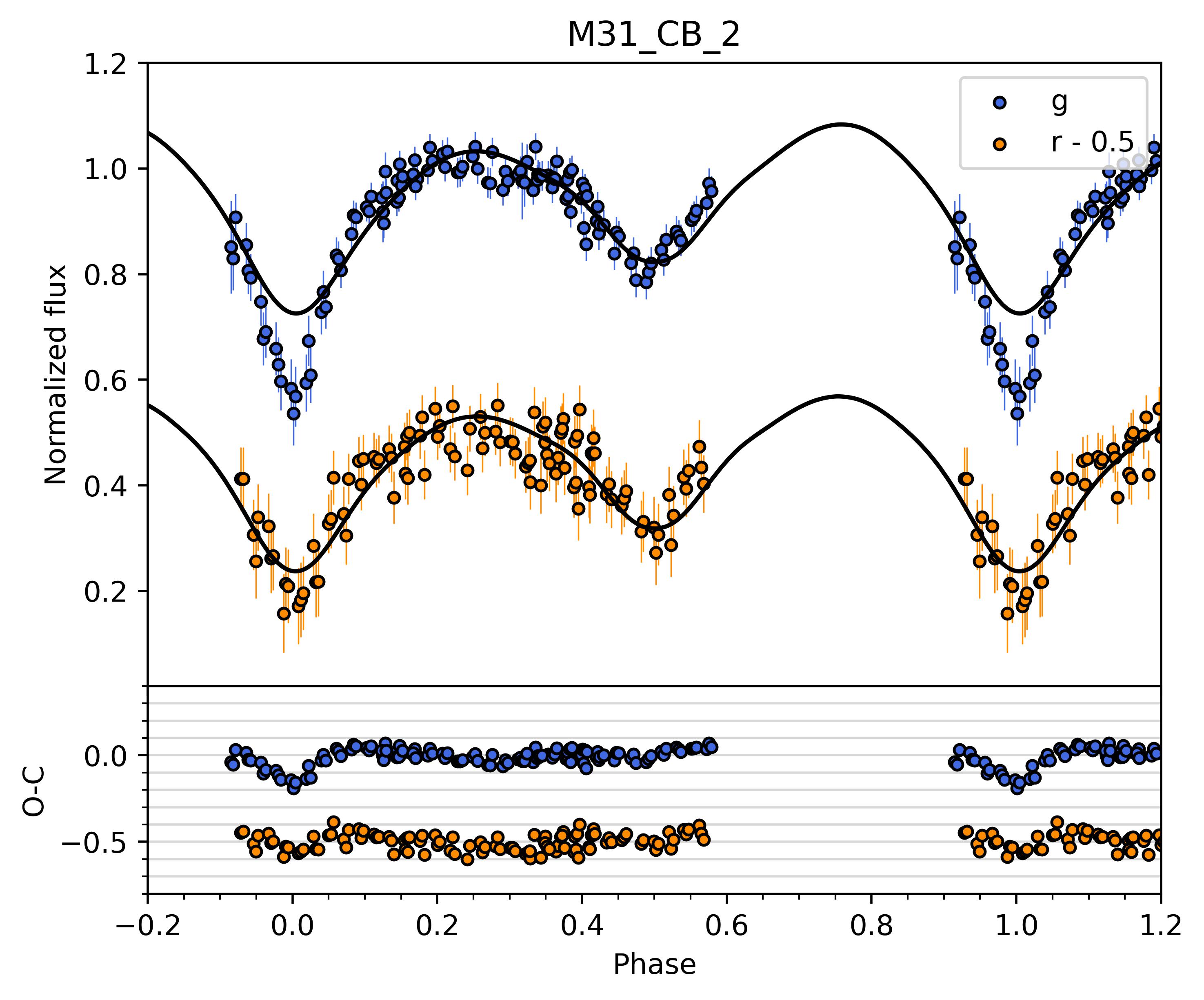}
    \end{subfigure}
    \begin{subfigure}
        \centering
        \includegraphics[width=0.24\textwidth]{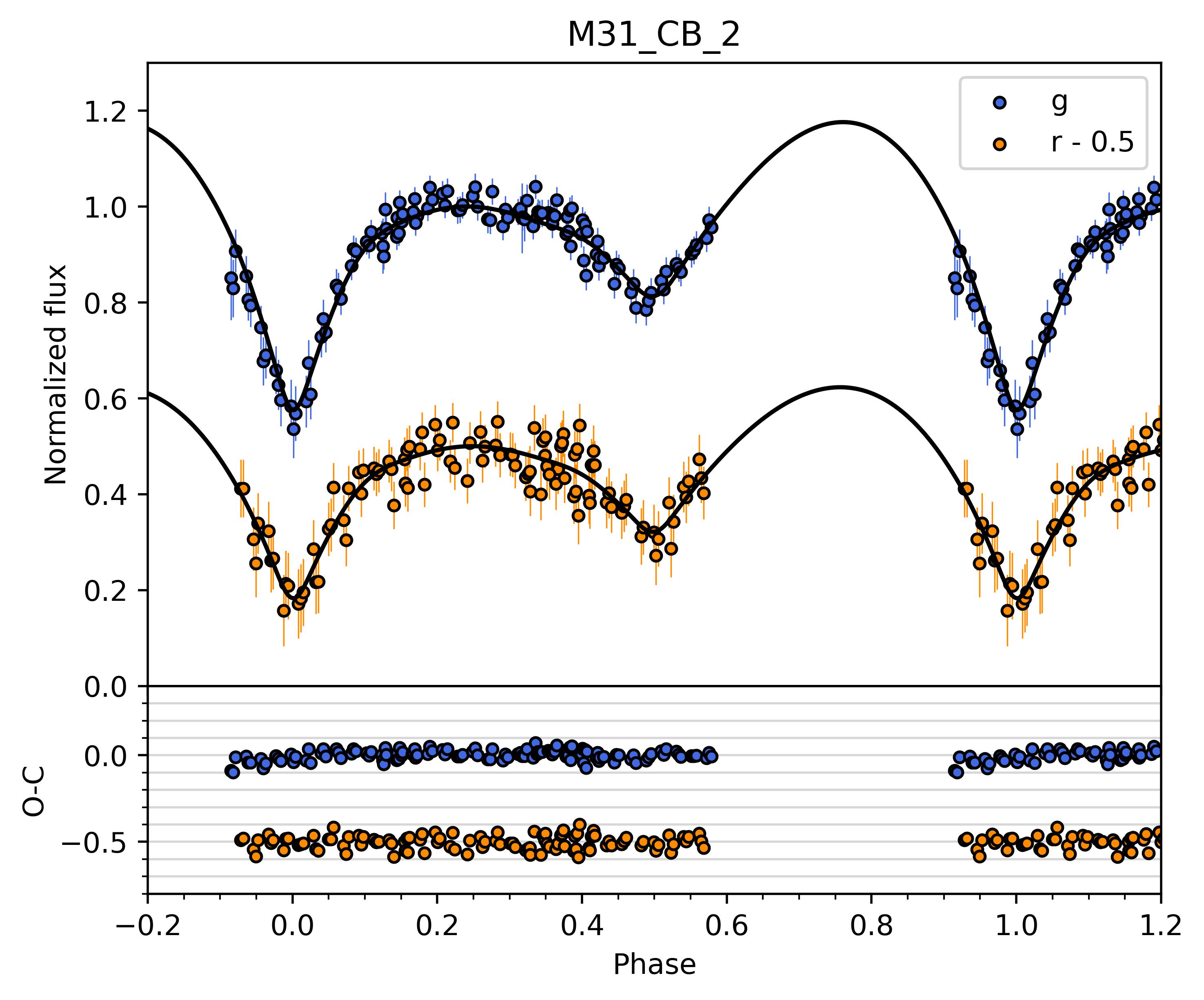}
    \end{subfigure}
    
    \begin{subfigure}
        \centering
        \includegraphics[width=0.24\textwidth]{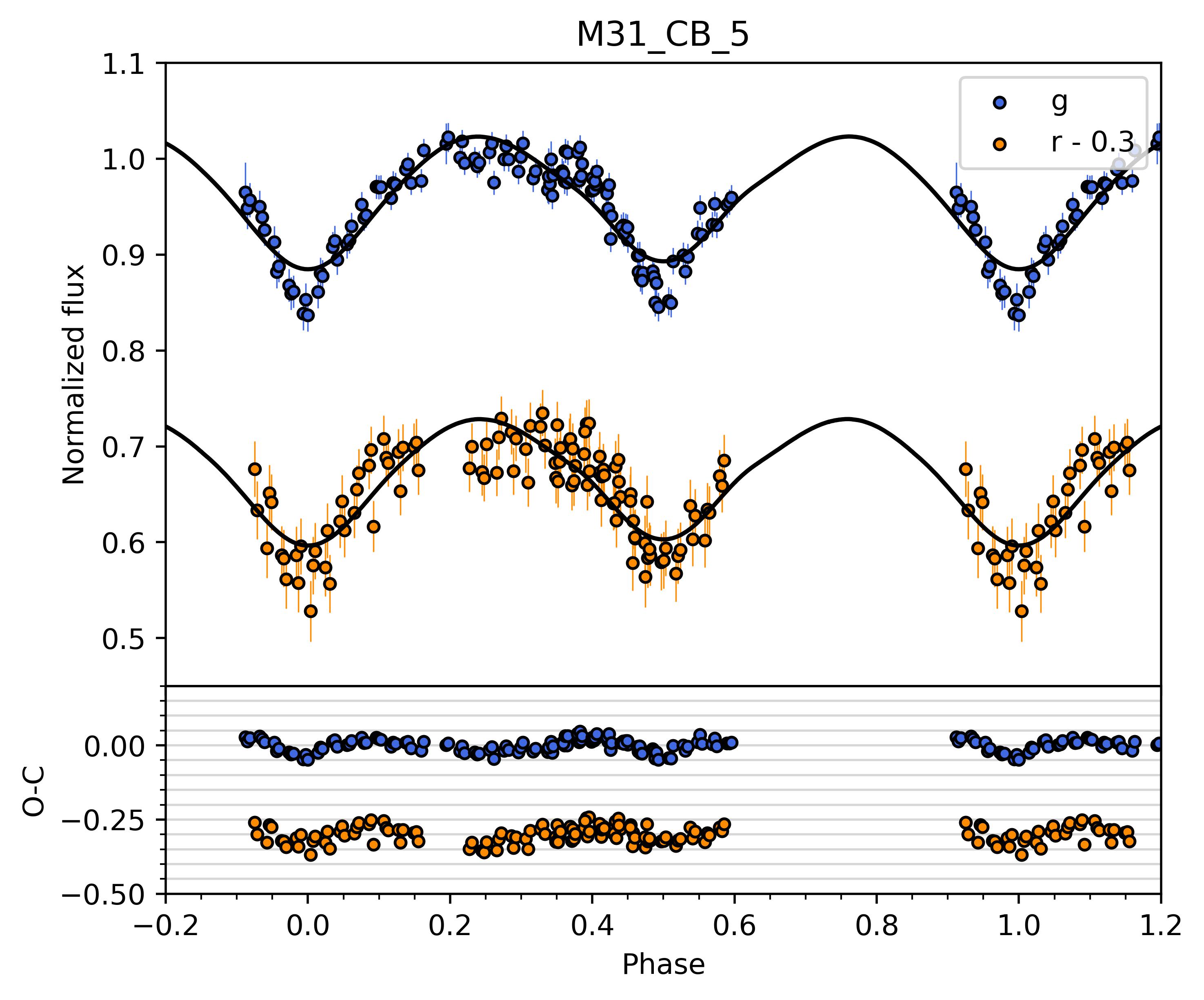}
    \end{subfigure}
    \begin{subfigure}
        \centering
        \includegraphics[width=0.24\textwidth]{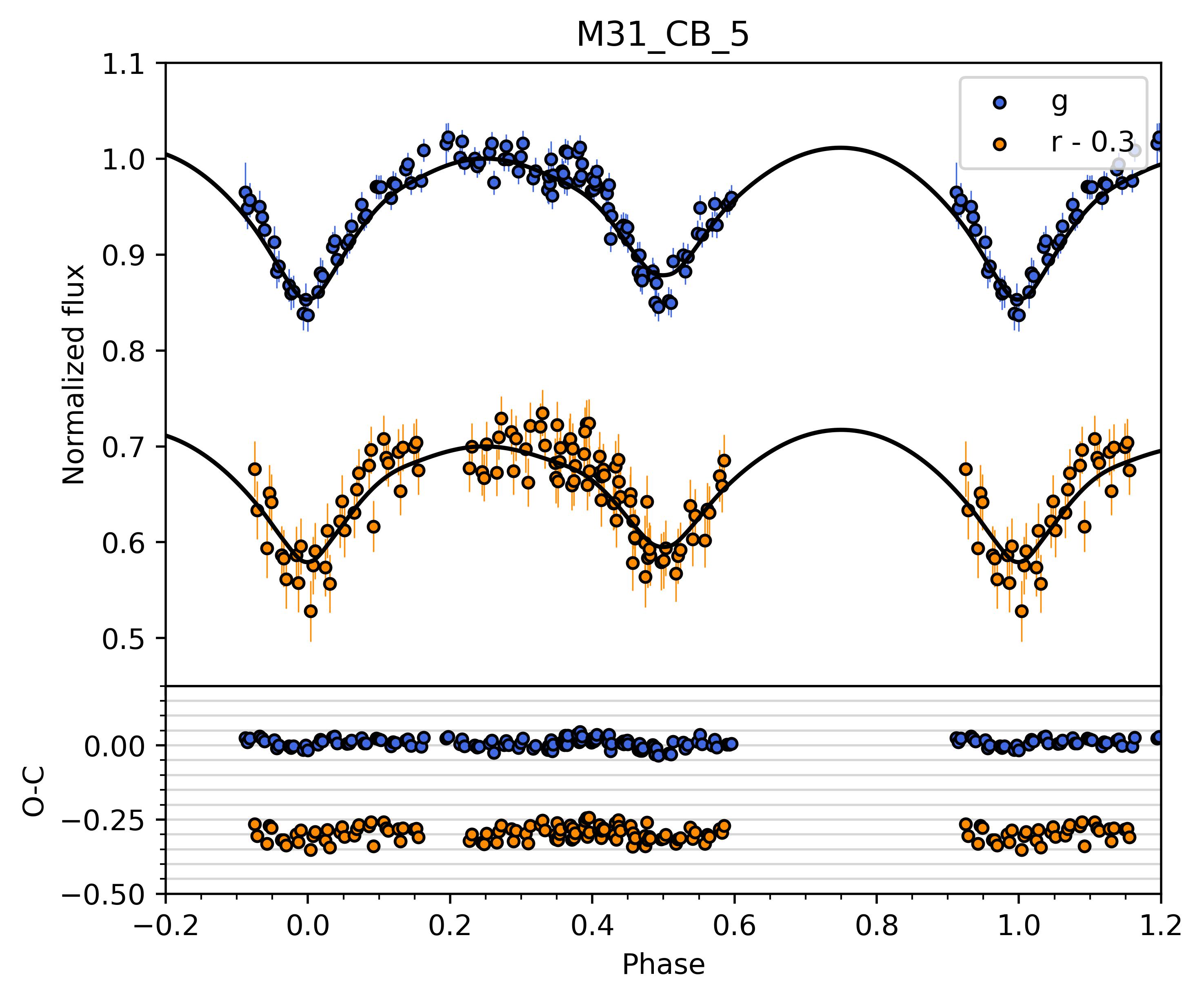}
    \end{subfigure}
    \begin{subfigure}
        \centering
        \includegraphics[width=0.24\textwidth]{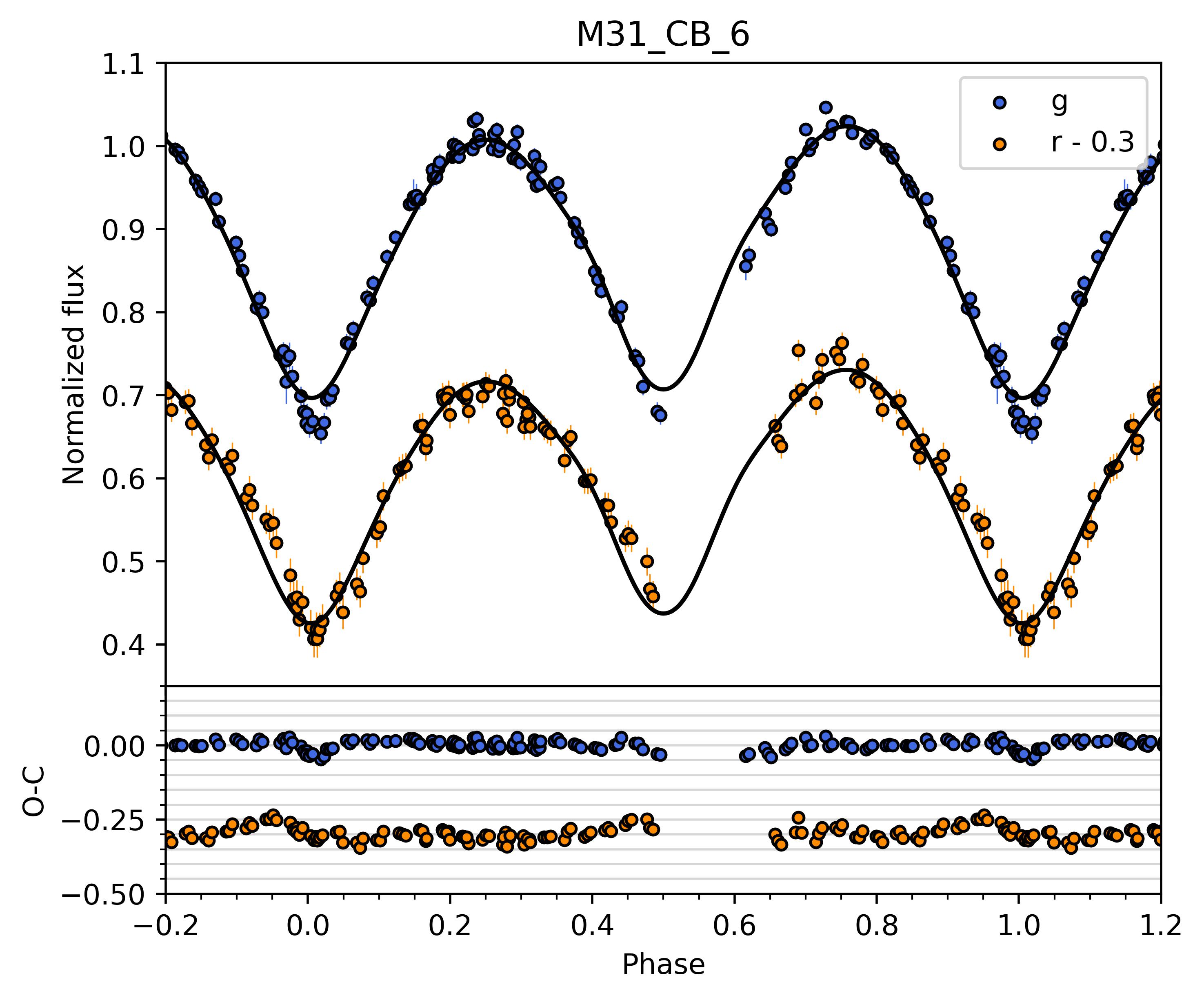}
    \end{subfigure}
    \begin{subfigure}
        \centering
        \includegraphics[width=0.24\textwidth]{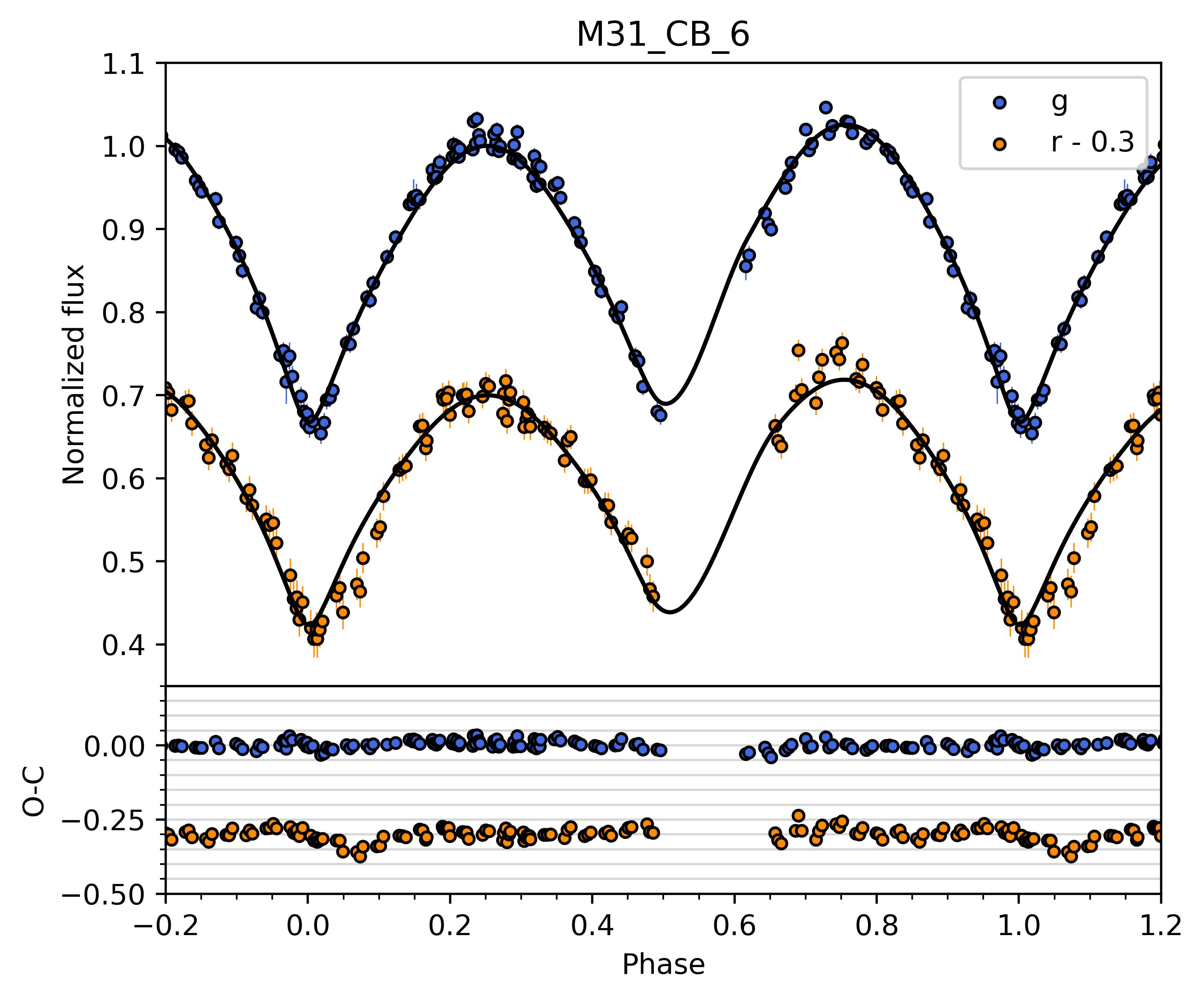}
    \end{subfigure}

    \caption{Comparison of the 14 targets without and with third light. For each target, the left shows the fitting result without the third light, and the right shows the fitting result with the third light. In each panel, the upper shows the fitting results, with blue dots for g-band data and orange dots for r-band data, and the black curve indicates the fitting curve. The bottom of the panel shows the O-C residuals.}
    \label{figA2}
\end{figure}

\begin{figure}[htbp]
    \centering 
    \addtocounter{figure}{-1}  
    
    \begin{subfigure}
        \centering
        \includegraphics[width=0.24\textwidth]{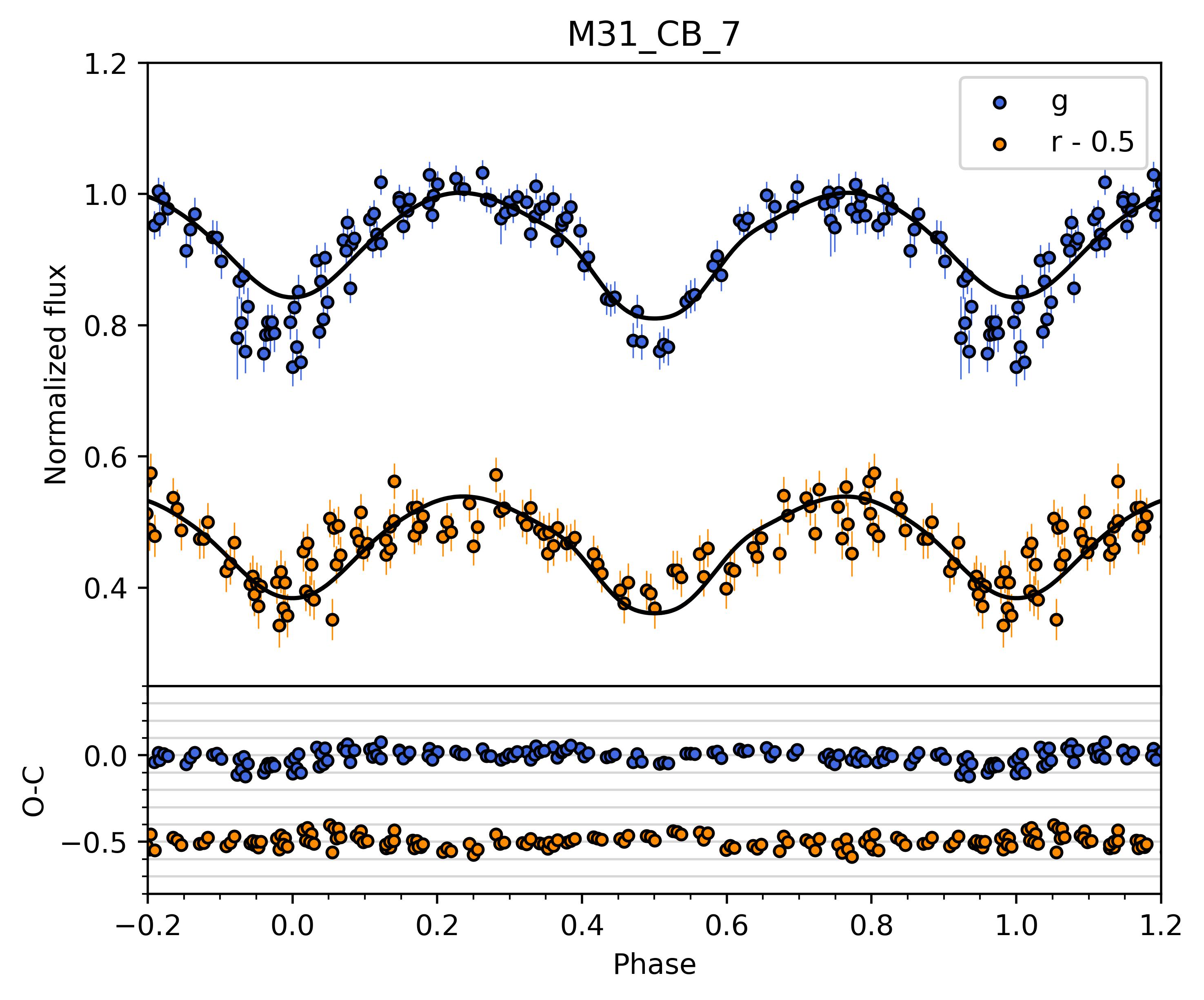}
    \end{subfigure}
    \begin{subfigure}
        \centering
        \includegraphics[width=0.24\textwidth]{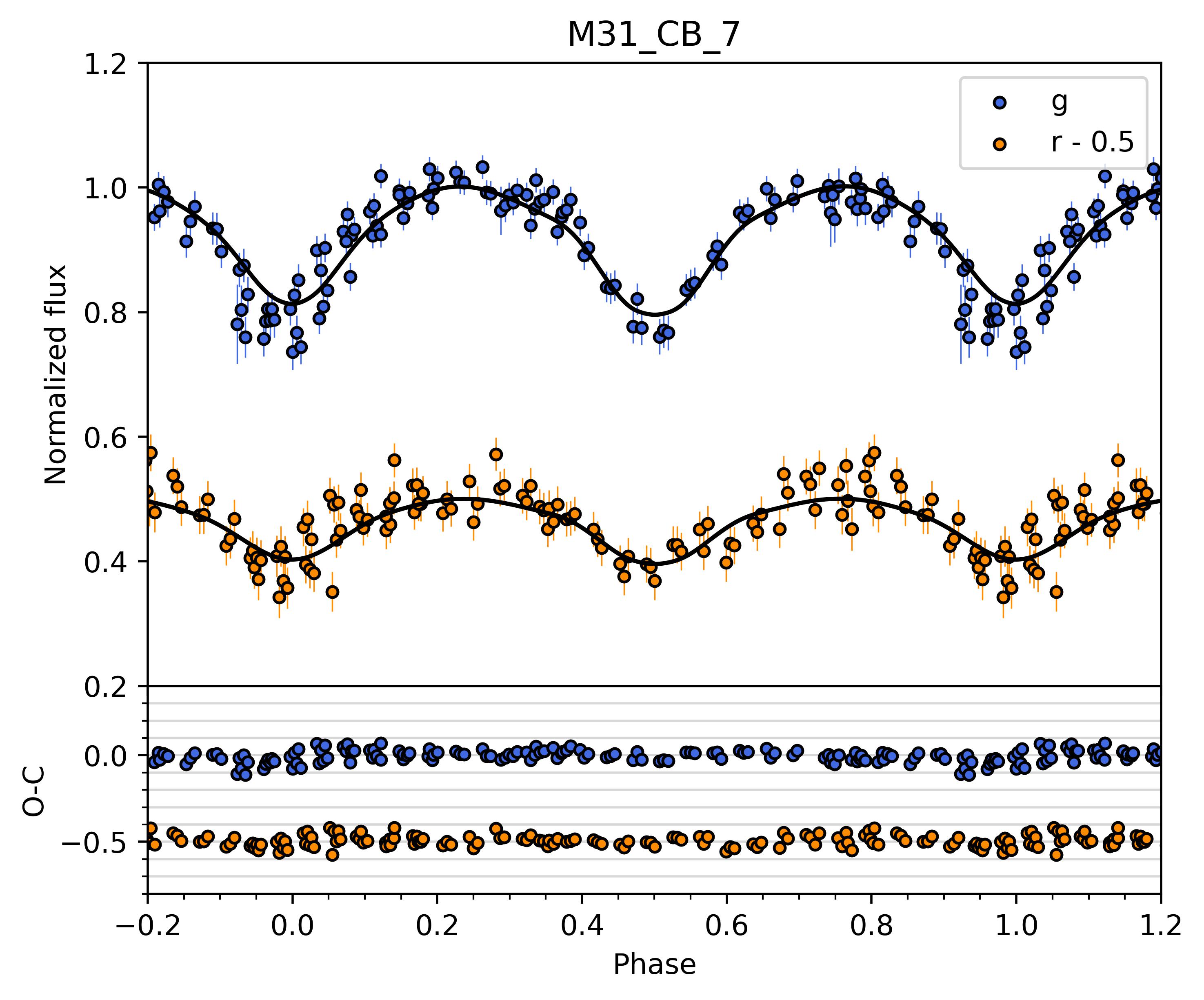}
    \end{subfigure}
    \begin{subfigure}
        \centering
        \includegraphics[width=0.24\textwidth]{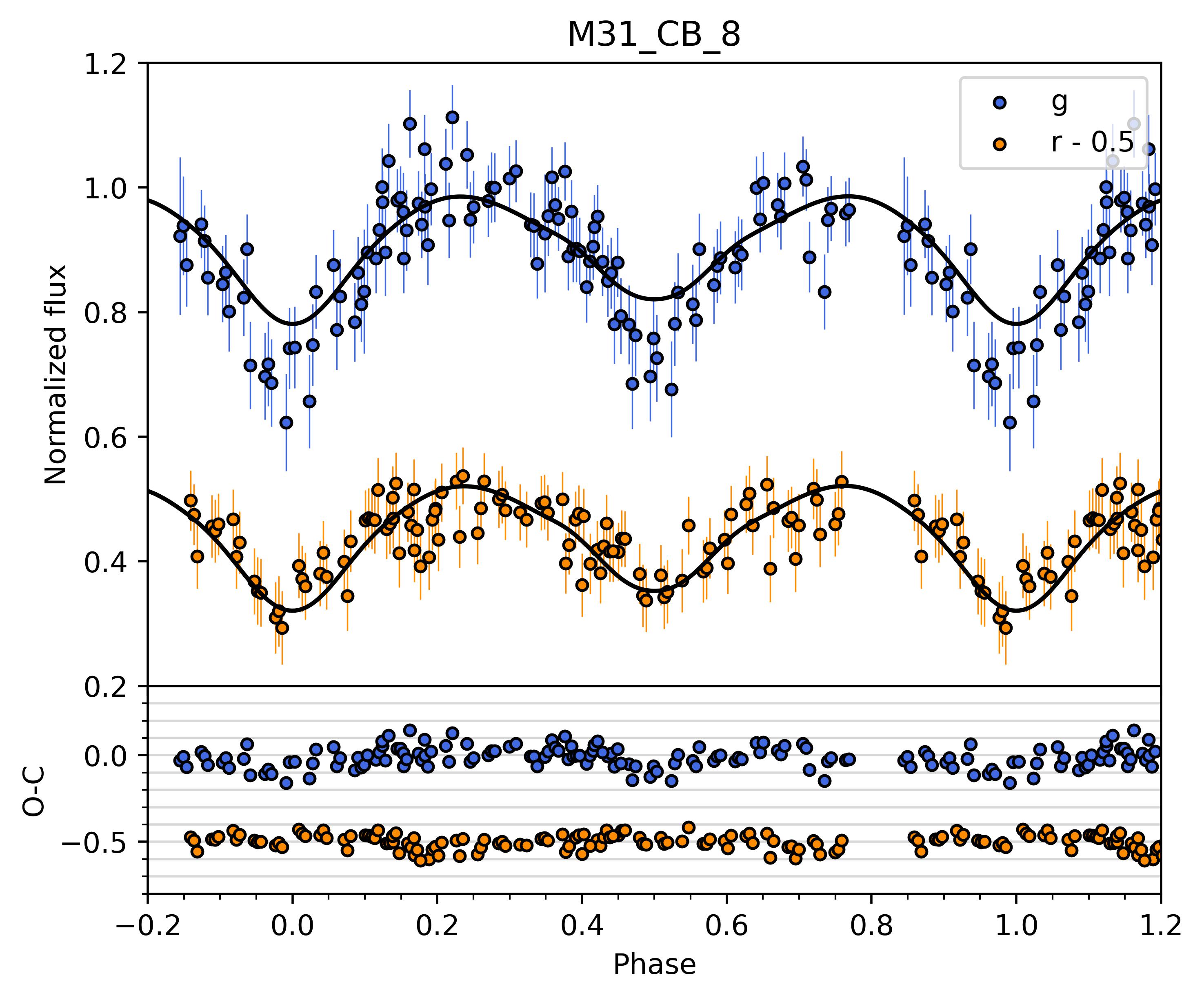}
    \end{subfigure}
    \begin{subfigure}
        \centering
        \includegraphics[width=0.24\textwidth]{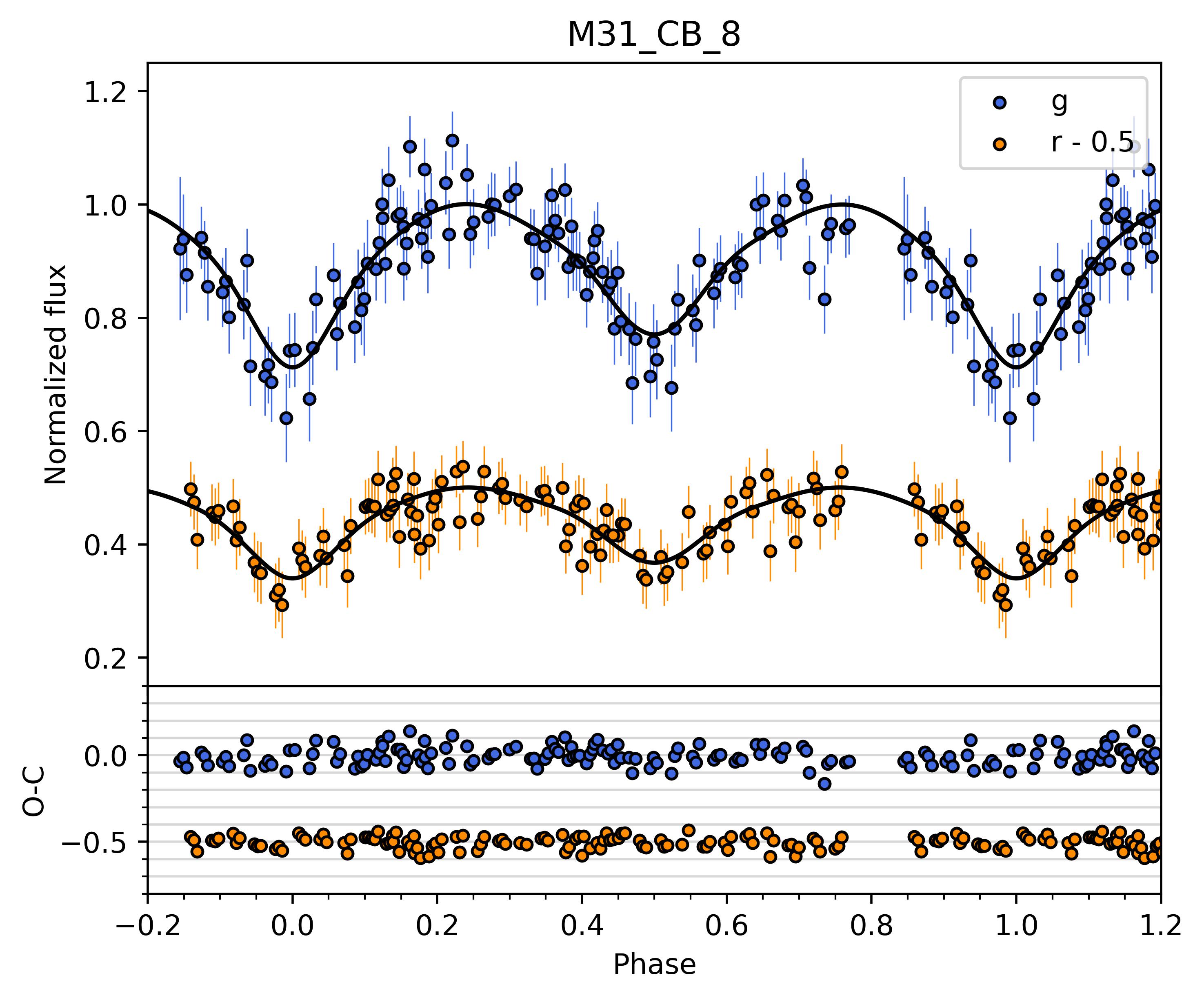}
    \end{subfigure}

    \begin{subfigure}
        \centering
        \includegraphics[width=0.24\textwidth]{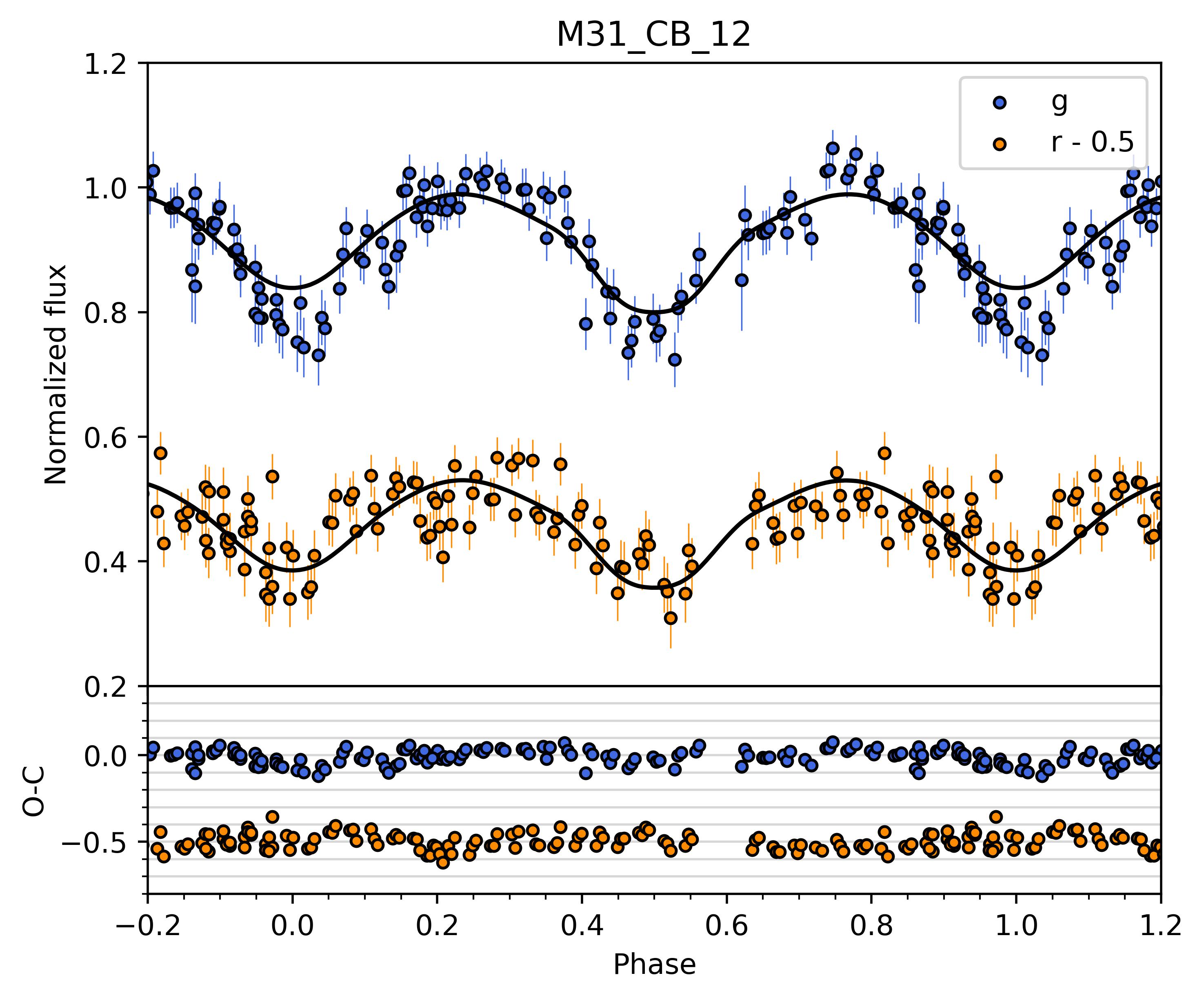}
    \end{subfigure}
    \begin{subfigure}
        \centering
        \includegraphics[width=0.24\textwidth]{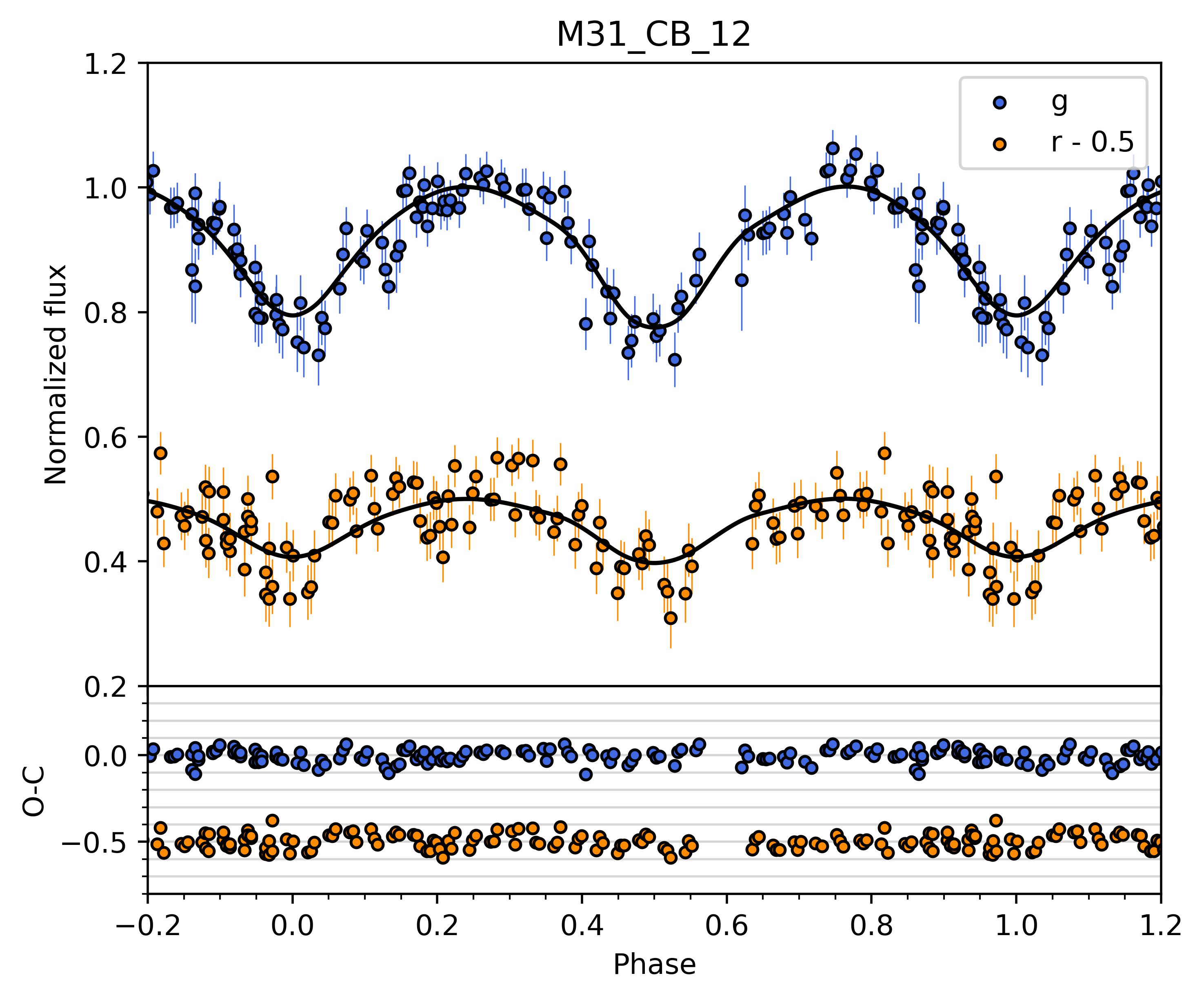}
    \end{subfigure}
    \begin{subfigure}
        \centering
        \includegraphics[width=0.24\textwidth]{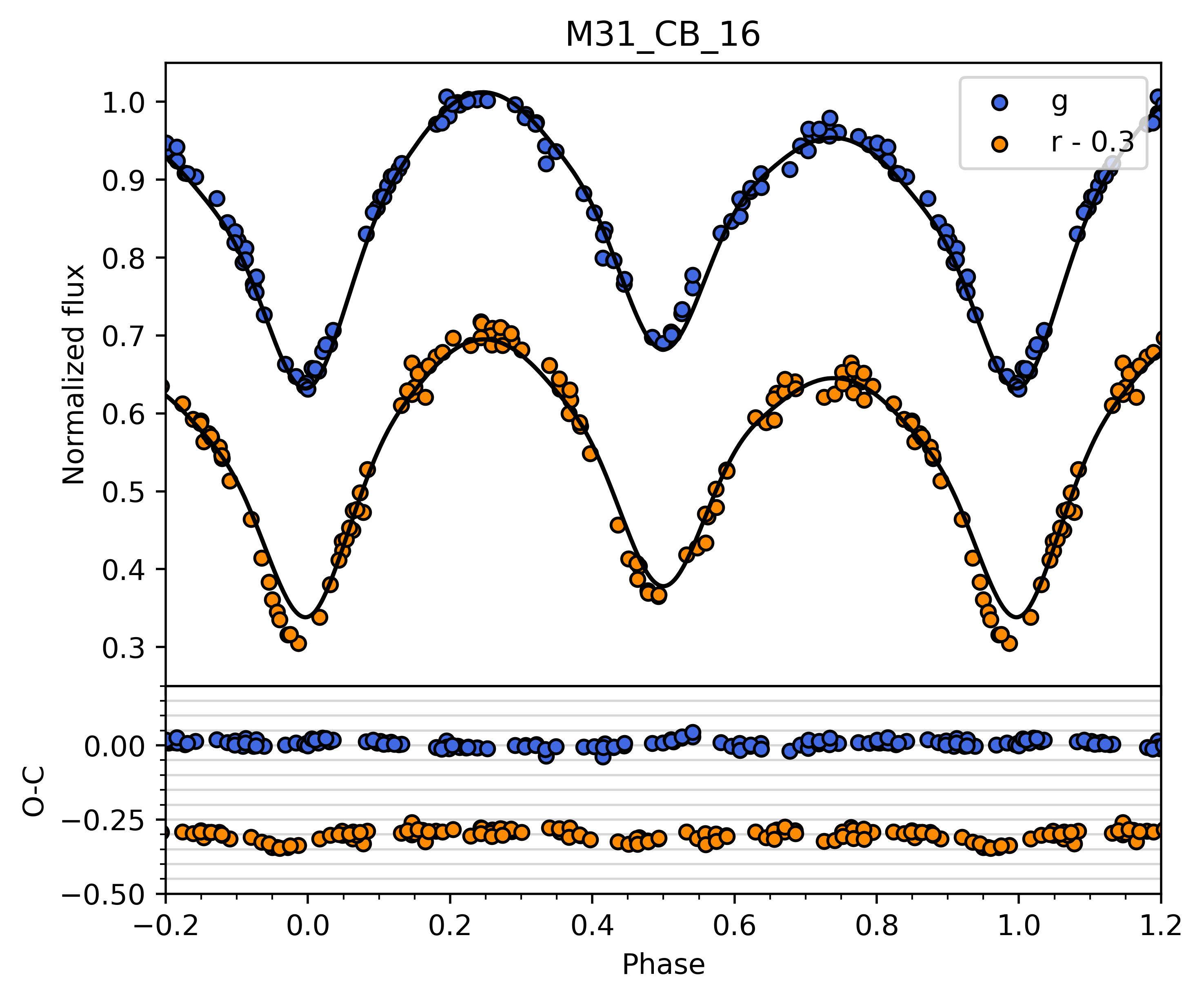}
    \end{subfigure}
    \begin{subfigure}
        \centering
        \includegraphics[width=0.24\textwidth]{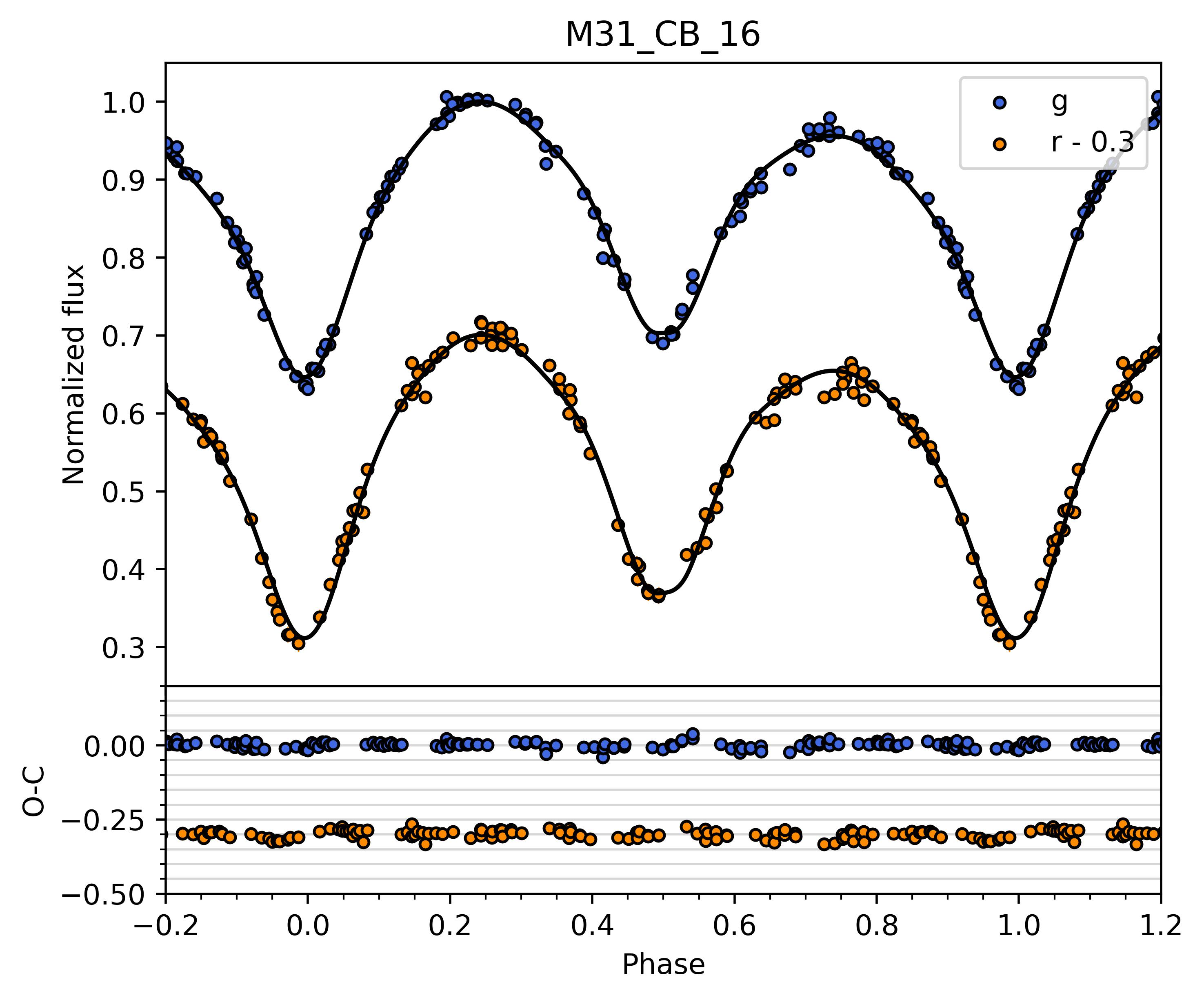}
    \end{subfigure}

    \begin{subfigure}
        \centering
        \includegraphics[width=0.24\textwidth]{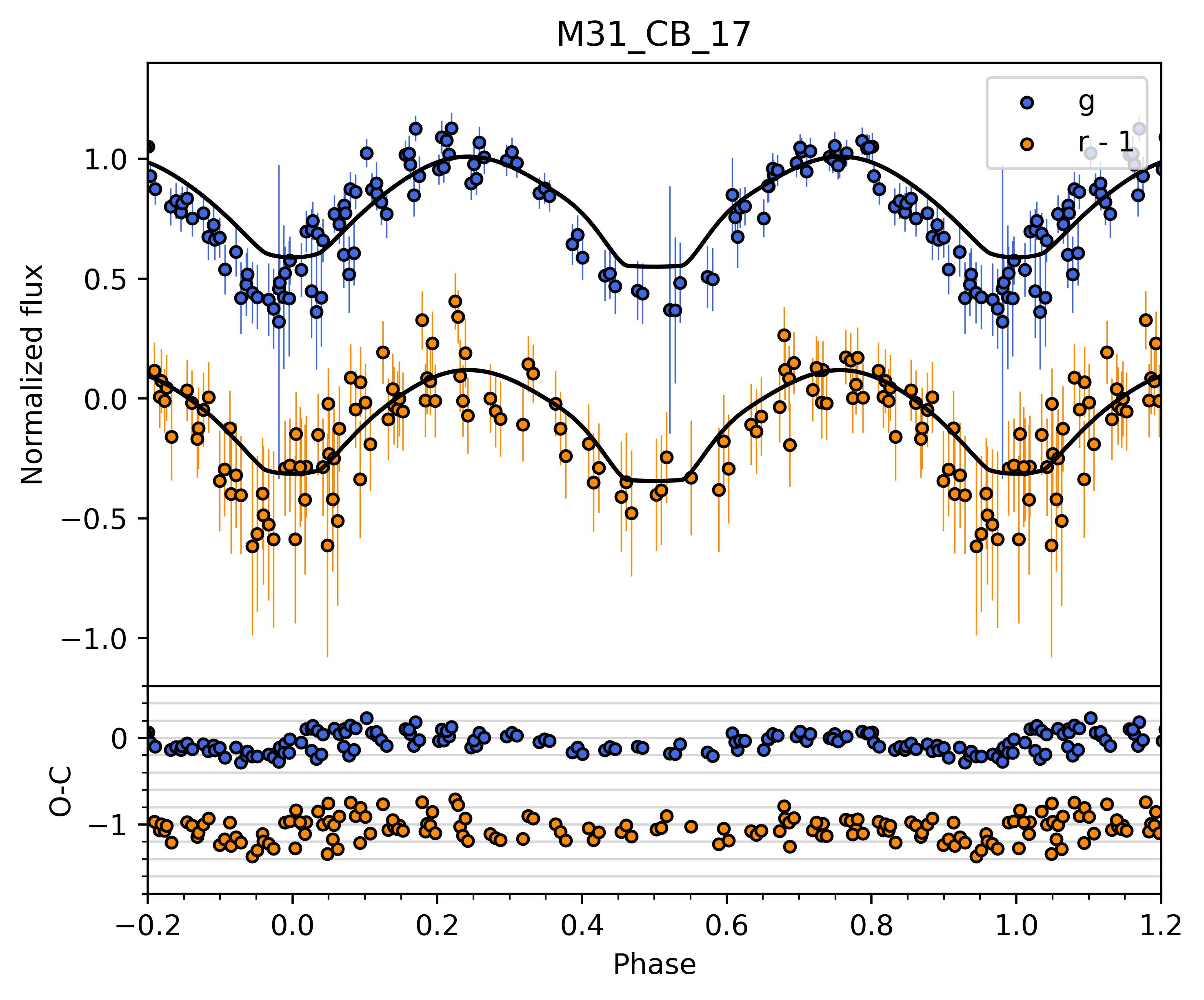}
    \end{subfigure}
    \begin{subfigure}
        \centering
        \includegraphics[width=0.24\textwidth]{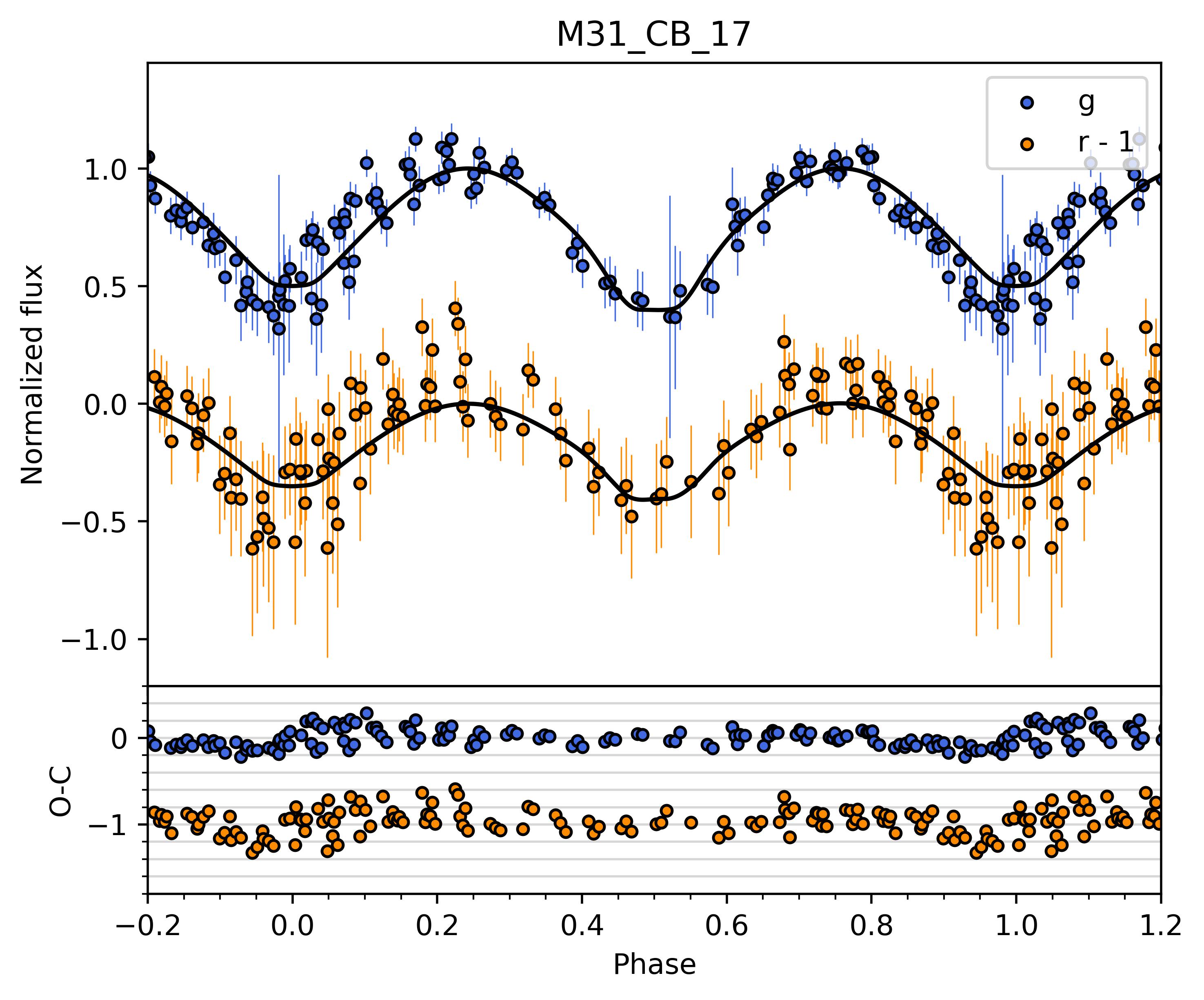}
    \end{subfigure}
    \begin{subfigure}
        \centering
        \includegraphics[width=0.24\textwidth]{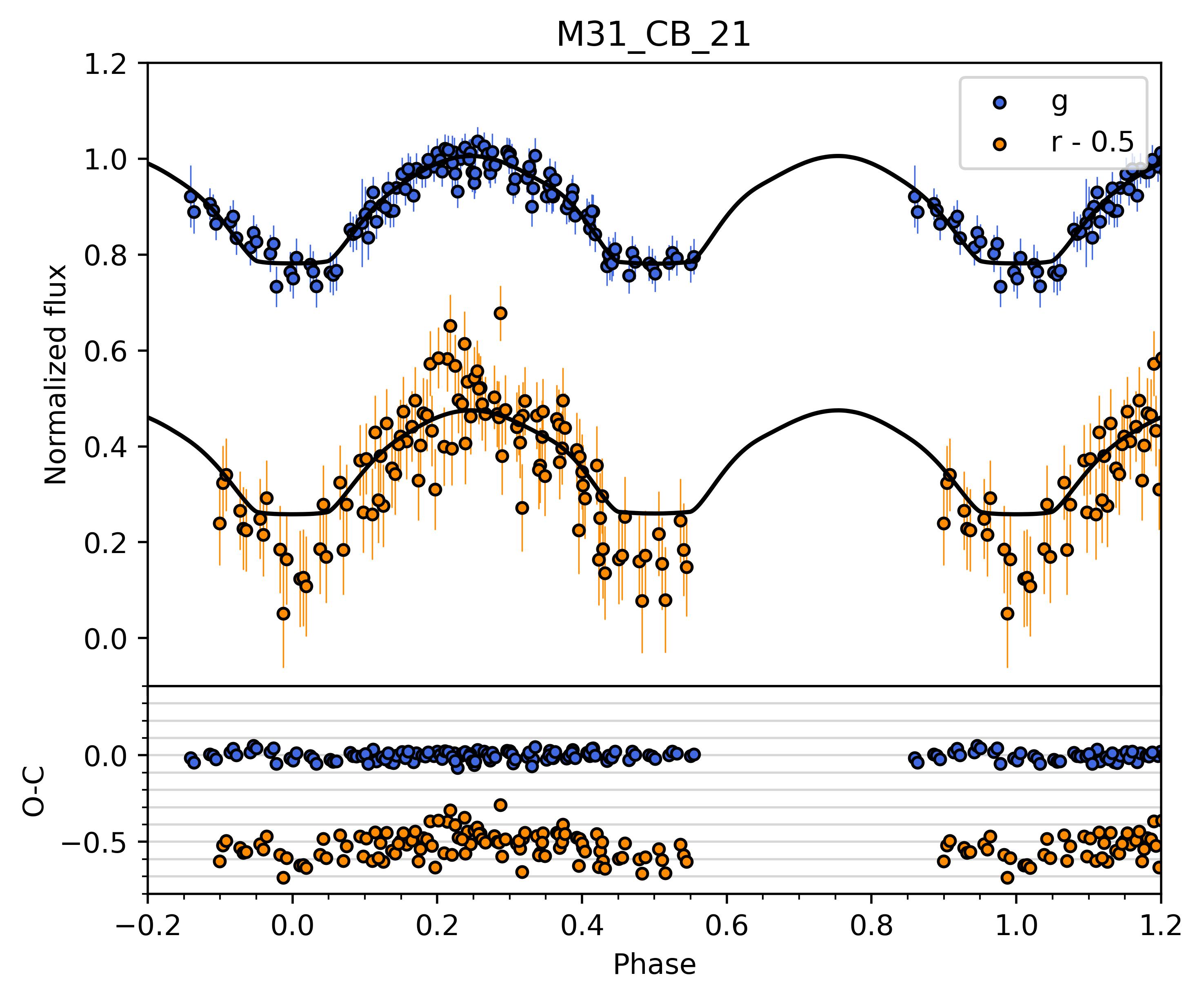}
    \end{subfigure}
    \begin{subfigure}
        \centering
        \includegraphics[width=0.24\textwidth]{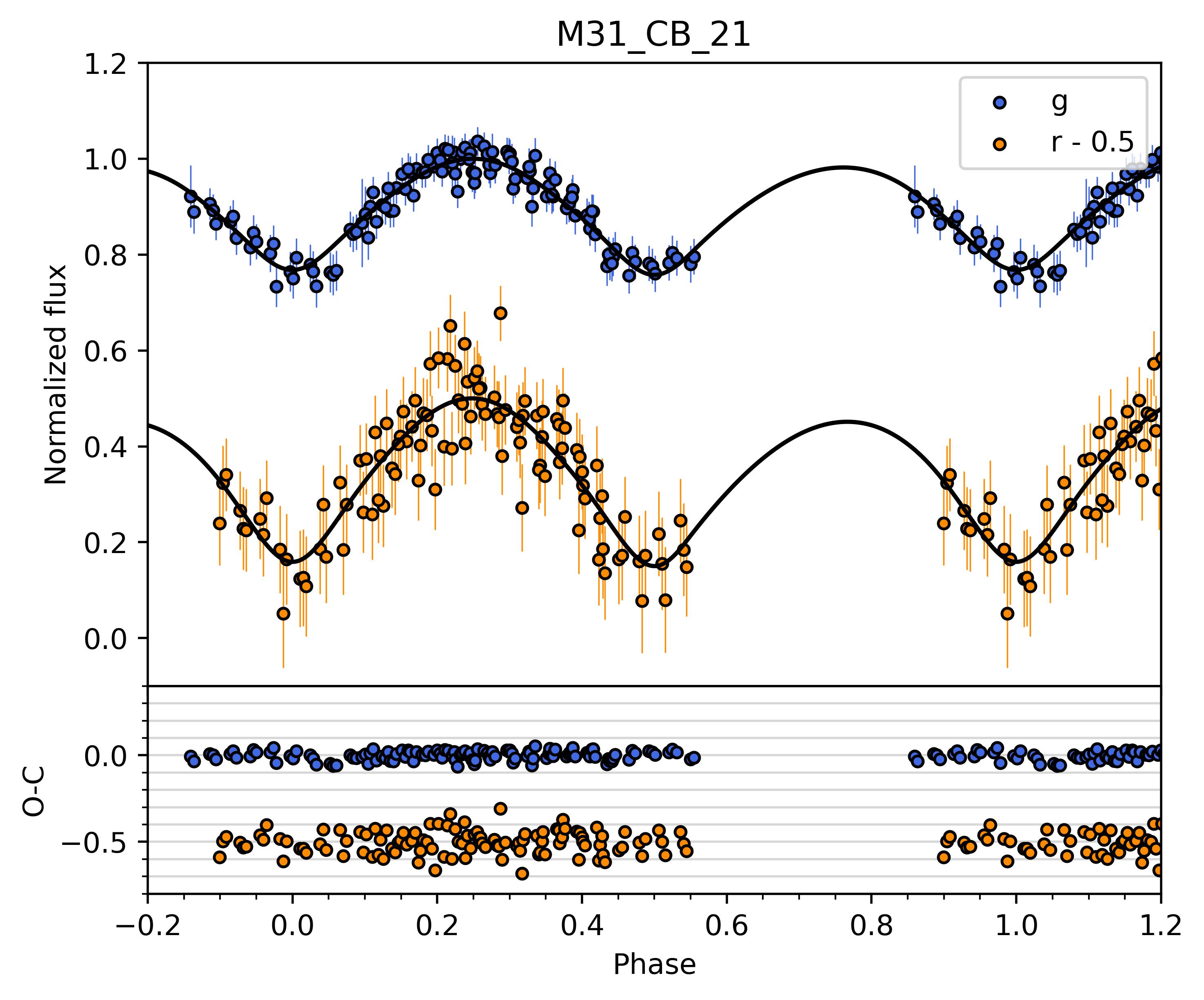}
    \end{subfigure}

    \begin{subfigure}
        \centering
        \includegraphics[width=0.24\textwidth]{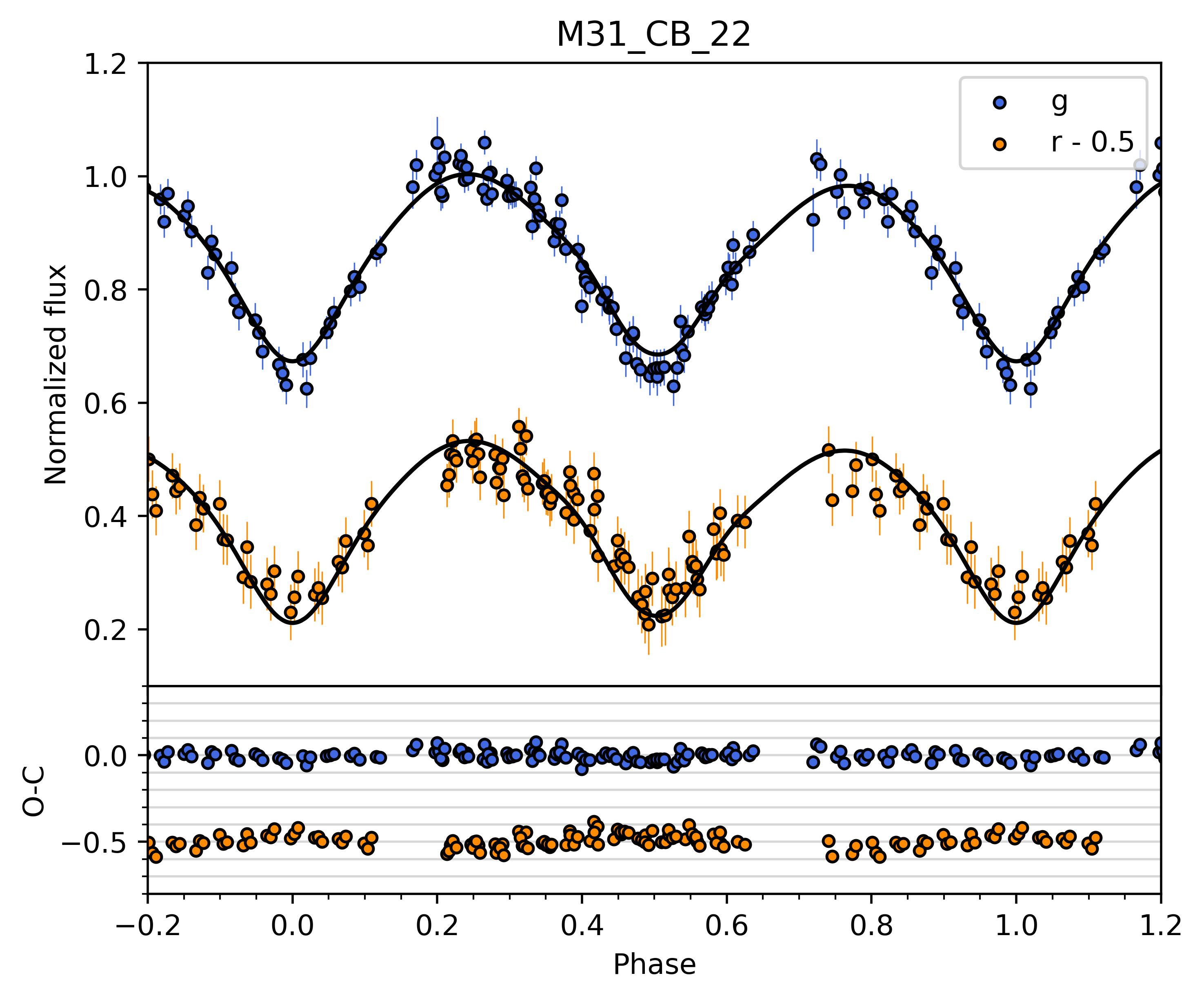}
    \end{subfigure}
    \begin{subfigure}
        \centering
        \includegraphics[width=0.24\textwidth]{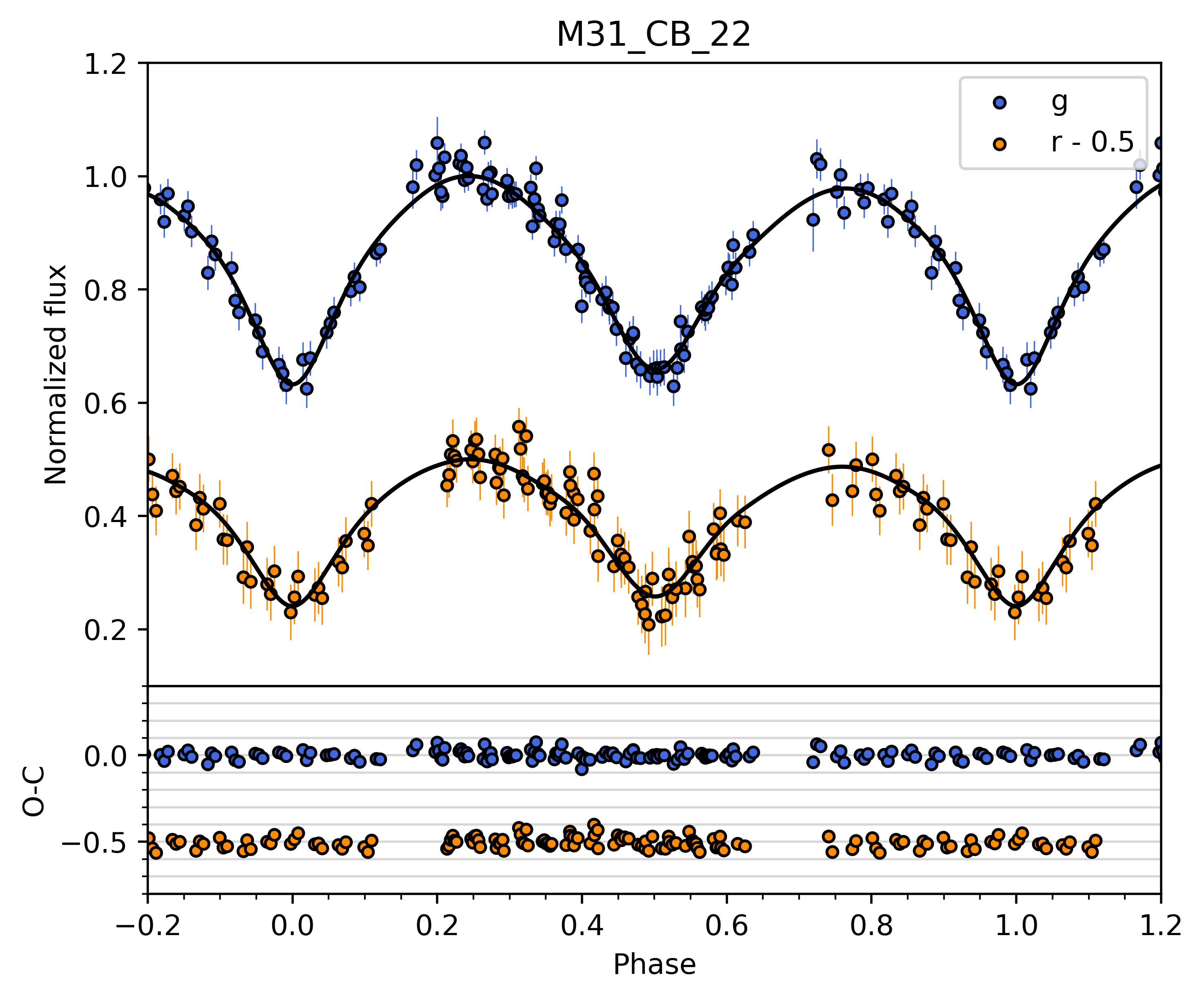}
    \end{subfigure}
    \begin{subfigure}
        \centering
        \includegraphics[width=0.24\textwidth]{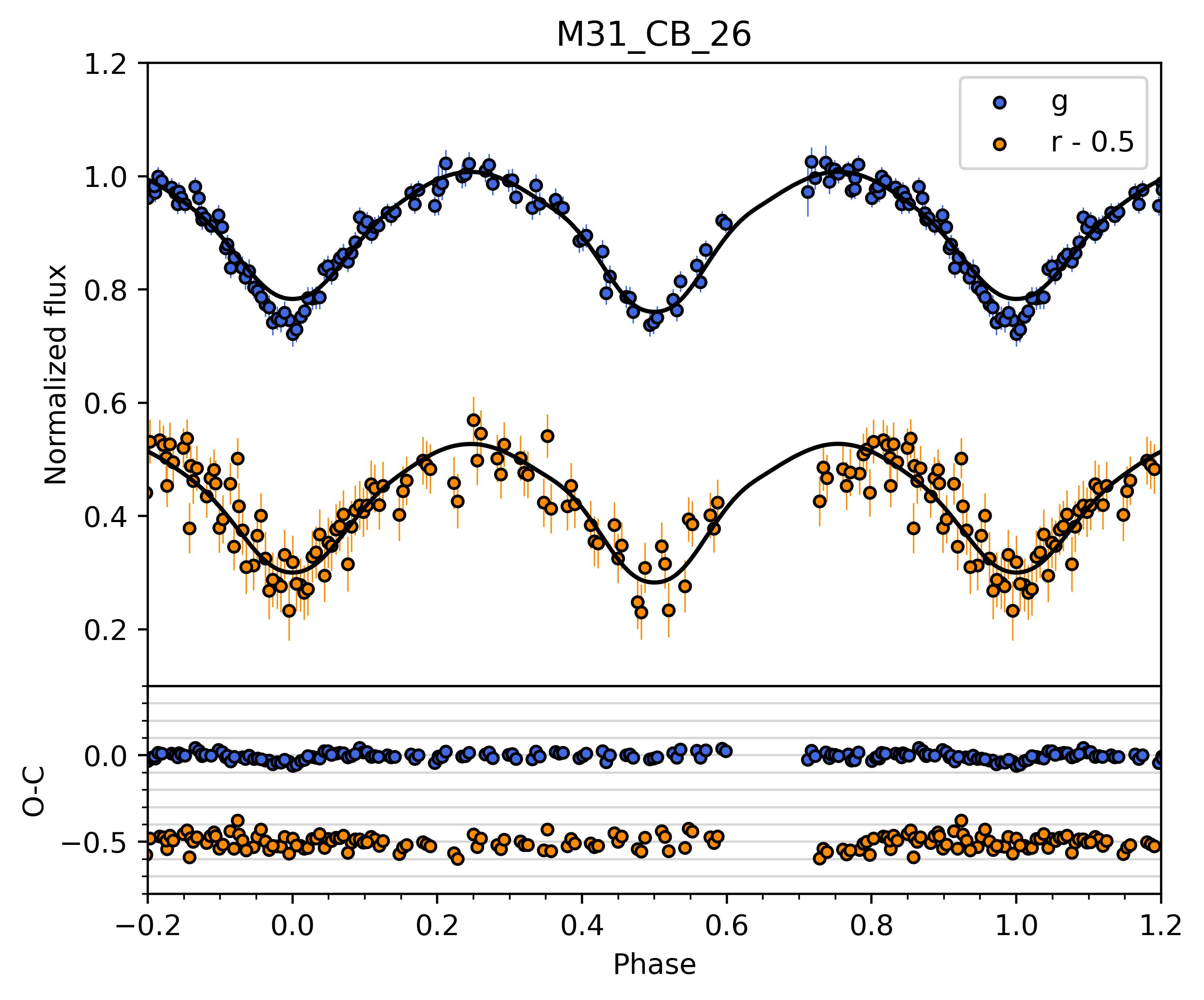}
    \end{subfigure}
    \begin{subfigure}
        \centering
        \includegraphics[width=0.24\textwidth]{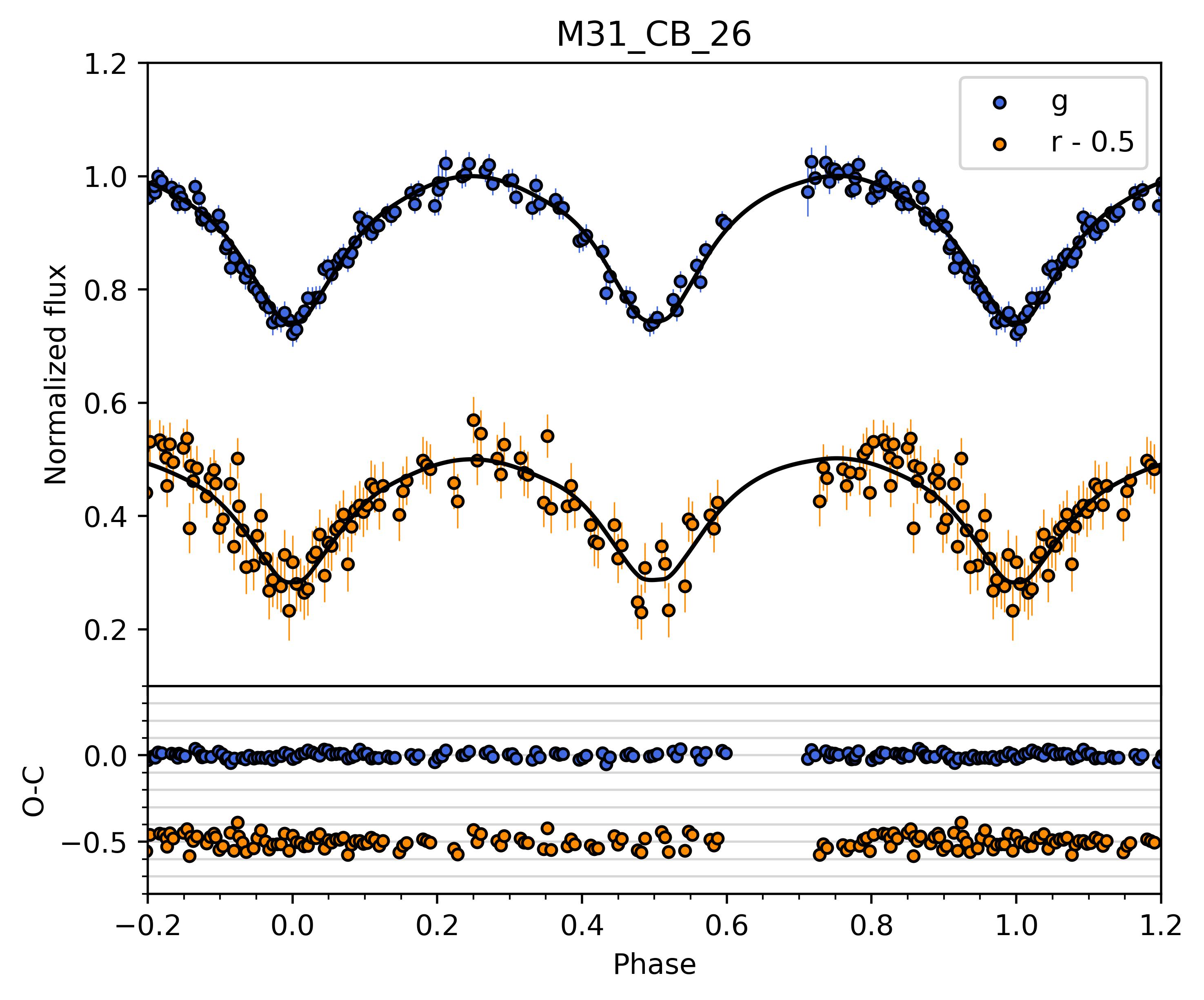}
    \end{subfigure}

    \begin{subfigure}
        \centering
        \includegraphics[width=0.24\textwidth]{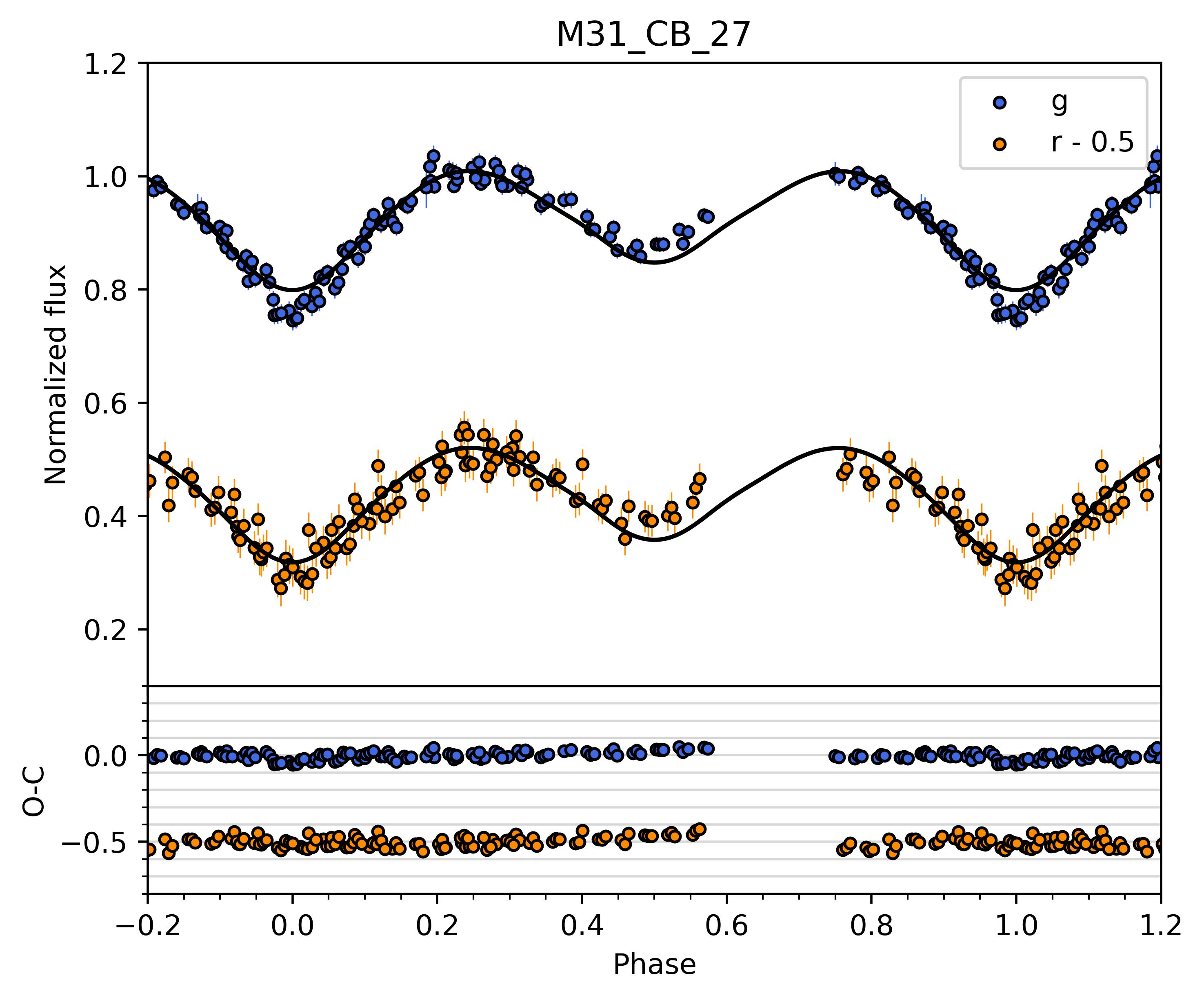}
    \end{subfigure}
    \begin{subfigure}
        \centering
        \includegraphics[width=0.24\textwidth]{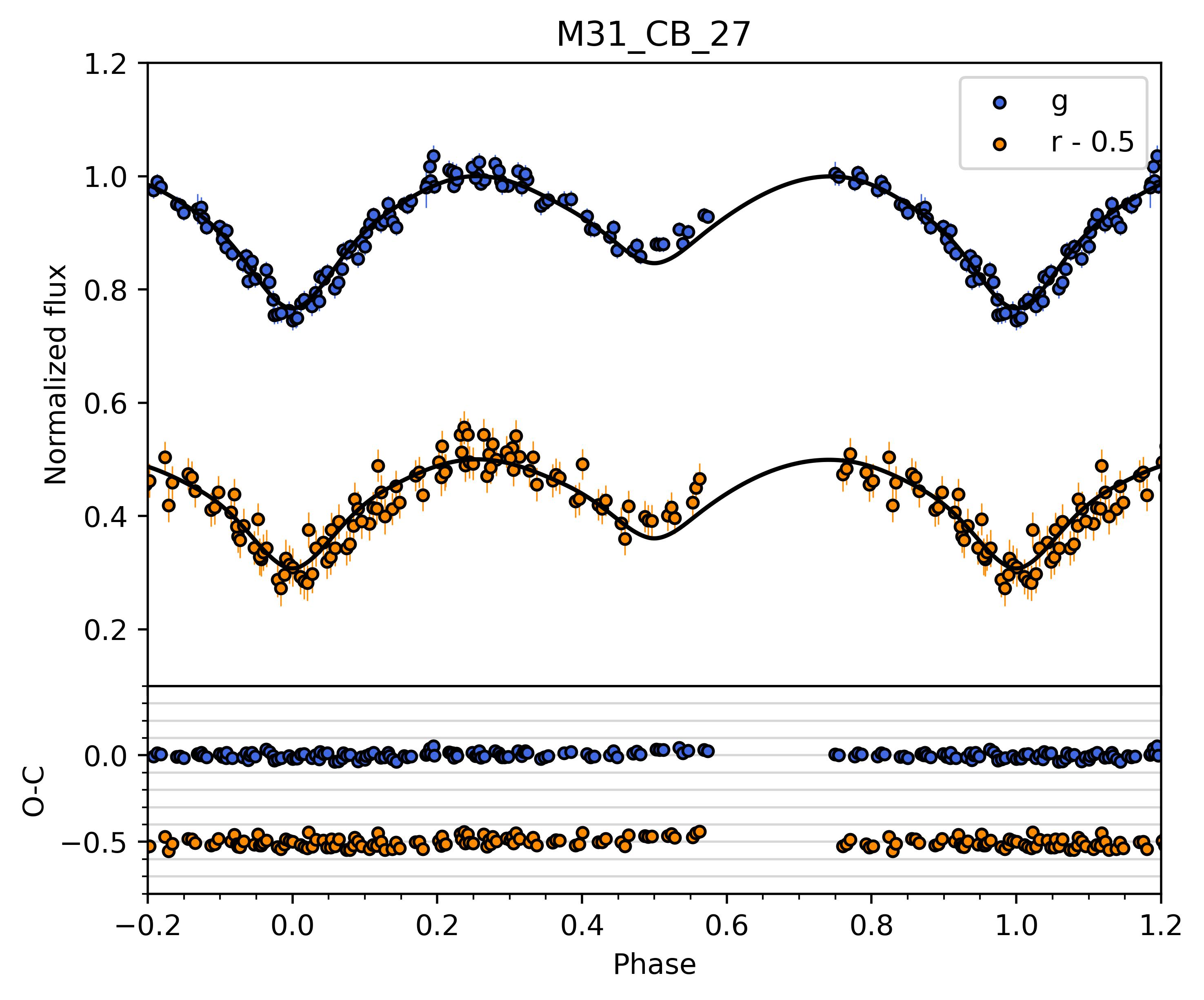}
    \end{subfigure}
    \begin{subfigure}
        \centering
        \includegraphics[width=0.24\textwidth]{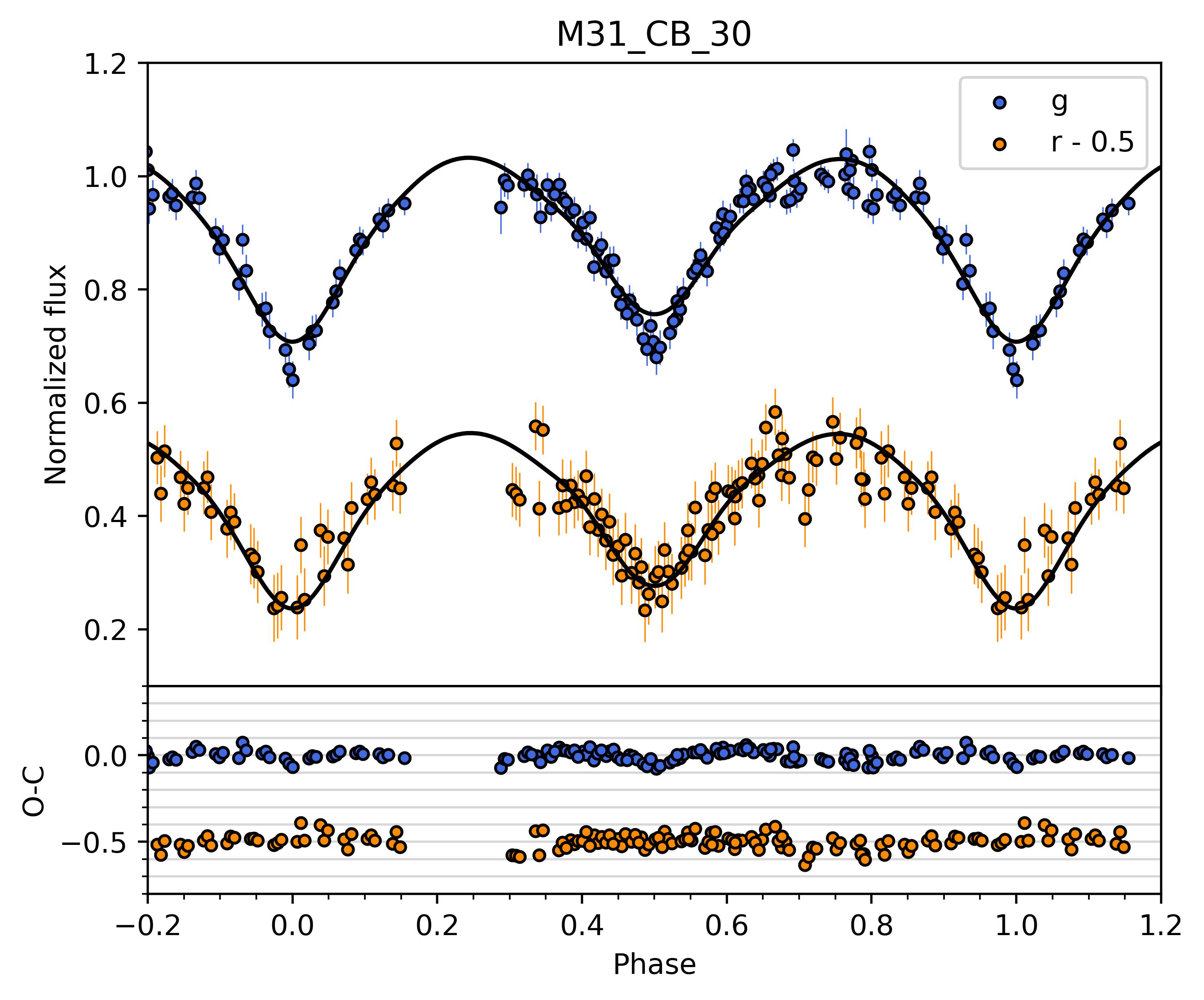}
    \end{subfigure}
    \begin{subfigure}
        \centering
        \includegraphics[width=0.24\textwidth]{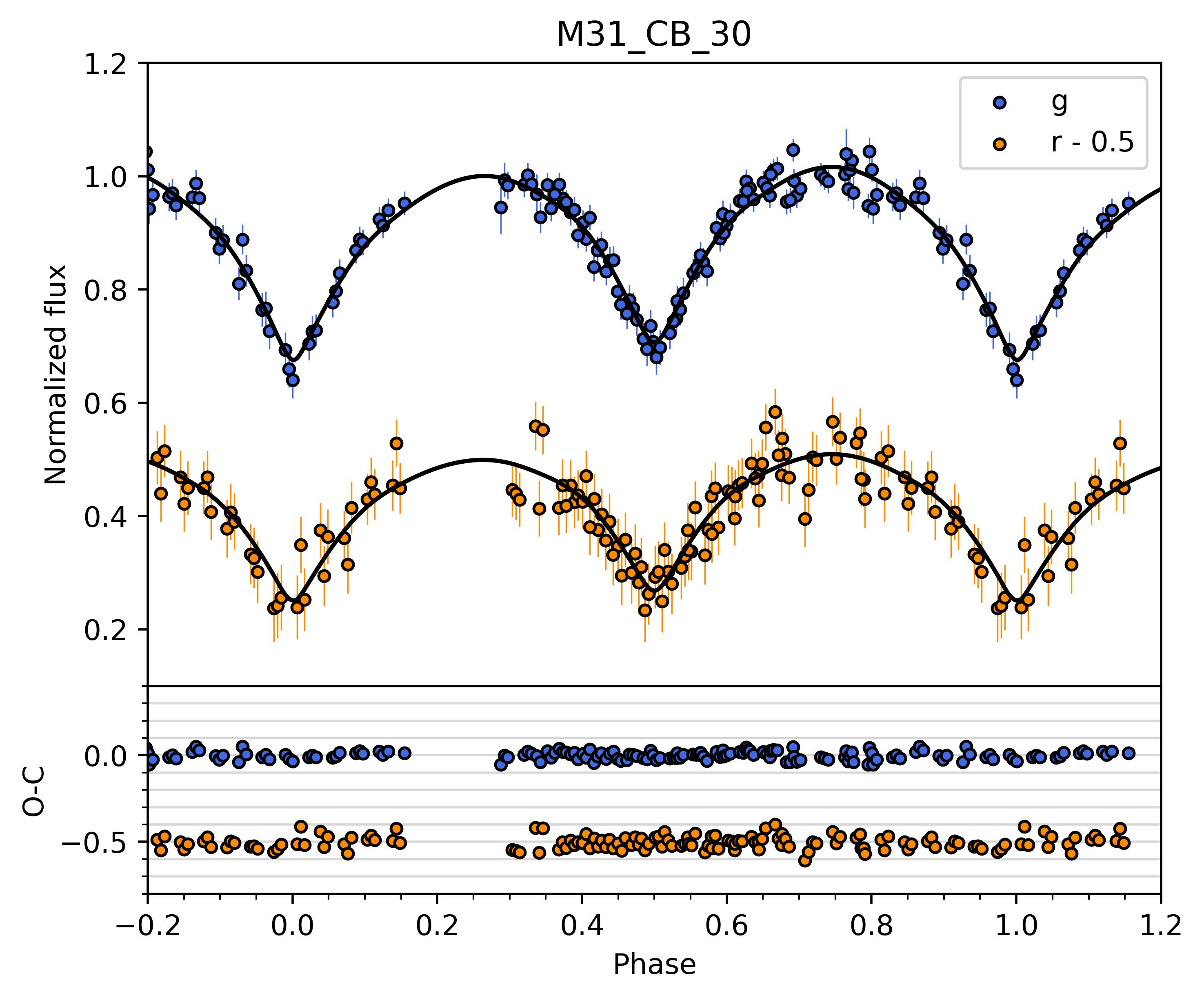}
    \end{subfigure}

    \caption{(continued)}
    \label{figA2-2}
\end{figure}

%




\clearpage
\bibliography{sample631}{}

\begin{thebibliography}{}
\expandafter\ifx\csname natexlab\endcsname\relax\def\natexlab#1{#1}\fi
\providecommand{\url}[1]{\href{#1}{#1}}
\providecommand{\dodoi}[1]{doi:~\href{http://doi.org/#1}{\nolinkurl{#1}}}
\providecommand{\doeprint}[1]{\href{http://ascl.net/#1}{\nolinkurl{http://ascl.net/#1}}}
\providecommand{\doarXiv}[1]{\href{https://arxiv.org/abs/#1}{\nolinkurl{https://arxiv.org/abs/#1}}}

\bibitem[{Abbott \& et~al.(2016)}]{Abbott2016}
Abbott, B.~P., \& et~al. 2016, Physical Review Letters, 116, 061102

\bibitem[{Abbott \& et~al.(2020)}]{Abbott2020}
Abbott, R., \& et~al. 2020, arXiv:2010.14527

\bibitem[{{Abramovici} {et~al.}(1992){Abramovici}, {Althouse}, {Drever}, {Gursel}, {Kawamura}, {Raab}, {Shoemaker}, {Sievers}, {Spero}, {Thorne}, {Vogt}, {Weiss}, {Whitcomb}, \& {Zucker}}]{1992Sci...256..325A}
{Abramovici}, A., {Althouse}, W.~E., {Drever}, R. W.~P., {et~al.} 1992, Science, 256, 325, \dodoi{10.1126/science.256.5055.325}

\bibitem[{{An} {et~al.}(2004){An}, {Evans}, {Hewett}, {Baillon}, {Calchi Novati}, {Carr}, {Cr{\'e}z{\'e}}, {Giraud-H{\'e}raud}, {Gould}, {Jetzer}, {Kaplan}, {Kerins}, {Paulin-Henriksson}, {Smartt}, {Stalin}, \& {Tsapras}}]{2004MNRAS.351.1071A}
{An}, J.~H., {Evans}, N.~W., {Hewett}, P., {et~al.} 2004, \mnras, 351, 1071, \dodoi{10.1111/j.1365-2966.2004.07853.x}

\bibitem[{{Arbutina} \& {Wadhwa}(2024)}]{2024SerAJ.208....1A}
{Arbutina}, B., \& {Wadhwa}, S. 2024, Serbian Astronomical Journal, 208, 1, \dodoi{10.2298/SAJ2408001A}

\bibitem[{{Castelli} \& {Kurucz}(2004)}]{2004A&A...419..725C}
{Castelli}, F., \& {Kurucz}, R.~L. 2004, \aap, 419, 725, \dodoi{10.1051/0004-6361:20040079}

\bibitem[{{Chambers} \& {Pan-STARRS Team}(2018)}]{2018AAS...23110201C}
{Chambers}, K., \& {Pan-STARRS Team}. 2018, in American Astronomical Society Meeting Abstracts, Vol. 231, 102.01

\bibitem[{{Christy} {et~al.}(2023){Christy}, {Jayasinghe}, {Stanek}, {Kochanek}, {Thompson}, {Shappee}, {Holoien}, {Prieto}, {Dong}, \& {Giles}}]{2023MNRAS.519.5271C}
{Christy}, C.~T., {Jayasinghe}, T., {Stanek}, K.~Z., {et~al.} 2023, \mnras, 519, 5271, \dodoi{10.1093/mnras/stac3801}

\bibitem[{{Conroy} {et~al.}(2020){Conroy}, {Kochoska}, {Hey}, {Pablo}, {Hambleton}, {Jones}, {Giammarco}, {Abdul-Masih}, \& {Pr{\v{s}}a}}]{2020ApJS..250...34C}
{Conroy}, K.~E., {Kochoska}, A., {Hey}, D., {et~al.} 2020, \apjs, 250, 34, \dodoi{10.3847/1538-4365/abb4e2}

\bibitem[{{Eker} {et~al.}(2006){Eker}, {Demircan}, {Bilir}, \& {Karata{\c{s}}}}]{2006MNRAS.373.1483E}
{Eker}, Z., {Demircan}, O., {Bilir}, S., \& {Karata{\c{s}}}, Y. 2006, \mnras, 373, 1483, \dodoi{10.1111/j.1365-2966.2006.11073.x}

\bibitem[{{Ekstr{\"o}m} {et~al.}(2012){Ekstr{\"o}m}, {Georgy}, {Eggenberger}, {Meynet}, {Mowlavi}, {Wyttenbach}, {Granada}, {Decressin}, {Hirschi}, {Frischknecht}, {Charbonnel}, \& {Maeder}}]{2012A&A...537A.146E}
{Ekstr{\"o}m}, S., {Georgy}, C., {Eggenberger}, P., {et~al.} 2012, \aap, 537, A146, \dodoi{10.1051/0004-6361/201117751}

\bibitem[{{Fliri} {et~al.}(2006){Fliri}, {Riffeser}, {Seitz}, \& {Bender}}]{2006A&A...445..423F}
{Fliri}, J., {Riffeser}, A., {Seitz}, S., \& {Bender}, R. 2006, \aap, 445, 423, \dodoi{10.1051/0004-6361:20042223}

\bibitem[{Gamerman \& Lopes(2006)}]{gamerman2006markov}
Gamerman, D., \& Lopes, H.~F. 2006, Markov Chain Monte Carlo: Stochastic Simulation for Bayesian Inference (Chapman \& Hall/CRC), \dodoi{10.1201/9781584885870}

\bibitem[{{Gazeas} \& {St{\c{e}}pie{\'n}}(2008)}]{2008MNRAS.390.1577G}
{Gazeas}, K., \& {St{\c{e}}pie{\'n}}, K. 2008, \mnras, 390, 1577, \dodoi{10.1111/j.1365-2966.2008.13844.x}

\bibitem[{{Gazeas} {et~al.}(2021){Gazeas}, {Zola}, {Liakos}, {Zakrzewski}, {Rucinski}, {Kreiner}, {Ogloza}, {Drozdz}, {Koziel-Wierzbowska}, {Stachowski}, {Siwak}, {Baran}, {Kjurkchieva}, {Marchev}, {Erdem}, \& {Szalankiewicz}}]{2021MNRAS.501.2897G}
{Gazeas}, K., {Zola}, S., {Liakos}, A., {et~al.} 2021, \mnras, 501, 2897, \dodoi{10.1093/mnras/staa3753}

\bibitem[{{Girardi} {et~al.}(2000){Girardi}, {Bressan}, {Bertelli}, \& {Chiosi}}]{2000A&AS..141..371G}
{Girardi}, L., {Bressan}, A., {Bertelli}, G., \& {Chiosi}, C. 2000, \aaps, 141, 371, \dodoi{10.1051/aas:2000126}

\bibitem[{{Gu} {et~al.}(2024){Gu}, {Yuan}, {Dong}, {Zheng}, {Cui}, {Ren}, {Fu}, {Huang}, \& {Fan}}]{2024ApJS..273....9G}
{Gu}, H., {Yuan}, H., {Dong}, S., {et~al.} 2024, \apjs, 273, 9, \dodoi{10.3847/1538-4365/ad45f9}

\bibitem[{{Heitmann}(1996)}]{1996dmcq.conf..347H}
{Heitmann}, H. 1996, in Dark Matter in Cosmology Quantam Measurements Experimental Gravitation, ed. R.~{Ansari}, Y.~{Giraud-Heraud}, \& J.~{Tran Thanh Van}, 347

\bibitem[{{Horvat} {et~al.}(2018){Horvat}, {Conroy}, {Pablo}, {Hambleton}, {Kochoska}, {Giammarco}, \& {Pr{\v{s}}a}}]{2018ApJS..237...26H}
{Horvat}, M., {Conroy}, K.~E., {Pablo}, H., {et~al.} 2018, \apjs, 237, 26, \dodoi{10.3847/1538-4365/aacd0f}

\bibitem[{{Hrivnak} {et~al.}(2006){Hrivnak}, {Lu}, {Eaton}, \& {Kenning}}]{2006AJ....132..960H}
{Hrivnak}, B.~J., {Lu}, W., {Eaton}, J., \& {Kenning}, D. 2006, \aj, 132, 960, \dodoi{10.1086/505691}

\bibitem[{{Ivezi{\'c}} {et~al.}(2019){Ivezi{\'c}}, {Kahn}, {Tyson}, {Abel}, {Acosta}, {Allsman}, {Alonso}, {AlSayyad}, {Anderson}, {Andrew}, {Angel}, {Angeli}, {Ansari}, {Antilogus}, {Araujo}, {Armstrong}, {Arndt}, {Astier}, {Aubourg}, {Auza}, {Axelrod}, {Bard}, {Barr}, {Barrau}, {Bartlett}, {Bauer}, {Bauman}, {Baumont}, {Bechtol}, {Bechtol}, {Becker}, {Becla}, {Beldica}, {Bellavia}, {Bianco}, {Biswas}, {Blanc}, {Blazek}, {Blandford}, {Bloom}, {Bogart}, {Bond}, {Booth}, {Borgland}, {Borne}, {Bosch}, {Boutigny}, {Brackett}, {Bradshaw}, {Brandt}, {Brown}, {Bullock}, {Burchat}, {Burke}, {Cagnoli}, {Calabrese}, {Callahan}, {Callen}, {Carlin}, {Carlson}, {Chandrasekharan}, {Charles-Emerson}, {Chesley}, {Cheu}, {Chiang}, {Chiang}, {Chirino}, {Chow}, {Ciardi}, {Claver}, {Cohen-Tanugi}, {Cockrum}, {Coles}, {Connolly}, {Cook}, {Cooray}, {Covey}, {Cribbs}, {Cui}, {Cutri}, {Daly}, {Daniel}, {Daruich}, {Daubard}, {Daues}, {Dawson}, {Delgado}, {Dellapenna}, {de Peyster}, {de Val-Borro}, {Digel}, {Doherty}, {Dubois},
  {Dubois-Felsmann}, {Durech}, {Economou}, {Eifler}, {Eracleous}, {Emmons}, {Fausti Neto}, {Ferguson}, {Figueroa}, {Fisher-Levine}, {Focke}, {Foss}, {Frank}, {Freemon}, {Gangler}, {Gawiser}, {Geary}, {Gee}, {Geha}, {Gessner}, {Gibson}, {Gilmore}, {Glanzman}, {Glick}, {Goldina}, {Goldstein}, {Goodenow}, {Graham}, {Gressler}, {Gris}, {Guy}, {Guyonnet}, {Haller}, {Harris}, {Hascall}, {Haupt}, {Hernandez}, {Herrmann}, {Hileman}, {Hoblitt}, {Hodgson}, {Hogan}, {Howard}, {Huang}, {Huffer}, {Ingraham}, {Innes}, {Jacoby}, {Jain}, {Jammes}, {Jee}, {Jenness}, {Jernigan}, {Jevremovi{\'c}}, {Johns}, {Johnson}, {Johnson}, {Jones}, {Juramy-Gilles}, {Juri{\'c}}, {Kalirai}, {Kallivayalil}, {Kalmbach}, {Kantor}, {Karst}, {Kasliwal}, {Kelly}, {Kessler}, {Kinnison}, {Kirkby}, {Knox}, {Kotov}, {Krabbendam}, {Krughoff}, {Kub{\'a}nek}, {Kuczewski}, {Kulkarni}, {Ku}, {Kurita}, {Lage}, {Lambert}, {Lange}, {Langton}, {Le Guillou}, {Levine}, {Liang}, {Lim}, {Lintott}, {Long}, {Lopez}, {Lotz}, {Lupton}, {Lust}, {MacArthur}, {Mahabal},
  {Mandelbaum}, {Markiewicz}, {Marsh}, {Marshall}, {Marshall}, {May}, {McKercher}, {McQueen}, {Meyers}, {Migliore}, {Miller}, \& {Mills}}]{2019ApJ...873..111I}
{Ivezi{\'c}}, {\v{Z}}., {Kahn}, S.~M., {Tyson}, J.~A., {et~al.} 2019, \apj, 873, 111, \dodoi{10.3847/1538-4357/ab042c}

\bibitem[{{Jiang} {et~al.}(2014){Jiang}, {Han}, \& {Li}}]{2014MNRAS.438..859J}
{Jiang}, D., {Han}, Z., \& {Li}, L. 2014, \mnras, 438, 859, \dodoi{10.1093/mnras/stt2252}

\bibitem[{{Jones} {et~al.}(2020){Jones}, {Conroy}, {Horvat}, {Giammarco}, {Kochoska}, {Pablo}, {Brown}, {Sowicka}, \& {Pr{\v{s}}a}}]{2020ApJS..247...63J}
{Jones}, D., {Conroy}, K.~E., {Horvat}, M., {et~al.} 2020, \apjs, 247, 63, \dodoi{10.3847/1538-4365/ab7927}

\bibitem[{Kippenhahn {et~al.}(2012)Kippenhahn, Weigert, \& Weiss}]{Kippenhahn2012}
Kippenhahn, R., Weigert, A., \& Weiss, A. 2012, Stellar Structure and Evolution, 2nd edn. (Springer), \dodoi{10.1007/978-3-642-30304-3}

\bibitem[{Langer(2012)}]{Langer2012}
Langer, N. 2012, Annual Review of Astronomy and Astrophysics, 50, 107, \dodoi{10.1146/annurev-astro-081811-125534}

\bibitem[{{Latkovi{\'c}} {et~al.}(2021){Latkovi{\'c}}, {{\v{C}}eki}, \& {Lazarevi{\'c}}}]{2021ApJS..254...10L}
{Latkovi{\'c}}, O., {{\v{C}}eki}, A., \& {Lazarevi{\'c}}, S. 2021, \apjs, 254, 10, \dodoi{10.3847/1538-4365/abeb23}

\bibitem[{{Lee} {et~al.}(2014){Lee}, {Koppenhoefer}, {Seitz}, {Bender}, {Riffeser}, {Kodric}, {Hopp}, {Snigula}, {G{\"o}ssl}, {Kudritzki}, {Burgett}, {Chambers}, {Hodapp}, {Kaiser}, \& {Waters}}]{2014ApJ...797...22L}
{Lee}, C.~H., {Koppenhoefer}, J., {Seitz}, S., {et~al.} 2014, \apj, 797, 22, \dodoi{10.1088/0004-637X/797/1/22}

\bibitem[{{Li} {et~al.}(2022{\natexlab{a}}){Li}, {Qian}, {Jiao}, \& {Ma}}]{2022ApJ...932...14L}
{Li}, F.~X., {Qian}, S.~B., {Jiao}, C.~L., \& {Ma}, W.~W. 2022{\natexlab{a}}, \apj, 932, 14, \dodoi{10.3847/1538-4357/ac6c81}

\bibitem[{{Li} {et~al.}(2022{\natexlab{b}}){Li}, {Gao}, {Liu}, {Gao}, {Li}, {Chen}, \& {Sun}}]{2022AJ....164..202L}
{Li}, K., {Gao}, X., {Liu}, X.-Y., {et~al.} 2022{\natexlab{b}}, \aj, 164, 202, \dodoi{10.3847/1538-3881/ac8ff2}

\bibitem[{{Li} {et~al.}(2020){Li}, {Kim}, {Xia}, {Michel}, {Hu}, {Gao}, {Guo}, \& {Chen}}]{2020AJ....159..189L}
{Li}, K., {Kim}, C.-H., {Xia}, Q.-Q., {et~al.} 2020, \aj, 159, 189, \dodoi{10.3847/1538-3881/ab7cda}

\bibitem[{{Li} \& {Wang}(2025)}]{2025ApJS..277...51L}
{Li}, K., \& {Wang}, L.-H. 2025, \apjs, 277, 51, \dodoi{10.3847/1538-4365/adba63}

\bibitem[{{Li} {et~al.}(2021{\natexlab{a}}){Li}, {Xia}, {Kim}, {Gao}, {Hu}, {Guo}, {Gao}, {Chen}, \& {Guo}}]{2021AJ....162...13L}
{Li}, K., {Xia}, Q.-Q., {Kim}, C.-H., {et~al.} 2021{\natexlab{a}}, \aj, 162, 13, \dodoi{10.3847/1538-3881/abfc53}

\bibitem[{{Li} {et~al.}(2024){Li}, {Gao}, {Guo}, {Gao}, {Chen}, {Wang}, {Xin}, {Han}, {Kim}, \& {Jeong}}]{2024A&A...692L...4L}
{Li}, K., {Gao}, X., {Guo}, D.-F., {et~al.} 2024, \aap, 692, L4, \dodoi{10.1051/0004-6361/202451947}

\bibitem[{{Li} {et~al.}(2021{\natexlab{b}}){Li}, {Riess}, {Busch}, {Casertano}, {Macri}, \& {Yuan}}]{2021ApJ...920...84L}
{Li}, S., {Riess}, A.~G., {Busch}, M.~P., {et~al.} 2021{\natexlab{b}}, \apj, 920, 84, \dodoi{10.3847/1538-4357/ac1597}

\bibitem[{{Li} \& {White}(2008)}]{2008MNRAS.384.1459L}
{Li}, Y.-S., \& {White}, S. D.~M. 2008, \mnras, 384, 1459, \dodoi{10.1111/j.1365-2966.2007.12748.x}

\bibitem[{{LSST Science Collaboration} {et~al.}(2009){LSST Science Collaboration}, {Abell}, {Allison}, {Anderson}, {Andrew}, {Angel}, {Armus}, {Arnett}, {Asztalos}, {Axelrod}, {Bailey}, {Ballantyne}, {Bankert}, {Barkhouse}, {Barr}, {Barrientos}, {Barth}, {Bartlett}, {Becker}, {Becla}, {Beers}, {Bernstein}, {Biswas}, {Blanton}, {Bloom}, {Bochanski}, {Boeshaar}, {Borne}, {Bradac}, {Brandt}, {Bridge}, {Brown}, {Brunner}, {Bullock}, {Burgasser}, {Burge}, {Burke}, {Cargile}, {Chandrasekharan}, {Chartas}, {Chesley}, {Chu}, {Cinabro}, {Claire}, {Claver}, {Clowe}, {Connolly}, {Cook}, {Cooke}, {Cooray}, {Covey}, {Culliton}, {de Jong}, {de Vries}, {Debattista}, {Delgado}, {Dell'Antonio}, {Dhital}, {Di Stefano}, {Dickinson}, {Dilday}, {Djorgovski}, {Dobler}, {Donalek}, {Dubois-Felsmann}, {Durech}, {Eliasdottir}, {Eracleous}, {Eyer}, {Falco}, {Fan}, {Fassnacht}, {Ferguson}, {Fernandez}, {Fields}, {Finkbeiner}, {Figueroa}, {Fox}, {Francke}, {Frank}, {Frieman}, {Fromenteau}, {Furqan}, {Galaz}, {Gal-Yam}, {Garnavich},
  {Gawiser}, {Geary}, {Gee}, {Gibson}, {Gilmore}, {Grace}, {Green}, {Gressler}, {Grillmair}, {Habib}, {Haggerty}, {Hamuy}, {Harris}, {Hawley}, {Heavens}, {Hebb}, {Henry}, {Hileman}, {Hilton}, {Hoadley}, {Holberg}, {Holman}, {Howell}, {Infante}, {Ivezic}, {Jacoby}, {Jain}, {R}, {Jedicke}, {Jee}, {Garrett Jernigan}, {Jha}, {Johnston}, {Jones}, {Juric}, {Kaasalainen}, {Styliani}, {Kafka}, {Kahn}, {Kaib}, {Kalirai}, {Kantor}, {Kasliwal}, {Keeton}, {Kessler}, {Knezevic}, {Kowalski}, {Krabbendam}, {Krughoff}, {Kulkarni}, {Kuhlman}, {Lacy}, {Lepine}, {Liang}, {Lien}, {Lira}, {Long}, {Lorenz}, {Lotz}, {Lupton}, {Lutz}, {Macri}, {Mahabal}, {Mandelbaum}, {Marshall}, {May}, {McGehee}, {Meadows}, {Meert}, {Milani}, {Miller}, {Miller}, {Mills}, {Minniti}, {Monet}, {Mukadam}, {Nakar}, {Neill}, {Newman}, {Nikolaev}, {Nordby}, {O'Connor}, {Oguri}, {Oliver}, {Olivier}, {Olsen}, {Olsen}, {Olszewski}, {Oluseyi}, {Padilla}, {Parker}, {Pepper}, {Peterson}, {Petry}, {Pinto}, {Pizagno}, {Popescu}, {Prsa}, {Radcka}, {Raddick},
  {Rasmussen}, {Rau}, {Rho}, {Rhoads}, {Richards}, {Ridgway}, {Robertson}, {Roskar}, {Saha}, {Sarajedini}, {Scannapieco}, {Schalk}, {Schindler}, \& {Schmidt}}]{2009arXiv0912.0201L}
{LSST Science Collaboration}, {Abell}, P.~A., {Allison}, J., {et~al.} 2009, arXiv e-prints, arXiv:0912.0201, \dodoi{10.48550/arXiv.0912.0201}

\bibitem[{{Lucy}(1967)}]{1967ZA.....65...89L}
{Lucy}, L.~B. 1967, \zap, 65, 89

\bibitem[{Maeder(2009)}]{Maeder2009}
Maeder, A. 2009, Physics, Formation and Evolution of Rotating Stars (Springer), \dodoi{10.1007/978-3-540-76949-1}

\bibitem[{{Marchant} \& {Bodensteiner}(2024)}]{2024ARA&A..62...21M}
{Marchant}, P., \& {Bodensteiner}, J. 2024, \araa, 62, 21, \dodoi{10.1146/annurev-astro-052722-105936}

\bibitem[{{Massey} {et~al.}(2016){Massey}, {Neugent}, \& {Smart}}]{2016AJ....152...62M}
{Massey}, P., {Neugent}, K.~F., \& {Smart}, B.~M. 2016, \aj, 152, 62, \dodoi{10.3847/0004-6256/152/3/62}

\bibitem[{{Menzel} {et~al.}(2023){Menzel}, {Davis}, {Parrish}, {Lawrence}, {Stewart}, {Cooper}, {Irish}, {Mosier}, {Levine}, {Pitman}, {Walsh}, {Maghami}, {Thomson}, {Wooldridge}, {Boucarut}, {Feinberg}, {Turner}, {Kalia}, \& {Bowers}}]{2023PASP..135e8002M}
{Menzel}, M., {Davis}, M., {Parrish}, K., {et~al.} 2023, \pasp, 135, 058002, \dodoi{10.1088/1538-3873/acbb9f}

\bibitem[{Mockus(1974)}]{mockus1975bayesian}
Mockus, J. 1974, Kybernetes, 3, 103, \dodoi{10.1108/eb005359}

\bibitem[{{Nelson}(2021)}]{2021NewA...8601565N}
{Nelson}, R.~H. 2021, \na, 86, 101565, \dodoi{10.1016/j.newast.2020.101565}

\bibitem[{{O'Connell}(1951)}]{1951PRCO....2...85O}
{O'Connell}, D.~J.~K. 1951, Publications of the Riverview College Observatory, 2, 85

\bibitem[{{Pecaut} \& {Mamajek}(2013)}]{2013ApJS..208....9P}
{Pecaut}, M.~J., \& {Mamajek}, E.~E. 2013, \apjs, 208, 9, \dodoi{10.1088/0067-0049/208/1/9}

\bibitem[{{Pellouin} {et~al.}(2025){Pellouin}, {Dvorkin}, \& {Lehoucq}}]{2025A&A...693A.283P}
{Pellouin}, C., {Dvorkin}, I., \& {Lehoucq}, L. 2025, \aap, 693, A283, \dodoi{10.1051/0004-6361/202450422}

\bibitem[{{Poro} {et~al.}(2024){Poro}, {Li}, {Michel}, {Castro}, {Fern{\'a}ndez Laj{\'u}s}, {Wang}, {Coliac}, {Alada{\u{g}}}, {Alizadehsabegh}, \& {Alicavus}}]{2024AJ....168..272P}
{Poro}, A., {Li}, K., {Michel}, R., {et~al.} 2024, \aj, 168, 272, \dodoi{10.3847/1538-3881/ad8345}

\bibitem[{{Poro} {et~al.}(2025){Poro}, {Li}, {Paki}, {Baudart}, {Michel}, {Wang}, {Fern{\'a}ndez Laj{\'u}s}, {Alicavus}, {Foschino}, {Aceves}, {Tamayo}, \& {Chavez}}]{2025MNRAS.537.3160P}
{Poro}, A., {Li}, K., {Paki}, E., {et~al.} 2025, \mnras, 537, 3160, \dodoi{10.1093/mnras/staf222}

\bibitem[{{Pr{\v{s}}a} \& {Zwitter}(2005)}]{2005ApJ...628..426P}
{Pr{\v{s}}a}, A., \& {Zwitter}, T. 2005, \apj, 628, 426, \dodoi{10.1086/430591}

\bibitem[{{Pr{\v{s}}a} {et~al.}(2016){Pr{\v{s}}a}, {Conroy}, {Horvat}, {Pablo}, {Kochoska}, {Bloemen}, {Giammarco}, {Hambleton}, \& {Degroote}}]{2016ApJS..227...29P}
{Pr{\v{s}}a}, A., {Conroy}, K.~E., {Horvat}, M., {et~al.} 2016, \apjs, 227, 29, \dodoi{10.3847/1538-4365/227/2/29}

\bibitem[{Rasmussen \& Williams(2006)}]{rasmussen2006gaussian}
Rasmussen, C.~E., \& Williams, C. K.~I. 2006, Gaussian Processes for Machine Learning (MIT Press), \dodoi{10.7551/mitpress/3206.001.0001}

\bibitem[{{Riess} {et~al.}(2016){Riess}, {Macri}, {Hoffmann}, {Scolnic}, {Casertano}, {Filippenko}, {Tucker}, {Reid}, {Jones}, {Silverman}, {Chornock}, {Challis}, {Yuan}, {Brown}, \& {Foley}}]{2016ApJ...826...56R}
{Riess}, A.~G., {Macri}, L.~M., {Hoffmann}, S.~L., {et~al.} 2016, \apj, 826, 56, \dodoi{10.3847/0004-637X/826/1/56}

\bibitem[{{Ruci{\'n}ski}(1973)}]{1973AcA....23...79R}
{Ruci{\'n}ski}, S.~M. 1973, \actaa, 23, 79

\bibitem[{{Schlafly} \& {Finkbeiner}(2011)}]{2011ApJ...737..103S}
{Schlafly}, E.~F., \& {Finkbeiner}, D.~P. 2011, \apj, 737, 103, \dodoi{10.1088/0004-637X/737/2/103}

\bibitem[{Shahriari {et~al.}(2016)Shahriari, Swersky, Wang, Adams, \& de~Freitas}]{7352306}
Shahriari, B., Swersky, K., Wang, Z., Adams, R.~P., \& de~Freitas, N. 2016, Proceedings of the IEEE, 104, 148, \dodoi{10.1109/JPROC.2015.2494218}

\bibitem[{{Shokry} {et~al.}(2018){Shokry}, {Saad}, {Hamdy}, {Beheary}, {Abolazm}, {Gadallah}, {El-Depsey}, \& {Al-Gazzar}}]{2018NewA...59....8S}
{Shokry}, A., {Saad}, S.~M., {Hamdy}, M.~A., {et~al.} 2018, \na, 59, 8, \dodoi{10.1016/j.newast.2017.08.005}

\bibitem[{{Stellingwerf}(1978)}]{1978ApJ...224..953S}
{Stellingwerf}, R.~F. 1978, \apj, 224, 953, \dodoi{10.1086/156444}

\bibitem[{{Stepien}(1995)}]{1995MNRAS.274.1019S}
{Stepien}, K. 1995, \mnras, 274, 1019, \dodoi{10.1093/mnras/274.4.1019}

\bibitem[{{Terrell} \& {Wilson}(2005)}]{2005Ap&SS.296..221T}
{Terrell}, D., \& {Wilson}, R.~E. 2005, \apss, 296, 221, \dodoi{10.1007/s10509-005-4449-4}

\bibitem[{{von Zeipel}(1924)}]{1924MNRAS..84..665V}
{von Zeipel}, H. 1924, \mnras, 84, 665, \dodoi{10.1093/mnras/84.9.665}

\bibitem[{{Williams} {et~al.}(2017){Williams}, {Dolphin}, {Dalcanton}, {Weisz}, {Bell}, {Lewis}, {Rosenfield}, {Choi}, {Skillman}, \& {Monachesi}}]{2017ApJ...846..145W}
{Williams}, B.~F., {Dolphin}, A.~E., {Dalcanton}, J.~J., {et~al.} 2017, \apj, 846, 145, \dodoi{10.3847/1538-4357/aa862a}

\bibitem[{{Wilsey} \& {Beaky}(2009)}]{2009SASS...28..107W}
{Wilsey}, N.~J., \& {Beaky}, M.~M. 2009, Society for Astronomical Sciences Annual Symposium, 28, 107

\bibitem[{{Zechmeister} \& {K{\"u}rster}(2018)}]{2018ascl.soft07019Z}
{Zechmeister}, M., \& {K{\"u}rster}, M. 2018, {GLS: Generalized Lomb-Scargle periodogram}, Astrophysics Source Code Library, record ascl:1807.019

\end{thebibliography}
\bibliographystyle{aasjournal}



\end{document}